\documentclass[12pt,preprint]{aastex}

\slugcomment{To appear in the Astrophysical Journal}

\shorttitle{Interstellar Grain Alignment}
\shortauthors{Andersson & Potter}

\begin{document}


\title{Observational Constraints on Interstellar Grain Alignment}


\author{B-G Andersson}
\affil{Center for Astrophysical Sciences, The Johns Hopkins University,
 3400 N. Charles Street, Baltimore, MD 21218}
\email{bg@pha.jhu.edu}

\and

\author{S.B. Potter}
\affil{South African Astronomical Observatory, PO Box 9, Observatory 7935, Cape Town, South Africa}
\email{sbp@saao.ac.za}

\begin{abstract}
We present new multicolor photo-polarimetry of stars behind the Southern Coalsack.  Analyzed together with multiband polarization data from the literature, probing the Chamaeleon I, Musca, $\rho$~Opiuchus, R~CrA and Taurus clouds, we show that the wavelength of maximum polarization ($\lambda_{max}$) is linearly correlated with the radiation environment of the grains.  Using Far-Infrared emission data, we show that the large scatter seen in previous studies of $\lambda_{max}$ as a function of A$_V$ is primarily due to line of sight effects causing some A$_V$ measurements to not be a good tracer of the extinction (radiation field strength) seen by the grains being probed.

  The derived slopes in $\lambda_{max}$ vs. A$_V$, for the individual clouds, are consistent with a common value, while the zero intercepts scale with the average values of the ratios of total-to-selective extinction (R$_V$) for the individual clouds.  Within each cloud we do not find direct correlations between $\lambda_{max}$ and R$_V$.  

The positive slope in consistent with recent developments in theory and indicating alignment driven by the radiation field.  The present data cannot conclusively differentiate between direct radiative torques and alignment driven by H$_2$ formation.  However, the small values of $\lambda_{max}$(A$_V$=0), seen in several clouds, suggest a role for the latter, at least at the cloud surfaces.
 
The scatter in the $\lambda_{max}$ vs. A$_V$ relation is found to be associated with the characteristics of the embedded Young Stellar Objects (YSO) in the clouds.  We propose that this is partially due to locally increased plasma damping of the grain rotation caused by X-rays from the YSOs.

\end{abstract}



\section{Introduction}

Starlight seen through the interstellar medium usually ends up being slightly polarized (at the level of up to a few percent), even when the background source is not.  This interstellar polarization was first detected independently by J. Hall and W.A. Hiltner \citep{hall1949,hiltner1949a}.  Already \citet{hiltner1949b} suggested that the polarization was due to interstellar dust interacting with the magnetic field.
  Theoretical models attempting to explain interstellar polarization soon followed with the first quantitative one by \citet{davis1951} who identified grain alignment by the magnetic field through paramagnetic dissipation, as the origin of the polarization.
  This identification of dichroic extinction as the origin of the polarization has remained as the prime candidate up until this time, although the details of the mechanism has been modified by many authors over the intervening half century (see \citet{lazarian2003} for a recent review).
  For instance, while \citet{davis1951} assumed paramagnetic grains, spun up by gas-grain collisions, it is now recognized that this combination will not suffice to keep the grains spun up against the damping influence of those same gas-grain collisions.  The grains have to have either a stronger magnetic moment (super-paramagnetic grains \citep{mathis1986}) or be spun up to rotational velocities well above the thermal energy of the gas (suprathermal rotation \citep{purcell1979}).

The magnetic relaxation causing the rapidly spinning grain to align its axis of angular momentum with the local direction of the magnetic field is well understood and based on solid state and nuclear physics \citep{purcell1979,lazarian1999}.
  The remaining poorly understood link in the explanatory chain to understand the origin of interstellar polarization is the mechanism of grain spin-up.  Several mechanisms have been proposed on theoretical grounds, including the energy released from molecular hydrogen formation on the surface of the dust grain \citep{purcell1979} and direct torques from an anisotropic radiation field \citep{dolginov1976,weingartner2003,cho2005}.

It is generally agreed that, for the large grains involved in the polarization, the damping of the grain spin, in neural gas, is due to gas-grain collisions \citep{draine1998}.  For smaller grains or in regions with more extreme conditions, radiation from the grains or other collisional partners can dominate \citep{draine1998}.  Even so, the grain neutral damping is often used as a benchmark.  A damping time can be defined such that it corresponds to the time it takes for a grain to collide with its own mass in gas particles.  If we define the effective radius, \textit{a}, as that which a grain of a given volume would have if it were spherical we find that \citep{whittet1992}:

\begin{equation}
\label{damping_equ}
t_{damping}\propto\frac{a}{n\sqrt{T}}
\end{equation}

where n and T are the gas number density and temperature, respectively.  Hence, smaller grains have their rotation damped out more rapidly, and are therefore the hardest to keep spinning.  A hotter and/or denser gas also dampens the grain rotation faster.  Therefore, the size distribution of aligned grains can be used as a probe of the alignment mechanism.

The first attempts at determining the wavelength dependence of interstellar polarization (e.g. \citet{hiltner1949b}) were inconclusive. \citet{davis1951} made a qualitative prediction of the wavelength dependence of interstellar polarization, later expanded on by \citet{davis1959}.  Isolated single wavelength observations hinting at such variations were indirectly reported by  Str\"{o}mgren in the 1954-1955 annual report of the Yerkes and McDonald Observatories where he notes that Hiltner had made observations at 1 and 2 $\mu$m, showing a steep drop in polarization compared to "the blue" \citep{stromgren1956}.  
The first systematic attempt at measuring the "polarization curve" to be published was undertaken by A. Behr at G\"{o}ttingen in 1958, at the prodding of L. Davis, Jr. \citep{behr1959}.  While wavelength dependence was indeed detected for a few stars, these observations only covered the wavelength range 0.37 - 0.51 $\mu$m.  The first observations to span and locate the peak of the polarization curve were made by \citet{gehrels1960}.

When observed over the range of the optical and near-infrared spectrum, interstellar polarization takes on a characteristic wavelength dependence, which can be parameterized through the relation:

\begin{equation}
p(\lambda)/p_{max}\ =\ exp[-K\ ln^2(\lambda_{max}/\lambda)]
\end{equation}

usually referred to as the Serkowski relation \citep{serkowski1973}, if K is set to the fixed value 1.15, or the Wilking relation \citep{wilking1980} if K is used as a fitting parameter.  \citet{codina1976} first suggested that K should be related to the size of the dust grains and results from \citet{wilking1982} and \citet{whittet1992} indicate that K and $\lambda_{max}$ are likely correlated and thus not independent parameters.  (See \citet{whittet1992} for an excellent review of the development of this parameterization)

As has been shown by \citet{kim1994a,kim1995} the shape and variability of the polarization curve can be understood in terms of the size distribution of aligned grains and thus $\lambda_{max}$ provides a measure of the average size of the aligned grains.
  As noted by \citet{kim1995} only a very small fraction of the grains are likely to be aligned and hence a degeneracy exists when using $\lambda_{max}$ as a probe of grain alignment between the total grain size distribution and the fractional alignment in each size bin.
  This degeneracy is evident from the observational correlation of $\lambda_{max}$ with R$_V$, the ratio of total-to-selective extinction, first noted by \citet{serkowski1968} (cf \citet{serkowski1975, whittet1978}), and is to be expected since, as shown by \citet{kim1994b}, R$_V$ traces the size distribution of the total grain population.
  As has been shown by e.g. \citet{vrba1984,vrba1993,whittet2001}, R$_V$ is in turn, in general, correlated with the visual extinction, A$_V$.  This is usually interpreted as being due to grain coagulation causing the average grain size to increase at larger depths into the cloud \citep{bernard1993,whittet2001,wurm2002}.

To break these degeneracies it is important to measure each parameter to high precision and to seek out regions where the grain size distribution and alignment conditions vary over relatively small scales.  The latter should, particularly if radiative processes drive the alignment, most likely be found in the outer parts of the clouds where the radiation field seen by the grains varies relatively rapidly.
  For the former, we need regions where sightlines at similar observed visual extinctions show significant differences in the grain size distributions.

A further complication in studying the variation of polarization is introduced by possible changes in strength or orientation of the magnetic field.  To mitigate this concern it is important to use line of sight samples reliably constrained to probe single cloud.

While the visual extinction is a convenient and straightforward probe of the amount of material along the line of sight, it is important to remember that the line-of-sight extinction may not be a good probe of the relative radiation field seen by the material being probed, particularly at modest values of A$_V$.
  The three dimensional geometry of the interstellar cloud, its relation to surrounding clouds and stars, and possible clumpiness of the material can cause the line-of-sight extinction to either over- or under-estimate the "effective extinction", here defined as the minimum opacity, vis-\'{a}-vis the diffuse interstellar radiation field, experienced by the material on the line of sight.  As we will see below, this is an important consideration in interpreting optical polarimetry.

The Southern Coalsack is a good target for studying the wavelength dependence of interstellar polarization and other absorption-based probes.  The cloud is isolated and is well located in three dimensions.
  Its location, straddling the Galactic mid-plane also guarantees a large number of background sources, whether hot stars are required or not.  As shown by \citet{bga2004} the outer parts of the cloud show a wide range of values of R$_V$.
  We have argued that this likely reflects clumpy dust destruction in the cloud envelope caused by the hot gas in the Upper Centaurus-Lupus Super bubble which envelopes the cloud.  However, as we will discuss below, for emission tracers, and in particular continuum emission, the location of the Coalsack in the Galactic plane can be a major disadvantage, since foreground and background emission may be difficult, or impossible, to tell apart.

For the purpose of the current study the important diffuse emission is the far-infrared light attributable to dust grains heated by the interstellar radiation field.  Several authors have used color temperatures based on the IRAS 60$\mu$m-to-100$\mu$m ratio to show that observationally this ratio provides a tracer of the radiation field impinging on the dust clouds (e.g. \citet{langer1989}(B5), \citet{snell1989} (B18 and Heiles' Cloud 2) and \citet{jarrett1989} (Ophiuchus)).  While it is likely that the analyses in these studies are only qualitatively valid due to the admixture of "large" and very small grains, and might not provide accurate absolute temperatures, the usage of the 60$\mu$m-to-100$\mu$m ratio to trace the damping of the ISRF at different depths of a cloud is not in serious doubt (see e.g. \citet{draine2007}).  In appendix \ref{irrapp} we use data from the clouds under study here to further support this usage.

For those areas and line of sight where the interstellar extinction is dominated by a single cloud, or cloud complex (such as on high Galactic latitude sightlines) we can be relatively sure that the visual extinction and FIR emission are both caused by the same material.
  However, for low Galactic latitude clouds, where background emission is significant, and clouds with significant (high mass) star formation, where the internally generated radiation field is comparable to the ISRF, the correlation of the visual extinction and FIR emission can be expected to break down.

In this study we have therefore complemented new multi-band polarimetry of the Coalsack with archival polarimetry and other supporting data for five additional near-by interstellar clouds: Chamaeleon I (henceforth: Chamaeleon), Musca, $\rho$ Ophiuchus (henceforth: Ophiuchus), R~Corona Australis (henceforth: R~CrA) and Taurus to address the alignment mechanism of interstellar grains.

The remainder of this paper is organized as follows:  First we present our new observations of the Southern Coalsack.  In section \ref{datasec} we present and discuss the analysis of the polarimetry for both the new Coalsack data and the reanalysis of the polarization data for the other five clouds, and supporting data consisting of optical, near and far-infrared (FIR) photometry.  We show that the wavelength of maximum polarization is correlated with the visual extinction, albeit with many outliers.  Section \ref{cartoonsec} proposes that the outliers in both the FIR and polarimetry data are due to cloud geometry and/or the presence of embedded - or nearby - stars making the observed visual extinction a poor tracer of the radiation field seen by the dust.  Section \ref{FIRsec} shows that the FIR emission is anti-correlated with the visual extinction and that most of the outliers in the polarization plots can indeed be identified as outliers also in the FIR plots.  This is expanded on in Appendix \ref{irrapp}.  We then use the I(60$\mu$m)/I(100$\mu$m) vs. A$_V$ relations to identify those sightlines where this is the case.  Section \ref{screen} reanalyzes the $\lambda_{max}$ vs. A$_V$ relations for the different clouds after the screening performed in section \ref{FIRsec} and we find that there are now tight correlations for four of the six clouds.  No such correlations are seen for $\lambda_{max}$ vs. R$_V$.  However, the y-axis intercept for $\lambda_{max}$ at A$_V$=0 is correlated with $<R_V>$.  Section \ref{sfr} discusses the dispersion seen in the $\lambda_{max}$ vs. A$_V$ relation in terms of the characteristics of the embedded young stellar objects in the clouds.

\section{Observations and Data Reduction}

We used the University of Cape Town Polarimeter (UCTP; \citet{cropper1985}) on the 1.9m telescopes of the South African Astronomical Observatory (SAAO), during 2005 April 6-8 \& 13 to perform multi-band observations of interstellar polarization of stars background to the Southern Coalsack.  The UCTP was configured in the simultaneous linear and circular polarimetry mode, with a RCA31034A GaAs photomultiplier as the detector Measurements were performed using UBVRI filters in the Kron-Cousin system.
  We used HD~94851 and HD~98161 \citep{turnshek1990} as unpolarized standard stars and HD~155197, HD~154445 \citep{schmidt1992}, HD~93632 \citep{marraco1993} as polarized standard stars.  At least one star from each class was observed each night.  Measurements of the sky polarizations were acquired for each star immediately prior to and following the main observation.  The data were reduced using a custom software package \citep{cropper1985}.  None of the stars showed a statistically significant circular polarization in any band.  Target star information is given in Table \ref{CS_stars}.
Calibrated polarization for the Coalsack targets are summarized in Table \ref{pol_res}

\section{Analysis}

\subsection{Polarization and Spectro-Photometric Data\label{datasec}}

For each target in our sample we fitted the polarization data with both the Serkowski and Wilking relations.  We then performed a F-test \citep{lupton1993} to determine whether the additional parameter associated with the "Wilking" equation was justified.  Only for those data sets where the additional parameter was justified at the 90\% level did we accept it.  For those stars where it was not statistically justified we have left the column for "K" blank in the result tables.  The parameter uncertainties were calculated by the fitting routines and verified using Monte Carlo simulations of the fits (\citet{press1986}, p529ff).
  Importantly for the present study, the wavelength of maximum polarization was rarely affected beyond the one-sigma level by the assumed or fitted value of K.  This was also the case if we instead of a fixed K-value of 1.15 used the relation K=1.66$\lambda_{max}$+0.01 \citep{whittet1992}.

For all stars, the extinction parameters were recalculated and verified.  Visual photometry was, where available, extracted from the Tycho database \citep{tycho2000} while Near Infrared (NIR) photometry was extracted from the 2MASS survey, except where explicitly noted.  Spectral classification was checked and updated.  Intrinsic colors were estimated from \citet{cox2000} with uncertainties based on the reported uncertainties in the sources of the spectral classification and, where required, linear interpolation between the table entries.

\subsubsection{Coalsack Data}

\begin{figure}
\plotone{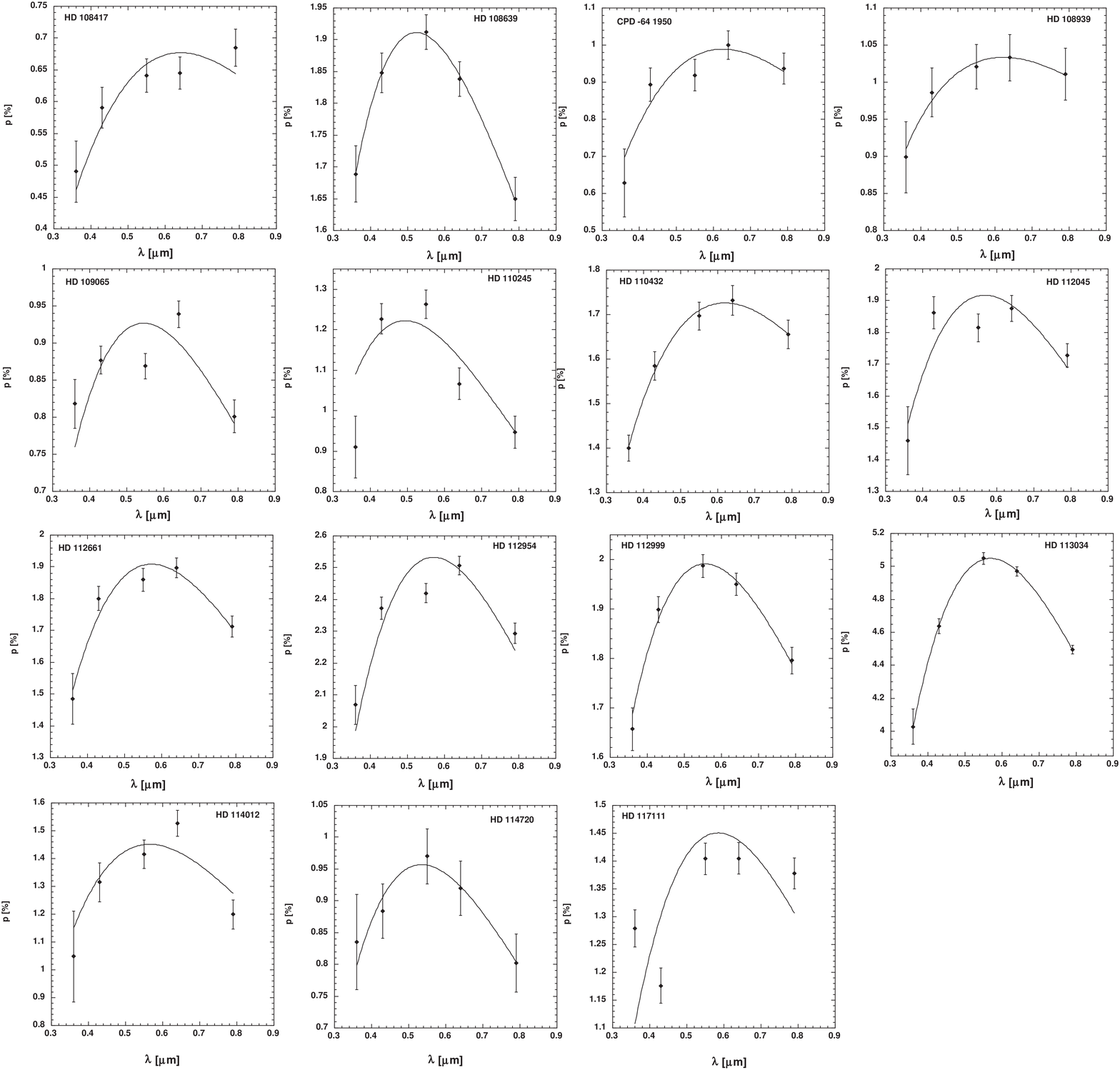}
\caption{Measured multi-band polarization for the Coalsack targets with best-fit polarization curve overlaid.  Polarization fit parameters are listed in Table \ref{pol_fit_CS}}
\label{CS_pol_curves}
\end{figure}

Table \ref{pol_fit_CS} list the best-fit parameters for the polarization curve fits for the Coalsack stars.  Since we do not have any Near Infrared polarimetry for the Coalsack stars, special care is needed to potentially justify the use of Wilking fits.  For these stars we show the reduced $\chi^2$ for the fits.  We note that while the F-test justifies the additional free parameter of the Wilking fit for several stars, in most cases the reduced $\chi^2$ for these fits are then less than unity.  The data and best-fit polarization curves are shown in Figure \ref{CS_pol_curves}.

\begin{figure}
\plotone{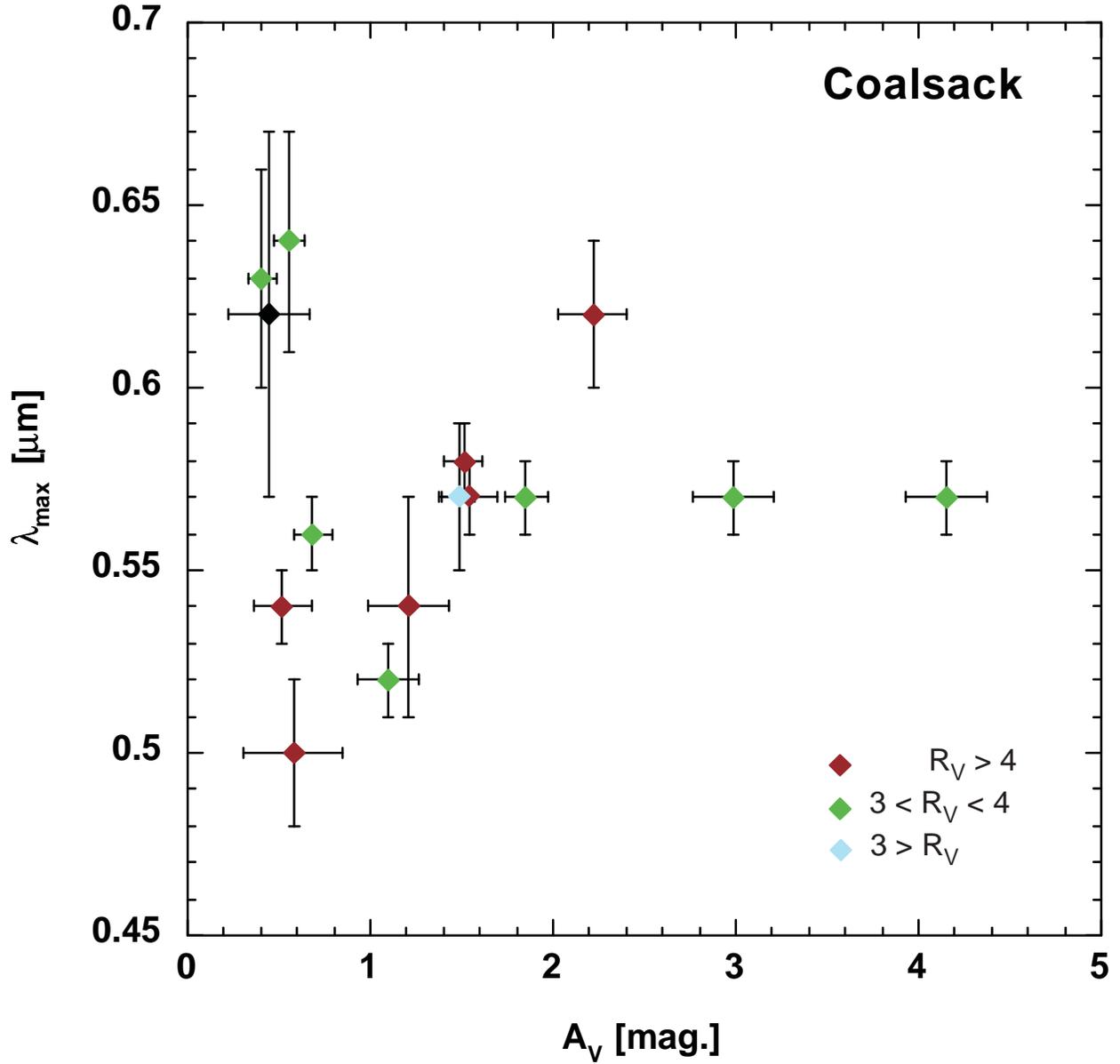}
\caption{ The wavelength of maximum polarization as a function of visual extinction for the newly observed stars behind the Southern Coalsack.  The data points have been color coded according to the measured values of the total-to-selective extinction along each line of sight.}
\label{CS_lmax_vs_av}
\end{figure}

For HD~110432, the 2MASS data is flagged as being of poor quality, we therefore used an average of the results from \citet{whittet1980}, \citet{dachs1982} and \citet{dachs1988} for this star.  In Figure \ref{CS_lmax_vs_av} we plot the location of the derived wavelength of maximum polarization ($\lambda_{max}$) as a function of visual extinction.  As noted above, earlier studies of the dependence of $\lambda_{max}$ on extinction parameters have found a relationship with the value of total-to-selective extinction (R$_V$; e.g. \citet{whittet1978}).  We do not find such a correlation for the Coalsack data.  In Figure \ref{CS_lmax_vs_av}, we have color-coded the sightlines according to their R$_V$ values.  While the two sightlines above A$_V$=2.5 show R$_V$ consistent of the value found at large visual extinction in this cloud (R$_V\approx$3.25), no systematic trends in R$_V$ are evident at smaller A$_V$.

\begin{figure}
\plotone{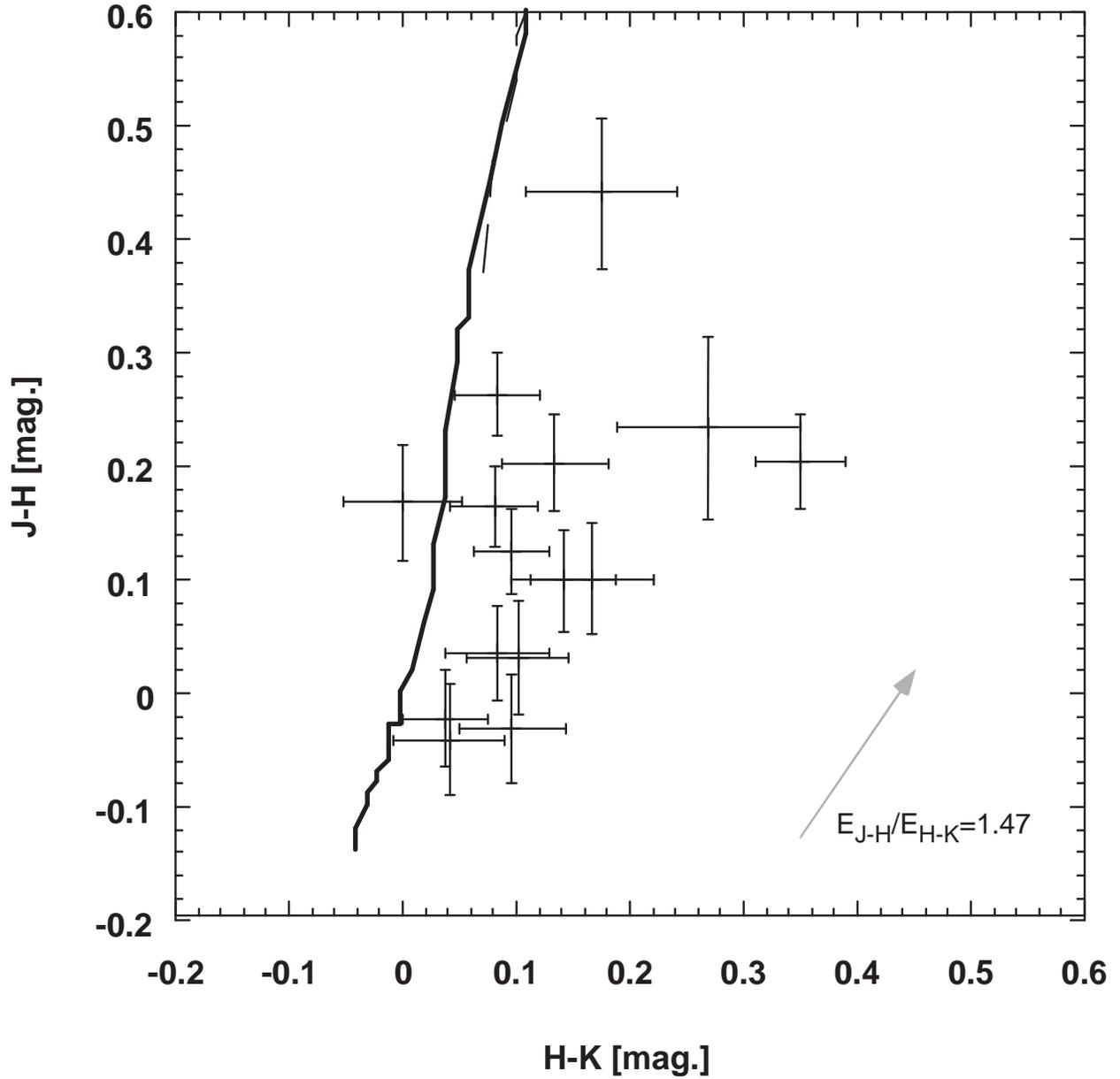}
\caption{The J-H vs. H-K color-color diagram for the Coalsack stars shows a consistent reddening for all stars (except HD~113034) with a reddening slope around 1.5 (see text).  The full drawn line shows the intrinsic colors of Main Sequence stars while the dashed line is for Giants.}
\label{JH_HK}
\end{figure}

\begin{figure}
\plotone{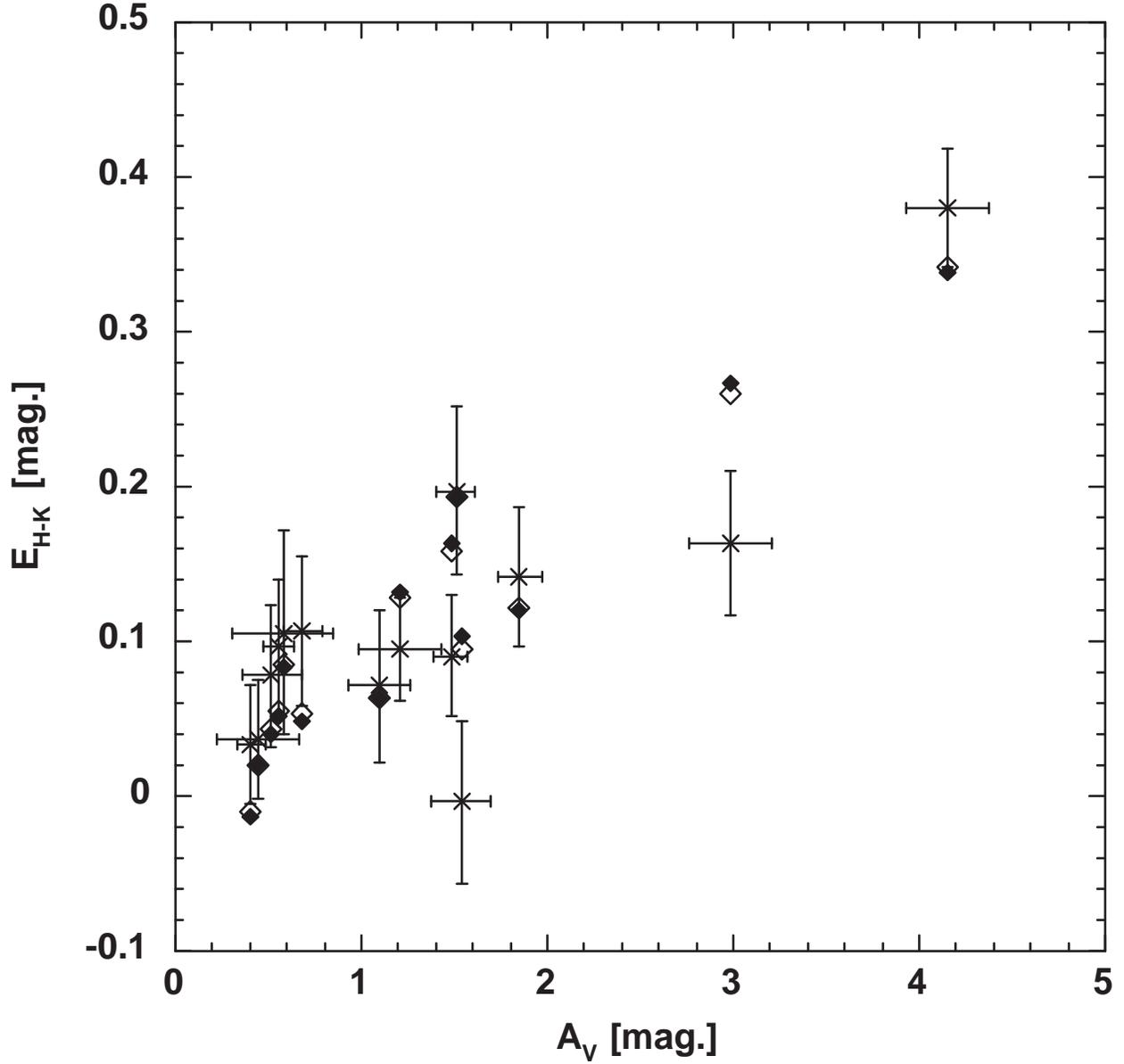}
\caption{The measured H-K color excess is plotted (x:es) together with the best fit color excesses assuming two reddening laws from the literature.  The largest offsets from the color excess for reddened intrinsic stellar colors are just above 2$\sigma$ and indicates at most minor contamination from circumstellar emission or binary companions.}
\label{Av_vs_E_HK}
\end{figure}

One possible caveat to the accuracy of the calculated R$_V$ values comes from the fact that several of the target stars are in some spectral classifications listed as either emission line stars or possible binaries.  To evaluate the possible influence of these complications for the Coalsack sample we used the J-H vs. H-K diagram, plotted in Figure \ref{JH_HK}.  Since the direction of the reddening vector in this color-color diagram varies between regions (e.g. \citet{racca2002}) we calculated the H-K color excess that, for each target gave the smallest offsets between measured colors and best-fit reddened colors, given the spectral classifications of the stars and direction of the reddening vector.  We used E$_{J-H}$/E$_{H-K}\approx$1.57 \citep{kenyon1998} and E$_{J-H}$/E$_{H-K}\approx$1.91 \citep{naoi2006} as test cases.  In Figure \ref{Av_vs_E_HK} we plot the H-K color excess versus the derived visual extinction.  The measured difference (H-K)-(H-K)$_0$ is plotted a x:es with error bars.  Open symbols represent the color excesses calculated assuming E$_{J-H}$/E$_{H-K}\approx$1.57, while filled diamonds represent the color excesses calculated assuming E$_{J-H}$/E$_{H-K}\approx$1.91.  Only for HD~112661 and HD~112045 are offsets of more than 2$\sigma$ seen between the measured and best-fit H-K color excesses (2.1 and 2.2$\sigma$, respectively for E$_{J-H}$/E$_{H-K}\approx$1.91).  While both of these stars are listed in SIMBAD as multiple, neither shows an exceptional R$_V$ value.  Even if these sightlines are suppressed, no correlation between $\lambda_{max}$ and R$_V$ is seen.

\subsubsection{Chamaeleon, Musca, Ophiuchus, R~CrA and Taurus Cloud Data}

Figure \ref{CS_lmax_vs_av} shows a tight correlation of $\lambda_{max}$ with A$_V$ over the limited range of A$_V\sim$1.0-2.5 but with outliers both at smaller and larger values of A$_V$.  A similar weak correlation was noted by \citet{whittet2001} in their Taurus data, but with a more pronounced scatter.
  To investigate whether the outlier points in the Coalsack plot are truly outliers in a real correlation, or whether the perceived correlation is instead a statistical fluke, we searched the literature for high quality, multi-wavelength polarimetry in sightlines penetrating well defined interstellar clouds.

\begin{figure}
\plotone{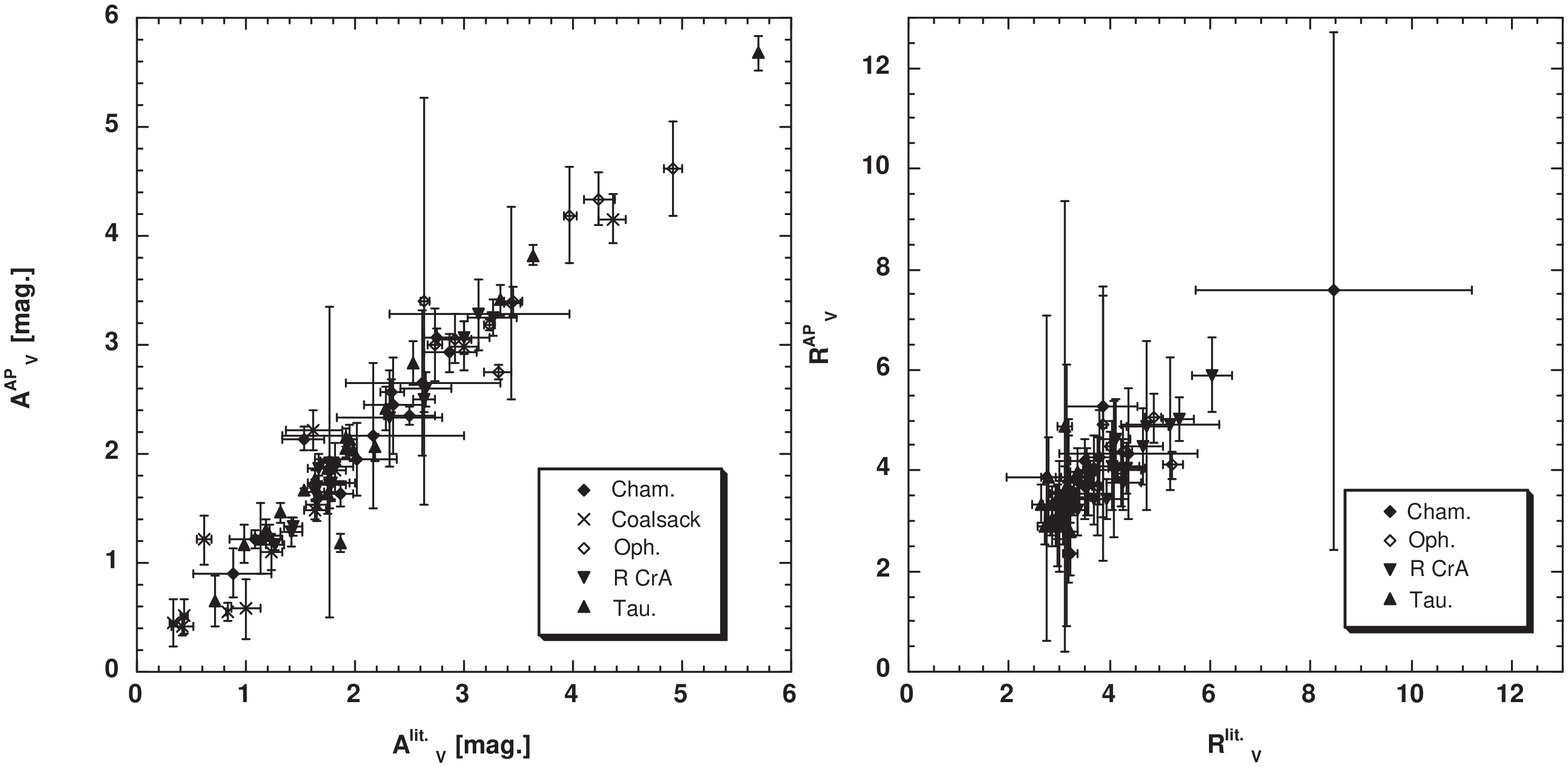}
\caption{The values of A$_V$ and R$_V$ derived here are compared with those extracted from the literature.  Data for Chamaeleon are from \citet{vrba1984}, for Ophiuchus from \citet{vrba1993}, for R~CrA from \citet{vrba1984} and for Taurus from \citet{whittet2001}.  \citet{arnal1993} did not publish visual extinctions and their quoted R$_V$ values were calculated from $\lambda_{max}$ fits, hence Musca is not included here.  Good agreement is seen for both parameters.}
\label{RvAv_us_vs_them}
\end{figure}

We selected five additional well-studied clouds with high quality published multi-wavelength polarimetry.  Polarimetry data were extracted from the studies by: Chamaeleon; \citet{whittet1992} \& \citet{covino1997}, Musca; \citet{arnal1993}, Ophiuchus; \citet{whittet1992} \& \citet{vrba1993}, R~CrA; \citet{whittet1992} and Taurus; \citet{whittet1992}, \citet{whittet2001}.
  To ensure as uniform a dataset as possible in term of photometry and stellar characteristics, we extracted visual (B,V) and near-infrared data from the Tycho and 2MASS archives.  We verified, and where possible updated the spectral classification, and then assigned intrinsic colors as for the Coalsack stars.  The resultant stellar characteristics are listed in Table \ref{stars_data}.  In a couple of cases unphysical values of - in particular - R$_V$ are encountered, likely indicating problems with the spectral classification.  These sightlines were therefore excluded from the subsequent analysis.  In Figure \ref{RvAv_us_vs_them} we compare the values of R$_V$ and A$_V$ derived here with those extracted from the literature.  For both parameters, good agreement is seen.  We note that the R$_V$ values quoted in \citet{arnal1993} are based on polarimetry and not on photometry and we have therefore not included them in these plots.

As for the Coalsack sightlines, we fitted Serkowski or Wilking functions to the polarization data, but selected only those stars with polarization measurements in at least 4 filters (for this reason we also chose not to include the Chamaeleon data from \citet{whittet1994}).  The polarization fit parameters are given in Table \ref{pol_fit}.  In most cases the $\lambda_{max}$ derived here agree with those in the original papers within 1$\sigma$ of the mutual uncertainties and in all but one cases within 2$\sigma$.  The one exception is the sight line towards HD~107875 in Musca (star number 7 in the nomenclature of \citet{arnal1993}), where the original paper reports $\lambda_{max}$=0.641$\pm$0.009~$\mu$m  while we find $\lambda_{max}$=0.49$\pm$0.01~$\mu$m using their stated polarization curve parametrization.  Given this discrepancy we have excluded this sightline form the analysis.

\begin{figure}
\plotone{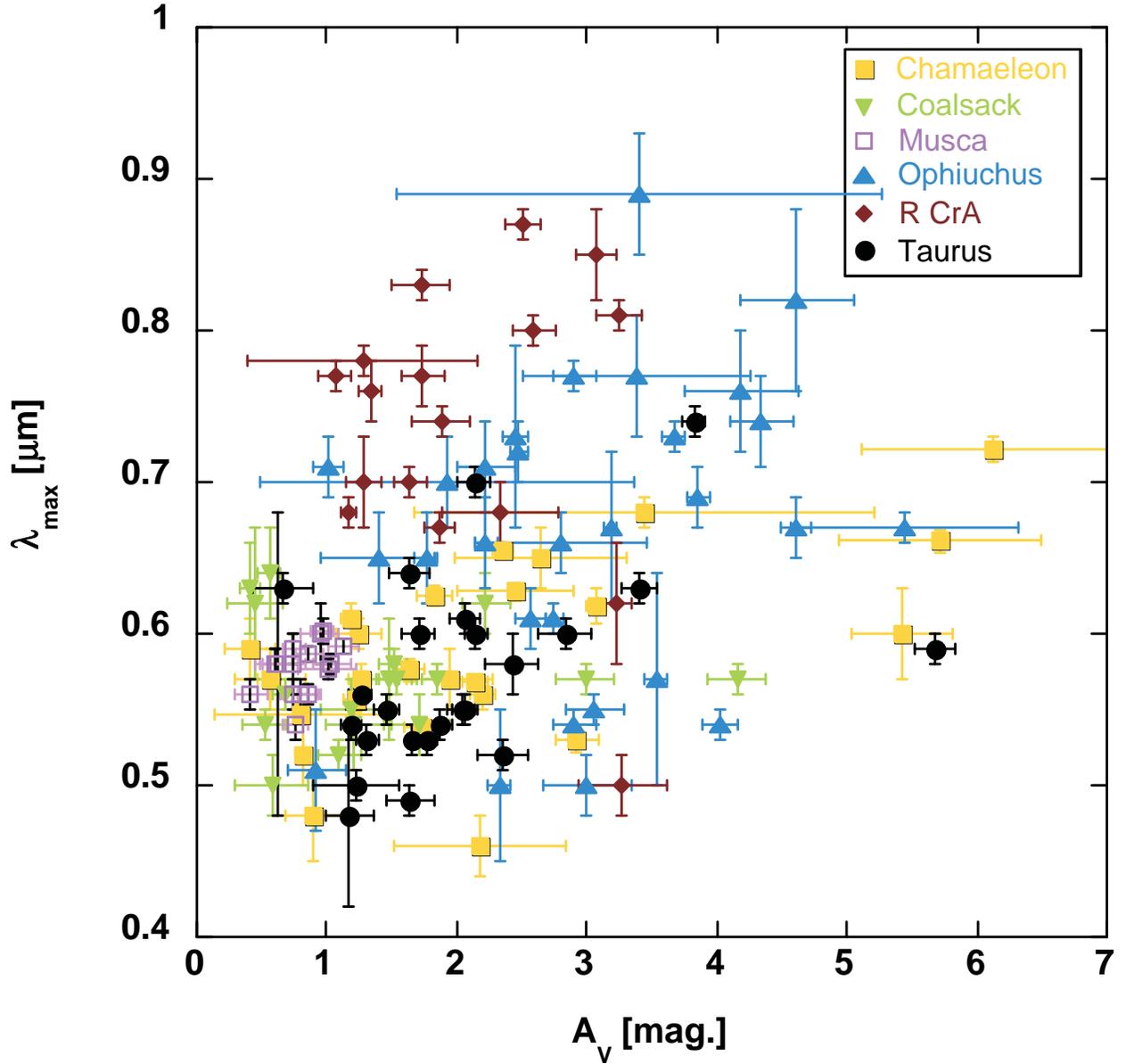}
\caption{The wavelength of maximum polarization as a function of visual extinction for the sightlines through the Chamaeleon, Musca, Ophiuchus, R~CrA and Taurus clouds.  For Chamaeleon Musca and Taurus a similar structure is seen in the distribution of points as that noted for the Coalsack.}
\label{pol_fig_other}
\end{figure}

Plotting $\lambda_{max}$ vs. A$_V$ for all six clouds (Figure \ref{pol_fig_other}), we see that, particularly for Chamaeleon, Musca and Taurus, a very similar structure is evident as for the Coalsack with a main grouping of point lying along what seems like a linear correlation from A$_V\sim$1 to $\sim$2.5 mag. but with outliers both at small and large A$_V$.  A critical issue, then, is whether the points we have here designated "outliers" are indeed that in a statistical sense, or whether, the suggested correlation of $\lambda_{max}$ with A$_V$ over approximately 1 to 3 magnitudes of extinction is illusory.

As noted above, two independent ways exist for estimating the extinction for a given parcel of gas and dust.  However, neither the directly measured visual extinction, nor the color-temperature of the FIR emission from the heated grains are immune to biases.  We will next discuss how the combination of the two tracers of extinction can be used to mitigate biasing in the determination of the effective opacity.

\subsection{Effective extinction vs. line-of-sight extinction\label{cartoonsec}}

The line-of-sight extinction might not be a good indicator of the extinction - or equivalently, radiation field - seen by the material sampled by an absorption (line or continuum) experiment  - the "effective extinction".  Figure \ref{cloud_cartoon} illustrates some of the ways in which such discrepancies might occur.  In this cartoon, we show a prolate cloud pointing towards the observer next to a roughly spherical cloud, with a "bridge" of material linking the two.  The gray zone is meant to illustrate the part of the cloud into which the interstellar radiation field penetrates.

For sightline 'A', which passes through the outskirts of the prolate cloud, the line of sight extinction is much larger than the effective extinction seen by the material probed.  The average radiation field seen by different parcels along the chord is similar and hence we'd expect this kind of sightline to show a large A$_V$ while retaining a relatively large 60/100 ratio.  A similar effect might arise if the sightline passes within the sphere of influence of an embedded, or nearby, stellar source contributing to the radiation field (A').

For sightline 'B' the line of sight extinction, again, is larger than that seen by the material, but here, the different parcels along the chord see very different radiation fields.  For this kind of sightline we would again expect a large A$_V$ but here - since the FIR radiation traces booth radiatively heated dust and dust in the dark part of the cloud - expect a low 60/100 ratio.

For sightline 'C', which passes through a region between clouds (or in an inter-clump region of a clumpy medium) the line of sight extinction instead underestimates the effective extinction experienced by the material due to shadowing effects by the surrounding clouds.

Finally, in sightline 'D' the observed and effective extinctions are similar and the different parcels of material along the chord experience similar radiation fields.

Similar effects and observational discrepancies might also arise in a clumpy medium with the sightlines passing through predominantly clump or inter-clump material.

\subsection{Far Infrared Data \label{FIRsec}}

Empirically an anti-correlation is seen between column density and dust temperature based on the 60-to-100 $\mu$m ratio in interstellar clouds (see, for instance figure 8 of \citet{snell1989}).  Since this ratio therefore indirectly traces the radiation field seen by the dust, we can use it as a probe of the "effective extinction" seen by the grain.  While this emission is also likely prone to biases, such as variations in the fraction of small grains, we show in Appendix \ref{irrapp} that we can use the combination of visual extinction and Far Infrared (FIR) emission to screen the data sets for anomalous sightlines.  We note that in this study we only do this in a relative sense for each cloud.  We are not attempting to find absolute dust temperatures.  While we will limit our analysis to the 60-to-100 $\mu$m ratio, we will use the 25-to-100 $\mu$m and 12-to-100 $\mu$m ratios to argue in Appendix \ref{irrapp} that the local changes seen in the relation between the 60-to-100 $\mu$m ratio and A$_V$ are indeed likely due to irradiation differences.

We extracted FIR data from the IRIS re-reduction \citep{IRIS2005} of the IRAS all-sky photometry.
  The spatial resolution of the IRIS maps are 3.8, 3.8, 4.0 \& 4.2 arc-minutes respectively for the 12, 25, 60 and 100$\mu$m bands, but the maps are pixelized on a 1.5' scale and hence over-sampled by about a factor of 2.5-2.8.  For the current study we used a 3x3 pixel average that allows us to both lessen the impact of any emission from the background star and to identify it, if significant.
  Comparing to the IRAS point (faint) source catalog, we find that, in most cases, the stars - when at all detectable - make only a minor contribution to the total FIR light on our lines of sight.  This is especially true of the longer wavelength bands.

Since the nominal, single pixel, photometric uncertainty in the IRIS maps is much smaller than the typical pixel-to-pixel variations (cf. \citet{IRIS2005}), we found that the standard deviation in the 3x3 pixel averages can be used as a sensitive indicator of stellar contamination of the FIR flux densities, complementing the IRAS point source and faint source catalogs.  For those sightlines where the background star - or a nearby star - contributes to the average, the standard deviation increases dramatically compared to the norm.

\subsubsection{Nominal A$_V$ vs. I(60$\mu$m)/I(100$\mu$m) relations }

We can probe the difference between line of sight and effective extinction by comparing the observed visual extinction to the amount of emission due to heated grain, as probed by the ratio of 60$\mu$m to 100$\mu$m flux densities.  To find a nominal relationship for each cloud, we used sightlines with reliably determined visual extinctions and compared these to the FIR data.  We used two partially overlapping data sets for this analysis:  Field stars were first selected from the Hipparcos catalog, providing trigonometric parallaxes and thus allowing a clean separation of the stars into groups foreground and background to the cloud, but with a fairly limited total number of sightlines.  We also used the catalog of Tycho stars with known spectral classifications \citep{wright2003} to maximize the number of sightlines used.  In this case, we usually do not have explicit distance information and we therefore imposed somewhat more stringent selection criteria for which stars to include in the analysis.

For the first field-star samples we selected all stars in the Hipparcos catalog within a three degree radius of \textit{l,b}=(297,-15.5), (354, 15), (0,-19.5), (174,-14) and for the Chamaeleon,  Ophiuchus, R~CrA and  Taurus clouds respectively.  For the Musca cloud we used a two degree radius centered on \textit{l,b}=(301, -8).  For the Coalsack, we used the target list of \citet{seidensticker1989a} as input in our Hipparcos search.  
From these original lists we then selected stars at distances (based on the Hipparcos trigonometric parallaxes) greater than those estimated for the clouds.  Complementing the Tycho photometry with 2MASS NIR photometry we calculated visual extinctions.  We rejected stars with negligible extinction and were left with samples of 39, 30, 23, 34 and 27 field stars for Chamaeleon, Musca, Taurus, R~CrA and Ophiuchus clouds respectively.

For the second sample we selected stars as above from the catalog of \citet{wright2003} for Chamaeleon, Coalsack, Musca, Ophiuchus and R~CrA.  For Taurus - because of the elongated shape of the cloud - we selected stars within a three degree radius of two centers at \textit{RA,Dec}=(04:40:00,25:30:00) and (04:15:00,28 00 00).  As with the Hipparcos sample, we extracted Tycho and 2MASS photometry, and calculated visual extinctions and ratios of total-to-selective extinctions.  Based on A$_V$ vs. distance diagrams from the Hipparcos samples, we then selected only those stars with A$_V>$\{0.3, 0.4, 0.4, 0.4, 0.3 and 0.5\} for Chamaeleon, Coalsack, Musca, Ophiuchus, R~CrA and Taurus respectively.  We used the IRAS point- and faint-source catalogs to screen out stars detected as point sources and finally screened for unreasonable values of R$_V$, which likely reflect unreliable spectral classifications.  Finally, we eliminated small regions (usually 0.5 or 1 degree radii) around stars where the I$_{60}$/I$_{100}$ ratio showed localized influence from those stars (see Appendix \ref{irrapp}).  We were then left with 90, 231, 60, 58, 96 and 154 stars respectively for the six clouds.

Because of the relatively bright limiting magnitude of the Hipparcos database, the size of the area used to select the field stars involves a trade-off between the largest acceptable distance from the cloud center and number statistics.  While the polarization samples do cover smaller areas on the sky, the three-degree radius (two degrees for Musca) was chosen based on emission-line and extinction maps (e.g. \citet{cambresy1999}).  Although we do not have kinematic data tracing the material giving rise to the extinction in the field star samples, we can use the measured A$_V$ values to see that this sample likely does probe the outer layers of the molecular clouds under study.  Our Hipparcos field star samples have visual extinctions of A$_V$\{min, max,mean\}=\{0.3,2.4,1.0\}~mag. for Chamaeleon, \{0.1,2.4,0.9\}~mag., for the Coalsack, \{0.4,1.7,0.77\}~mag. for Musca, \{0.4,4.5,1.6\}~mag. for Ophiuchus, \{0.1,1.7,0.6\}~mag. for R~CrA and \{0.4,3.3,1.3\}~mag. for Taurus.  As shown by \citet{savage1977} and \citet{rachford2002} the transition to molecular hydrogen occurs already at E$_{B-V}\sim$0.1 (A$_V\sim$0.3) and hence most of our field star sightlines probe molecular material.  In addition, as shown by e.g. \citet{pgw1983}, \citet{vanderwerf1989} and \citet{bga1993}, molecular clouds are surrounded by extensive atomic envelopes.  Figure 3 of \citet{boulanger1998} also illustrates this for the Chamaeleon complex.  Particularly for the high-latitude clouds in our study, it is thus unlikely that the field stars probe material unrelated to the clouds probes by the polarimetry.

As we are explicitly searching out, and expecting to find, points where the systematic errors dominate the random ones, we used a robust fitting algorithm (\cite{press1986}, p 539ff) to find the nominal I(60$\mu$m)/I(100$\mu$m) vs. A$_V$ relations for each cloud.  This algorithm uses an iterative procedure based on Tukey's Biweight weighting with a limit of 6 outlier-resistant standard deviations \citep{press1986}, as implemented in the IDL routine "robust\_linefit", available in the "astron" library\footnote{Available at http://idlastro.gsfc.nasa.gov/}.  For the larger Tycho sample we also used weighted linear fits.  The two algorithms yield very similar fitting parameters in this case.

For the Hipparcos samples, the robust fits for Chamaeleon, Musca, Ophiuchus, R CrA and Taurus yield:

\begin{equation}
I_{60}/I_{100}=\begin{array}{c}  (0.241\pm0.005)\\ (0.29\pm0.01)\\ (0.27\pm0.02)\\ (0.21\pm0.01)\\ (0.196\pm0.008)\end{array}-A_V\times\begin{array}{l} (0.022\pm0.005)\ [Cham]\\ (0.026\pm0.010)\ [Musc]\\ (0.01\pm0.02)\ [Oph]\\ (0.035\pm0.01)\ [R~CrA]\\(0.013\pm0.006)\ [Tau]\end{array}
\end{equation}

For the Tycho samples, the robust fits yield:
\begin{equation}
I_{60}/I_{100}=\begin{array}{c}  (0.240\pm0.004)\\ (0.29\pm0.1)\\ (0.27\pm0.1)\\ (0.244\pm0.006)\\ (0.179\pm0.003)\end{array}-A_V\times\begin{array}{l} (0.018\pm0.004)\ [Cham]\\ (0.030\pm0.007)\ [Musc]\\ (0.024\pm0.010)\ [Oph]\\ (0.023\pm0.007)\ [R~CrA]\\(0.012\pm0.002)\ [Tau]\end{array}
\end{equation}

In Figure \ref{fir_vs_av} we show the best linear fits (solid lines, top panel) using the robust fitting algorithm to the Tycho field star samples only (i.e. the polarization targets were not included in the fits).

For Chamaeleon, Musca, Ophiuchus, R CrA and Taurus well-defined linear correlations are found (for Ophiuchus, the uncertainties on the fit coefficients for the Hipparcos sample, are too large for the fit to be significant).  For the Coalsack, which straddles the Galactic plane, this technique fails to provide a reasonable correlation, presumably because of the strong influence of diffuse background sources in the FIR data.

The middle panels of Figure \ref{fir_vs_av} shows the locations of the polarization targets in the FIR vs. A$_V$ diagrams.

\begin{figure}
\plotone{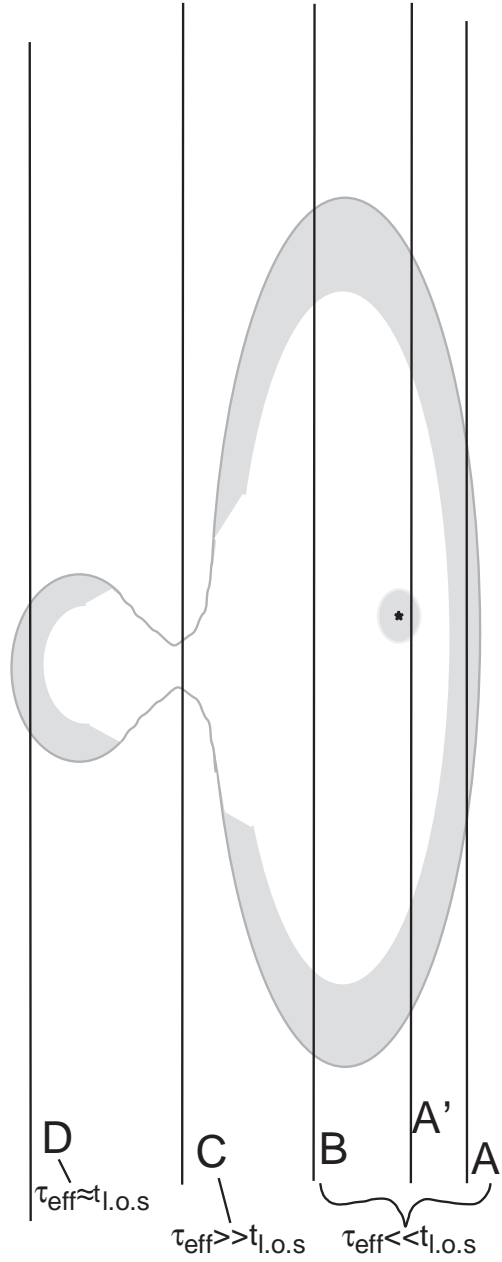}
\caption{Possible origins of differences between line-of-sight extinction and "effective" extinction.  For a asymmetrical cloud with its long axis close to the line of sight, the measured visual extinction, A$_V$ can both over and under estimate the extinction experienced by the gas and dust on the line of sight.  See text for details.}
\label{cloud_cartoon}
\end{figure}

In the lower panels of Figure \ref{fir_vs_av}, we show the distributions of offsets in I$_{60}$/I$_{100}$ from each data point to the best fit line at the same value of A$_V$ (dashed histograms for the field stars and solid histograms for the polarization targets), and the best fit Gaussians to these distributions.  The widths of the distributions are similar for all clouds ($\sigma$=0.012, 0.014, 0.036, 0.026 and 0.016, respectively).  Based on similar geometrical arguments as for Figure \ref{cloud_cartoon}, it is likely that these widths trace the porosity of the outer parts of the clouds where the radiative grain heating occurs.  In Appendix \ref{irrapp} we show that the widths of these dispersions are correlated with the SFR in the clouds.

The dashed lines in Figure \ref{fir_vs_av} show the 2$\sigma$ distance from the best-fit solution.  Points below and to the left of the left-hand dashed lines are likely to have characteristics of sightlines 'B' in Figure \ref{cloud_cartoon}, for high values of A$_V$, or 'C', for low values of A$_V$, while points above and to the right of the right-hand dashed lines are likely to have characteristics of sightlines 'A'.

We note that with the exception of the identified outliers, the polarization sample targets and the field stars overlap in the well-defined Gaussian distributions in offset from the fits, providing additional support for the assumption that the two observational samples are drawn from a common parent population.

While the above screening method fails for the Coalsack, inspecting the $\lambda_{max}$ vs. A$_V$ plot for the Coalsack and comparing it to those of Chamaeleon and Taurus, we can see that the points at relatively high extinction and relatively low $\lambda_{max}$ in the Coalsack have corresponding sightlines in both Chamaeleon and Taurus and that these sightlines in the Coalsack fall in the equivalent "A" region in the I$_{60}$/I$_{100}$ vs. A$_V$ plot as their counterparts for Chamaeleon and Taurus.  Similarly, the points at high $\lambda_{max}$ and low A$_V$ are similar to points in region "C" for Chamaeleon both in the $\lambda_{max}$ vs. A$_V$ and I$_{60}$/I$_{100}$ vs. A$_V$ plots.  While based on a more indirect and less satisfactory procedure, we therefore screen these points also in the Coalsack sample.  In Figure \ref{CS_FIR_vs_Av_select} we show the resulting I$_{60}$/I$_{100}$ vs. A$_V$ plot with the sources thus deselected marked.   The width of the $\Delta$I$_{60}$/I$_{100}$ distribution for the remaining sightlines for the Coalsack is $\sigma$=0.04 also similar to the values found for the other clouds.

\section{Results and Discussion}

\subsection{Origin of the Outliers}

We argue that the outliers in figure \ref{fir_vs_av} are likely primarily due to localized differences in the irradiation of the dust.  Appendix \ref{irrapp} provides several direct lines of evidence in favor of this interpretation.   For Chamaeleon and R~CrA \textbf{all} the sightlines identified as type "A" can be seen to be due to proximity to HD~97300 or HD175362, respectively.  Similarly, for Ophiuchus almost all type "A" sightlines can be seen to be close to either $\sigma$~Sco or $\rho$~Oph D (Figure \ref{DFIR_all}).  Additionally, as also shown in appendix \ref{irrapp}, the ratio I$_{60}$/I$_{100}$ shows a stronger response to the vicinity of a hot star than I$_{25}$/I$_{100}$ while the I$_{12}$/I$_{100}$ ratio shown little or no response to the proximity of hot stars.  These results provide strong support for an origin in irradiation differences.  As shown in Appendix \ref{irrapp} the dispersion around the best-fit lines in the I$_{60}$/I$_{100}$ vs. A$_V$ plots are correlated with the characteristics of the star formation in the cloud (see below).  We interpret this as due to increasing porosity (clumpiness) in the clouds produced by increasing star formation activity, which changes the radiative transfer of the light heating the dust grains.  A quantitative analysis of this clumpy transfer is beyond the scope of the present paper.

It should be noted that the screening procedure used in figure \ref{fir_vs_av} does not rely on this interpretation.  The screening as such is a straightforward numerical procedure and uses a explicit numerical limit for which sightlines to label "anomalous".  The use of the I(60$\mu$m)/I(100$\mu$) vs. A$_V$ ratio to screen the polarimetry data should also not introduce any biases in the latter.  This is as only a very small fraction of the grains contribute to the polarization \citep{kim1995} while all grains contribute to the FIR emission.  Also, even if the outliers in figure \ref{fir_vs_av} were due to significant enhancements in the population of very small grains, the size distributions of the VSGs and the grains responsible for the visual polarization have, at most, a minute overlap \citep{kim1995,desert1990}.

\subsection{Debiased $\lambda_{max}$ vs. A$_V$ plots \label{screen}}

In Figure \ref{lmax_vs_av_masked_TYC} we show the plots of $\lambda_{max}$ vs. A$_V$ which result if we reject the points in Figure \ref{fir_vs_av} which falls more than 2$\sigma$ from the robust fits for the Tycho sample.  Overlaid are the best linear fits for each cloud (full drawn lines).  The dashed lines represent the relations based on the average slope from all the clouds and a zero intercept based on the mean value of R$_V$ of the cloud (see below).  Very similar plots result for the Hipparcos based screening, with the main difference that the R~CrA sample shows a broader scatter.  We show the $\lambda_{max}$ vs. A$_V$ plot for R~CrA using the Hipparcos screening in Figure \ref{lmax_vs_av_masked_RCrA}.  In what follows, we will, except were explicitly noted, discuss these "de-biased" data sets.

Three conclusions can immediately be drawn from these new plots: 

1) While the plot of $\lambda_{max}$ vs. A$_V$ including all the data shows only a very weak correlation, the "screened" plot shows a distinct correlation of $\lambda_{max}$ with A$_V$ for Chamaeleon, Coalsack, Musca and Taurus and consistent results for Ophiuchus and R~CrA.  We can quantify this latter statement by performing Spearman rank order tests on the two groups of data sets.  For Chamaeleon and Taurus both the unscreened and Hipparcos screened data sets show small ($<$6\%) probabilities for being uncorrelated (the Tycho screened sets yield 2\% and 14\% probabilities).  For the Coalsack the probability for an accidental correlation drops from 44\% to 6\% after screening.  For Musca (where no sightlines are screened) we find a 18\% probability of being uncorrelated.

2) Within each individual cloud sample, no obvious correlations are seen between $\lambda_{max}$ and R$_V$.  

3) It now is clear to the eye that the slopes of $\lambda_{max}$ vs. A$_V$ in the individual clouds are very similar, but the three groups are offset from each other in $\lambda_{max}$.  We used robust fits to quantify this and find that for the Hipparcos screened sample:

\begin{equation}
\lambda_{max}=\begin{array}{c}  (0.54\pm0.02)\\ (0.50\pm0.02)\\ (0.55\pm0.02)\\ (0.60\pm0.05)\\ (0.72\pm0.10)\\ (0.48\pm0.02)\end{array}+A_V\times\begin{array}{l} (0.04\pm0.01)\ [Cham]\\ (0.05\pm0.01)\ [CS]\\ (0.04\pm0.02)\ [Musc]\\(0.01\pm0.03)\ [Oph]\\(0.01\pm0.08)\ [R~CrA]\\(0.05\pm0.01)\ [Tau]\end{array}
\end{equation}

While for the Tycho screened sample:

\begin{equation}
\lambda_{max}=\begin{array}{c}  (0.54\pm0.01)\\ (0.50\pm0.02)\\ (0.55\pm0.02)\\ (0.64\pm0.05)\\ (0.75\pm0.03)\\ (0.51\pm0.02)\end{array}+A_V\times\begin{array}{l} (0.024\pm0.006)\ [Cham]\\ (0.05\pm0.01)\ [CS]\\ (0.04\pm0.02)\ [Musc]\\(0.01\pm0.03)\ [Oph]\\(0.03\pm0.02)\ [R~CrA]\\(0.03\pm0.02)\ [Tau]\end{array}
\end{equation}

The slopes in $\lambda_{max}$ vs. A$_V$ are all very close and, in particular, those for Chamaeleon, Coalsack, Musca and Taurus are all both very close and have small error bars, indicating that the slope is universal (Figure \ref{rvvslmax}, left panel).  A weighted average of the slopes yields d$\lambda_{max}$/dA$_V$=0.028$\pm$0.005 (0.038$\pm$0.007) for the Tycho (Hipparcos) screened samples.

As has been shown by \citet{whittet1978} $\lambda_{max}$ is generally found to be correlated with R$_V$.  This is not surprising as, as has been shown by \citet{kim1995} and \citet{kim1994a}, changes in both $\lambda_{max}$ and R$_V$ are most sensitive to changes in the population (total and aligned part) of the smaller grains.  In the right-hand panels of Figure \ref{lmax_vs_av_masked_TYC} we plot $\lambda_{max}$ vs. R$_V$ for the screened samples.  No correlations are seen even for the clouds where correlation in the $\lambda_{max}$ vs. A$_V$ plots are evident.

We calculated weighted averages of R$_V$ for the full polarization targets samples and find R$_V$=3.6$\pm$0.1, 3.6$\pm$0.2, 3.6$\pm$0.2, 3.7$\pm$0.7, 4.1$\pm$0.2 and 3.3$\pm$0.1 for Chamaeleon, the Coalsack, Musca, R~CrA and Taurus, respectively.  In Figure \ref{rvvslmax} (right panel) we plot the result with the best linear fit of the zero-intercept of $\lambda_{max}$ ($\lambda_{max}$(A$_V$=0)) for the Tycho screened samples with the average R$_V$.  The best, weighted, fit yields:

\begin{equation}
\lambda_{max}(A_V=0)=(-0.47\pm0.16) + (0.28\pm0.04)<R_V>
\end{equation}

or, if we impose a zero intercept:

\begin{equation}
\lambda_{max}(A_V=0)=(0.151\pm0.002)<R_V>
\end{equation}

for the Hipparcos screened sample the equivalent fits yield:

\begin{equation}
\lambda_{max}(A_V=0)=(-0.26\pm0.24) + (0.22\pm0.07)<R_V>
\end{equation}

or, if we impose a zero intercept:

\begin{equation}
\lambda_{max}(A_V=0)=(0.146\pm0.002)<R_V>
\end{equation}

In both cases the zero intercept fit is close to the $\lambda_{max}$=(0.18$\pm$0.01)R$_V$ found by \citet{whittet1978}.  

We interpret the correlation of $\lambda_{max}(A_V=0)$ with $<R_V>$ as being due to the underlying differences in the total grain size distribution between the clouds, and the (universal) slope in $\lambda_{max}$ vs. A$_V$ as due to varying degrees of grain alignment at different depths into the clouds caused by a common mechanism.

The data indicate that the lack of aligned small grains at larger visual extinctions, indicated by large values of $\lambda_{max}$, is due to loss of alignment rather than grain destructions.  This can be seen by comparing $\lambda_{max}$ as a function of p$_{max}$/A$_V$ and of R$_V$.  In Figure \ref{lmax_vs_pdAmax_Rv} we show $\lambda_{max}$ for the Tycho screened lines of sight, adjusted to Chamaeleon, as functions of either the alignment efficiency (p$_{max}$/A$_V$) or R$_V$.  The adjustment performed here consists of subtracting the difference in derived $\lambda_{max}(A_V=0)$ between the cloud and that for Chamaeleon for the sightlines in each cloud (i.e.: $\lambda_{max}^*$=$\lambda_{max}$-($\lambda_{max}$(A$_V$=0)$^{cloud}$-$\lambda_{max}$(A$_V$=0)$^{Cham}$.)).

A distinct anti-correlation (albeit, again, with outliers) is seen in the $\lambda_{max}$ vs. p$_{max}$/A$_V$ (Figure \ref{lmax_vs_pdAmax_Rv}, left panel), indicating that as the relative number of aligned grains increase the average size of the aligned grains decrease.  No correlation is seen in the $\lambda_{max}$ vs. R$_V$ plot (Figure \ref{lmax_vs_pdAmax_Rv}, right panel)  Very similar plots, again, result if we instead use the Hipparcos screened sample.

As equation \ref{damping_equ} shows, when all other parameters remain fixed, the smallest grains will have their rotation damped out by gas collisions the fastest.  Hence, when we find that smaller and smaller grains remain aligned either the damping is lessened or the driving mechanism for the spin-up in being enhanced.  The correlations of $\lambda_{max}$ with A$_V$ shows that as we get increasingly close to the cloud surface, smaller grains remain aligned.  As has been shown on large scales for low radiation intensity cloud envelopes by \citet{bga1993,wannier1999} and recently on small scales in the higher radiation field case of the Horse Head nebula by \citet{habart2005}, cloud envelopes show isobaric structures in the gas pressure and hence equation \ref{damping_equ} becomes

\begin{equation}
\label{damping_equ2}
t_{damping}\propto\frac{a\sqrt{T}}{P_{gas}}
\end{equation}

with P$_{gas}$ a constant for a given cloud.

Hence an increase in temperature (assuming an isobaric equation of state) should lead to a decrease in $\lambda_{max}$.  We searched the literature for Copernicus and FUSE measurements of the J=1/J=0 excitation temperature for the six clouds in our study.  Unfortunately, only for Ophiuchus and Chamaeleon are there presently multiple interstellar sightlines published and not all of those sightlines have been studied in polarimetry.  For Ophiuchus, five sightlines were studied by \citet{snow1980} with an additional two by \citet{cartledge2004}. These yield a range in 1-0 temperatures of 46 to 90~K.
  For Chamaeleon, \citet{gry2002} report 1-0 excitation temperatures for three stars between 60 and 66~K.
  \citet{rachford2001} report T=63~K for the line of sight towards HD110432 behind the Coalsack.  For the five Ophiuchus sightlines with H$_2$ data, measured $\lambda_{max}$ can be extracted from \citet{serkowski1975} while one of the Chamaeleon stars (HD~96675 $\equiv$ F24) and HD~110432 are included in the present data samples.  In Figure \ref{Dlmax_vs_T} we plot both the measured values of $\lambda_{max}$ and the offset from the best-fit relations in $\lambda_{max}$ vs. A$_V$ for each cloud.  No clear correlation is apparent.  We therefore conclude that variations in gas temperature are unlikely to explain the variations in the grain alignment.

Based on these observational results we conclude that the data support grain alignment driven by the radiation field.  This conclusion is consistent with recent developments in the theory of interstellar grain alignment.  Several authors \citep{draine1996, weingartner2003, cho2005} have shown that direct radiative torques are the theoretically most likely mechanism for explaining the spin-up of interstellar grains required to allow the magnetic alignment to take place.

We note, however, that our results cannot exclude alignment driven by molecular hydrogen formation (in steady state any formation of H$_2$ must be preceded by the radiative destruction of the molecule, and hence this mechanism is also more strongly active at smaller A$_V$s).  Indeed, if we use the models of \citet{kim1995} to estimate the smallest aligned grains at the surface of the clouds, using $\lambda_{max}$(A$_V$=0)$\approx0.5~\mu$m, we find that grains as small as 0.01-0.04~$\mu$m need to be at least partially aligned.  For silicate grains in a diffuse cloud environment, Table 5 of \citet{draine1996} shows that for such small grains, H$_2$ formation driving is dominant over direct radiative torques.  We note though that according to \citet{lazarian1999} very small grains should flip frequently due to Barnett and nuclear relaxation.  Thus under H$_2$ formation torques (fixed in the grain coordinate system) they are expected to become thermally trapped and not be able to achieve suprathermal spins.  Further observational and theoretical studies are clearly needed to clarify the origin of the very small $\lambda_{max}$(A$_V$=0) we observe.
 
\subsection{Influence of star formation \label{sfr}}

Figure \ref{lmax_vs_av_masked_TYC} indicates that the scatter in $\lambda_{max}$ is related to the star formation activity of the cloud.  We will here explore possible mechanism for this dependence, but start by noting that turbulence, and its associated line of sight variations in the magnetic field direction, is not a likely cause.  This is because $\lambda_{max}$ is not, to first order, dependent on the absolute amount of polarization and since turbulence should affect the polarization in all bands similarly, leaving $\lambda_{max}$ unaffected. 

In Figure \ref{scatterhist} we show histograms of the distance from the best fits in $\lambda_{max}$ with A$_V$ for the six clouds using the "universal slope" relations.
As is clear by a visual inspection, the scatter increases from the Coalsack, Chamaeleon and Musca through Taurus to Ophiuchus and R~CrA.

In Figure \ref{scattervsbsf} we plot the scatter around the best fit in $\lambda_{max}$ with A$_V$ as a function of the "bright star fraction" (bsf(L$_{bol}^{i}$)) of embedded Young Stellar Objects (YSO).    The bright star fraction is defined as the number fraction of objects brighter than a given bolometric luminosity (L$_{bol}^{i}$) to the total number of objects (cf \citet{chen1997}).  The error-bars on the bsf:s reflect counting statistics.  The data for Chamaeleon, Ophiuchus, R~CrA and Taurus are taken from \citet{chen1997}.  For the Coalsack, which does not show any star formation, we here use bsf$\equiv$0.  Musca has not been specifically studied for star formation activity although some T Tauri stars are detected in the general Chamaeleon-Musca complex \citep{mizuno1998}.  To separate the Coalsack and Musca points we have for plotting purposes assigned a bsf of 0.17 mid-way between Chamaeleon and the Coalsack.  Neither point was used the fitting.

We used power law and exponential fits for several choices of the break-point (L$_{bol}^{i}$; see above) in the bsf and find a best fit using an exponential for L$_{bol}$=1L$_\odot$, yielding a correlation coefficient of R=0.88 (for the Hipparcos screened sample and 0.79 for the Tycho screened sample) the resulting fit is over-plotted in Figure \ref{scattervsbsf}.  At both smaller and larger L$_{bol}^{i}$ the fit is worse.  While not conclusive, this is consistent with an origin of the correlation due to the influence of the YSOs since the contribution, per unit luminosity range, to the total luminosity of the YSOs peaks around 1-2 L$_\odot$.  This is illustrated in Figure \ref{YSO_lum} where we show the relative contribution to the bolometric and X-ray luminosity per unit L$_{bol}$ for different sub-samples.  The data for the bolometric luminosity are a combination of the data for Chamaeleon, Ophiuchus and Taurus, taken from \citet{chen1995}.  The individual cloud data give similar results but with bigger scatter.  The data for the X-ray luminosity are from \citet{grosso2000}.  We chose a variable binning aimed at providing good resolution while collecting enough targets per bin to achieve reasonable statistic per bin.  The relative binned bolometric luminosity peaks at $\sim$1~L$_\odot$ depending marginally on the exact binning, while the X-ray luminosity peaks at $\sim$1-2~L$_\odot$ with significant contributions from individual high luminosity stars. (The 10-20~L$_\odot$ bin however only contains two stars.  We have also excluded target A21 from \citet{grosso2000} with L$_{bol}$=1100$_\odot$ and log(L$_X$)=30.8 in the plot.)

In Taurus (and R~CrA if using the Hipparcos screening) the outliers in Figure \ref{scatterhist} are primarily located on the positive side of the plots. Depending on whether we use the Tycho or Hipparcos based screenings, the data for R~CrA either shows a fairly wide dispersion or a significant offset in the center of the distribution.  For Chamaeleon, the Coalsack, Musca and Ophiuchus, the average of the distances of points from fits are much less than their associated standard deviation.  Also for Taurus the average offset is smaller that the standard deviation, but here the distribution has a statistically significant positive skew (\citet{press1986}, p. 457) of 0.66$\pm$0.53 (for the Hipparcos screening; 0.57$\pm$0.53 for the Tycho screening).

The scatter is likely due to porosity in the clouds, introduced by the effects of star formation, as seen in the dispersions between I$_{60}$/I$_{100}$ vs. A$_V$ plots (Appendix \ref{irrapp}).  However, for variations in the radiation field intensity caused by clouds porosity alone we would expect that the scatter be symmetrical.  Particularly for Taurus and R~CrA we are thus led to consider possible mechanisms which will drive $\lambda_{max}$ selectively to larger values, and hence also possible sources of additional, localized, grain rotation damping.

Given the ubiquity of X-ray emission from YSOs, it is worth considering what effects X-rays from the embedded YSOs would have on the grain rotation. One important X-ray induced effect is the ejection of photoelectrons from the grains and both the subsequent heating of the gas and charging of the grains.  As shown by \citet{draine1998} the dominant rotation-damping mechanism in molecular cloud environments for very small grain (smaller than those discussed herein) is plasma drag due to the interaction between the ions in the gas and the electric dipole moment of the grain.
  Since the dipole moment of the grain is proportional to the grain charge \citep{draine1998}, enhanced localized grain charging would also mean enhanced localized rotation damping, in addition to the enhancement from gas heating.  If we extrapolate the molecular cloud (MC) panel of Figure 4 of \citet{draine1998} to the 0.05~$\mu$m range, we find that in the nominal model, neutral-gas drag is the dominant damping mechanism.
  However, plasma drag is only less important by a factor of a few.  Hence, an increase in the grain charge by an order of magnitude would invert this relative importance.  The typical X-ray (h$\nu$$\sim$0.5-5~keV) luminosity of embedded YSOs is 10$^{30}$-10$^{31}$~ergs~s$^{-1}$ with flares reaching significantly higher \citep{ozawa2005, grosso2000}.

To derive a first order estimate of whether X-rays from YSOs can significantly effect the grain rotation, we use the Ophiuchus cloud observations.  In the observations of \citet{grosso2000} for the inner part of cloud, 66 targets were found within an approximate radius of 15', or at a distance of 140~pc, 0.6~pc, yielding a characteristic projected source-to-source distance of 0.075~pc.  At this radius the X-rays from a single source contributes an energy density of about u=5$\times10^{-16}$~ergs~cm$^{-3}$.  \citet{draine1998} assumed a energy density in the radiation field of the 1\% of the nominal ISRF for their molecular cloud medium, or u$\sim$9$\times10^{-15}$~ergs~cm$^{-3}$ \citep{weingartner2001}.  Taking into account also the contributions of FUV emission from the YSOs, the high photoelectric yield at soft X-rays \citep{dwek1996} and the contributions from the diffuse X-ray emission, likely caused by stellar winds \citep{ezoe2006}, we feel that the influence of embedded X-ray sources is a viable cause of the localized rotation damping and warrants further study.  However, observational studies and modeling, beyond the scope of the present paper, are required to quantify the importance of X-ray induced grain rotation damping.

\section{Conclusions}

We have used new multi-band polarimetry of the Southern Coalsack to study the alignment of interstellar grains.  For this study the Coalsack has the advantage of not showing a systematic increase in the total-to-selective extinction, R$_V$ with increasing visual extinction and to show no star formation.  The former allows us to break the degeneracy between increasing values of the wavelength of maximum polarization ($\lambda_{max}$) due to the potentially lessened driving of the grain spin and due to grain growth.  We find a tight correlation of $\lambda_{max}$ with A$_V$ over the limited range of A$_V$=1-2.5, but with a significant amount of outliers.  Using polarization data from the literature, together with archival FIR emission data, for an additional five near-by clouds, we show that the outliers in the $\lambda_{max}$ vs. A$_V$ plot are primarily due to geometrical effects, causing the measured visual extinction to become a poor tracer of the extinction "seen" by the dust grains under study.
 
Screening the observations based on the I$_{60}$/I$_{100}$ vs. A$_V$ relationship we find that for clouds without, or with only low star formation activity, $\lambda_{max}$ is tightly correlated with A$_V$.  We performed the screening based on two somewhat overlapping data sets, namely field stars background to the clouds from, either the Hipparcos catalog, or from the catalog of Tycho targets with known spectral classifications \citep{wright2003}.  Screening with the two field-star samples yield consistent results but slightly different numerical parameters.  In both cases, the correlation between $\lambda_{max}$ and A$_V$ shows a universal slope and zero-A$_V$ intercepts proportional to the averages of R$_V$ in the clouds, such that

\begin{equation}
\lambda_{max}=(0.15\pm0.01)<R_V> + (0.038\pm0.007)A_V
\end{equation}
for the Hipparcos based screening and 
\begin{equation}
\lambda_{max}=(0.15\pm0.01)<R_V> + (0.028\pm0.005)A_V
\end{equation}
for the Tycho based screening, where we have imposed a zero intercept in the $\lambda_{max}^{A_V=0}$($<R_V>$) relation.

Within each cloud, we do not find correlations between $\lambda_{max}$ and R$_V$

We interpret the positive slope in $\lambda_{max}$ vs. A$_V$ as evidence for radiation induced grain spin-up.  Our data cannot conclusively differentiate between direct radiative torques and H$_2$ formation torques.  A possible role for the latter is however indicated by the small value of $\lambda_{max}$ at A$_V$=0, using the models by \citet{kim1995}.

Compensating for the different $\lambda_{max}$(A$_V$=0) between clouds, we find that $\lambda_{max}$ is anti-correlation between with the alignment efficiency, p$_{max}$/A$_V$, further supporting the conclusion that the systematic variation in $\lambda_{max}$ as a function of A$_V$ is indeed due to grain alignment rather than changes in the grain size distribution.

The scatter around the $\lambda_{max}$ vs. A$_V$ correlation  - as well as the I$_{60}$/I$_{100}$ vs. A$_V$ relationship - increases with the star formation rate in the cloud and is correlated with the number fraction of YSOs brighter than 1~L$_{bol}$.  In particular for the R~CrA and Taurus clouds the scatter shows asymmetries toward positive values, indicating possible evidence for localized enhancements in the grain rotation damping.  The implied patchiness of the excess damping together with the correlation of scatter in $\lambda_{max}$ with bright YSOs and the high X-ray output of such embedded sources lead us to propose a connection to photoelectric grain charging and associated plasma damping of the grain rotation.
 
Thus, to quantitatively understand interstellar polarization we must know the strength of the radiation field at the location of the grain, as well as the star formation environment of the region, likely including the X-ray flux seen by the grain.  Further surveys of the polarization curve for clouds with known star formation rates (SFR) as well as significant numbers of well-characterized line of sight at large (A$_V>3$~mag.) visual extinctions, both in the present sample of clouds and in new ones would allow the results presented in this paper to be tested and extended.  We are currently pursuing several such studies.

\acknowledgements
It is a pleasure to acknowledge many helpful discussions with Alex Lazarian, Joe Weingartner and Bruce Draine.   Kip Kuntz and Steve Drake provided helpful pointers to the X-ray properties of YSOs.  The referees provided several suggestions and challenges that help to significantly improve the paper.  This research has made extensive use of the VizieR catalogue access tool at CDS, Strasbourg, France, as well as of the NASA/ IPAC Infrared Science Archive, operated by the Jet Propulsion Laboratory, California Institute of Technology, under contract with the National Aeronautics and Space Administration.  B-G A. gratefully acknowledges travel support from the Center for Astrophysical Sciences at the Johns Hopkins University.

\appendix

\section{Does the 60-to-100$\mu$m ratio really trace the local radiation field intensity? \label{irrapp}}

As noted in the introduction, several authors (e.g. \citet{langer1989, snell1989, jarrett1989}) have shown that the I$_{60}$/I$_{100}$ ratio is anti-correlated with column density of CO and interpreted this as a temperature effect.  However, as shown by e.g. \citet{bernard1993} the relative abundance of the very small grains (VSG) decreases towards the center of clouds and we therefore need to ascertain what part of the variations seen in the FIR ratios is dominated by grain heating rather than by abundance effects.  First, we note that modern theories of the FIR emission \citep{draine2007} supports the use of the I$_{60}$/I$_{100}$ ratio as a irradiation tracer.  Figure 12 of that paper shows a very small effect from the variation of the PAH fraction, while figure 13 shows that the ratio is very sensitive to the strength of the radiation field.

Arguing in somewhat more detail, the models by \citet{desert1990} indicate that the 100 ~$\mu$m band emission is dominated by large grains with a minor contribution from the very small grains (VSG), the 60~$\mu$m band contains a significant contribution from both large grains and the VSG.  The 25~$\mu$m band emission predominantly samples the very small grains (with some emission from poly-aromatic hydrocarbons; PAH) and the 12~$\mu$m band predominantly samples the PAH:es with a contribution from the VSGs.  Hence, if the variability in the FIR ratios is primarily driven by VSG abundance effects, the I$_{25}$/I$_{100}$ ratio should show bigger deviations whereas if the variability is primarily due to temperature variations, the I$_{60}$/I$_{100}$ should be more affected.  Also, if irradiation (and hence heating) dominates the FIR emission, we should be able to detect effects due to physically near-by hot stars on the dust in each cloud.

In Figure \ref{FIR_both_vs_Av} we plot the ratios of I$_{60}$/I$_{100}$, I$_{25}$/I$_{100}$ and I$_{12}$/I$_{100}$ as functions of the on-the-sky distance between the locations of the Tycho selected field stars and early B stars located close to the clouds (Table \ref{app1tab}).  This figure shows both that the I$_{60}$/I$_{100}$ responds to the enhanced radiation field caused by localized sources and that the I$_{25}$/I$_{100}$ and I$_{12}$/I$_{100}$ ratios show successively less effect from the enhanced radiation field.

Figure \ref{disp_FIR_vs_Av} shows the dispersions around the best fit in I$_{60}$/I$_{100}$ vs. A$_V$ for the Tycho selected field stars for the clouds in our study as a function of the bright star fraction for each cloud (we have here also added the result from an equivalent analysis or the Lupus I field).  A clear correlation is seen, tightly fit by a power law function.  This can be qualitatively understood in terms of variations in the local radiative heating of the dust in a porous medium where the porosity is due to the influence of newly formed stars.

\begin{figure}
\plotone{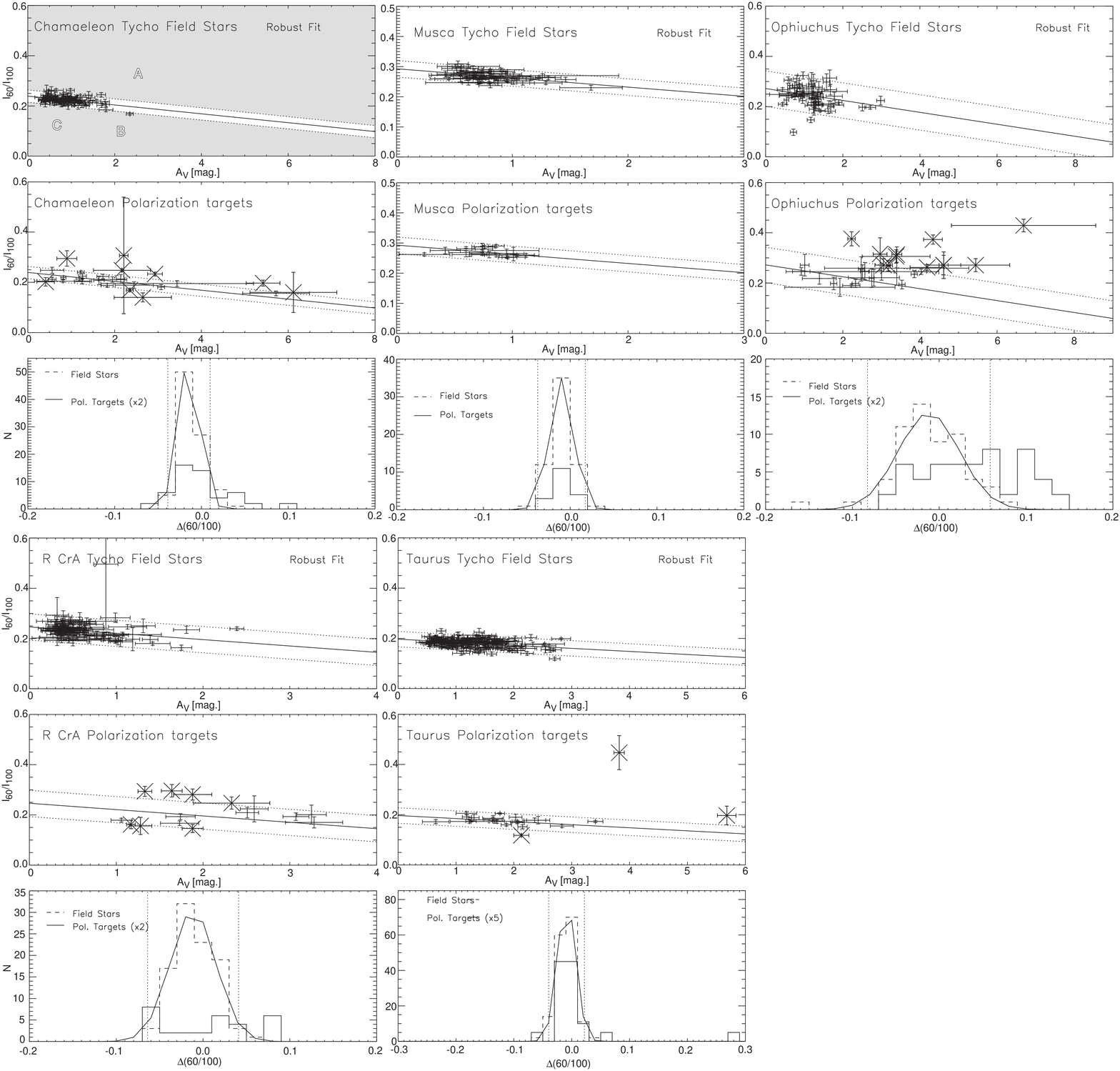}
\caption{Anomalous sightlines are found comparing the I$_{60}$/I$_{100}$ ratios to the measured visual extinction.  The top panel for each cloud shows the relationship found between I$_{60}$/I$_{100}$ ratios and A$_V$ for the field star samples selected from the \citet{wright2003} catalog.  The middle panels over-plot this best fit on the data for the polarization target samples.  The bottom panels show the distribution of fit residuals for the field star samples (dashed histograms) and polarization target samples (solid histograms).  Also plotted are the best-fit Gaussians for the field star samples (solid line) and the $\pm$2$\sigma$ limits used for screening targets as anomalous (dotted lines).  The shaded regions in the top plot for Chamaeleon illustrate the approximate areas corresponding to the "sightline types" discussed in the text and illustrated in Figure \ref{cloud_cartoon}.  This "debiasing" technique fails for the Coalsack sightlines, presumably due to the strong FIR background in the Galactic plane.  See text for details.}
\label{fir_vs_av}
\end{figure}

\begin{figure}
\plotone{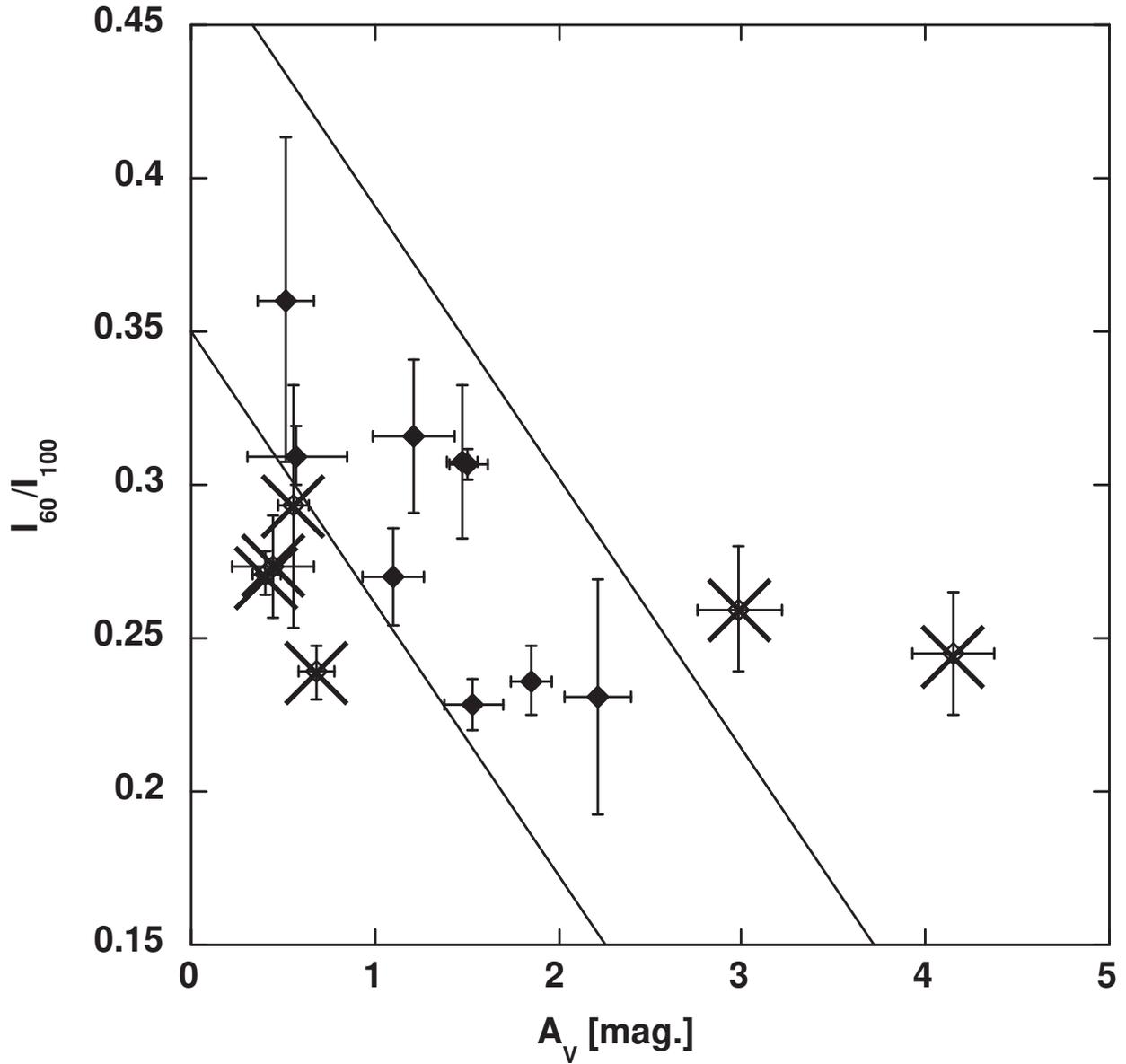}
\caption{For the Coalsack, establishing a nominal A$_V$ vs. I$_{60}$/I$_{100}$ slope fails, presumably due to the influence of background emission in the IRAS data.  We therefore screened the Coalsack data based on similarities with the Chamaeleon and Taurus plots.  The screened-out sources are marked with X-es.}
\label{CS_FIR_vs_Av_select}
\end{figure}

Finally, we must ask whether the field star samples provide accurate representations of the behavior of the dust sampled by the polarization targets.  The field star samples are generally somewhat shallower than the polarization samples.  However, as the lower panels of Figure \ref{fir_vs_av} shows - once anomalous sightlines are excluded - the distributions of the field stars and polarization targets around the best fit lines are very similar.  Unfortunately, background stars with well established spectral classifications and large visual extinctions are very rare.  However, in Figure \ref{tau_deep} we show the I$_{60}$/I$_{100}$ vs. A$_V$ plot for Taurus, with the field stars from the water ice surveys of \citet{murakawa2000} and \citet{teixeira1999} added.  Here, the Tycho field stars are plotted as black dots with gray error-bars, the polarization targets used by us as black diamonds (those found to be "anomalous sightlines" have been over-plotted with X:es), sightlines with $\tau$(H$_2$0)$<$0.05 as red open diamonds and those with $\tau$(H$_2$0)$>$0.1 as filled blue diamonds.  No qualitative change in I$_{60}$/I$_{100}$ vs. A$_V$ is seen at the onset of water ice mantles.  However, at the optical depth where the 60 and 100 $\mu$m intensities flatten out as a function of A$_V$ (insert) the linear I$_{60}$/I$_{100}$ vs. A$_V$ relation seems to break down, as would be expected if the I$_{60}$/I$_{100}$ ratio was primarily driven by irradiation.

Together, these results lead us to conclude that we can indeed use the I$_{60}$/I$_{100}$ ratio as a tracer of the strength of the radiation field seen by the dust grains.  We note that we are not attempting to derive physical dust temperatures here, which, due to the complexity of the dust population, would not be reliable from a single ratio.  We are simply using the FIR ratio as a tracer of the relative strength of the short wavelength radiation in different parts of each cloud.    In order to minimize any residual effects of variations in the grain size distribution, we additionally only use the FIR data to screen the visual extinction data for anomalous sightlines, as described in the main text.

\clearpage
\begin{deluxetable}{llrrrrrrrll}
\rotate
\tabletypesize{\footnotesize}
\tablecaption{Coalsack Target Stars. \label{CS_stars}}
\tablewidth{0pt}
\tablehead{
\colhead{Star} & \colhead{Sp. Class\tablenotemark{a}} & \colhead{V} & \colhead{B-V} & \colhead{(B-V)$_0$} & \colhead{V-K} & \colhead{(V-K)$_0$} & \colhead{R$_V$} & \colhead{A$_V$} & \colhead{ref.\tablenotemark{b}} & \colhead{screen?\footnotemark{c}}\\
\colhead{ }&   \colhead{ } & \colhead{[mag.]} & \colhead{[mag.]} & \colhead{[mag.]} & \colhead{[mag.]} & \colhead{[mag.]} & \colhead{ } & \colhead{[mag.]} &\colhead{ } & \colhead{ }}
\startdata
HD 108417 & A1~V (1) & 9.97$\pm$0.02 & 0.17$\pm$0.02 & 0.02$\pm$0.04 & 0.58$\pm$0.03 & 0.07$\pm$0.07 & 3.7$\pm$1.2& 0.6$\pm$0.1 & MSS & s\\

HD 108639 & B1~III (2) & 7.83$\pm$0.01 & 0.04$\pm$0.01 & -0.27$\pm$0.06 &  & -0.74$\pm$0.15 & 3.5$\pm$0.9 & 1.1$\pm$0.2 & S89 & \\

CPD -64 1950 & A8 (2) & 9.64$\pm$0.03 & 0.61$\pm$0.03 & 0.52$\pm$0.08 & 1.72$\pm$0.04 & 1.35$\pm$0.06 & 4.5$\pm$4.3& 0.4$\pm$0.1 & S89 & s\\

HD 108939 & B8 III (2) & 8.06$\pm$0.01 & 0.00$\pm$0.01 & -0.11$\pm$0.09 & 0.17$\pm$0.02 & -0.24$\pm$0.2 & 4.0$\pm$3.7 & 0.5$\pm$0.2 & MSS & s\\

HD 109065 & A1 V (2) & 8.15$\pm$0.01 & 0.15$\pm$0.01 & 0.02$\pm$0.08 & 0.54$\pm$0.03 & 0.07$\pm$0.14 & 4.1$\pm$2.9 & 0.5$\pm$0.2 & MSS & \\

HD 110245 & F8/G0 III (3) & 8.39$\pm$0.01 & 0.68$\pm$0.01 & 0.55$\pm$0.10 & 2.28$\pm$0.03 & 1.75$\pm$0.25 & 4.3$\pm$3.9 & 0.6$\pm$0.3 & MSS & \\

HD 110432 & B2 (2) & 5.317$\pm$0.003 & 0.19$\pm$0.01 & -0.24$\pm$0.06 & 1.36$\pm$0.05 & -0.66$\pm$0.16 & 5.2$\pm$0.9 & 2.2$\pm$0.2 & HGS69 & \\

HD 112045 & A1 IV/V (2) & 9.84$\pm$0.03 & 0.37$\pm$0.04 & 0.02$\pm$0.08 & 1.47$\pm$0.04 & 0.07$\pm$0.14 & 4.3$\pm$1.2 & 1.5$\pm$0.2 & MSS & \\

HD 112661 & B0/1 III/IV (3) & 9.26$\pm$0.02 & 0.55$\pm$0.02 & -0.24$\pm$0.1 & 1.94$\pm$0.03 & -0.78$\pm$0.2 & 3.8$\pm$0.6 & 3.0$\pm$0.2 & MSS & s\\

HD 112954 & B9 IV (1) & 8.42$\pm$0.01 & 0.46$\pm$0.02 & -0.06$\pm$0.06 & 1.56$\pm$0.03 & -0.13$\pm$0.1 & 3.6$\pm$0.6 & 1.8$\pm$0.1 & MSS & \\

HD 112999 & B6 III (1) & 7.38$\pm$0.01 & 0.04$\pm$0.01 & -0.15$\pm$0.02 & 0.26$\pm$0.02 & -0.36$\pm$0.09 & 3.7$\pm$0.7 & 0.7$\pm$0.1 & MSS & s \\

HD 113034 & B0/1 III (3) & 9.32$\pm$0.02 & 0.90$\pm$0.03 & -0.24$\pm$0.1 & 3.16$\pm$0.03 & -0.62$\pm$0.2 & 3.6$\pm$0.4 & 4.2$\pm$0.2 & MSS & s \\

HD 114012 & A0 V (1) & 9.10$\pm$0.02 & 0.48$\pm$0.02 & -0.02$\pm$0.04 & 1.35$\pm$0.03 & 0.00$\pm$0.07 & 2.9$\pm$0.3 & 1.5$\pm$0.1 & MSS & \\

HD 114720 & B8 V (2) & 9.65$\pm$0.03 & 0.07$\pm$0.03 & -0.11$\pm$0.04 & 0.86$\pm$0.04 & -0.24$\pm$0.2 & 6.6$\pm$2.2 & 1.2$\pm$0.2 & MSS & \\

HD 117111 & B2V (1) & 7.72$\pm$0.01 & 0.04$\pm$0.01 & -0.24$\pm$0.02 & 0.71$\pm$0.04 & -0.66$\pm$0.09 & 5.4$\pm$0.6 & 1.5$\pm$0.1 & MSS & \\
\enddata
\tablenotetext{a}{Estimated uncertainties, in subclasses, are given in parenthesis}
\tablenotetext{b}{Spectral classes from: S89: \citet{seidensticker1989b}; HGS69: \citet{hiltner1969}; MSS: \citet{mss1975}}
\tablenotetext{c}{Sightlines that are screened out in the Hipparcos field star based analysis. (see section \ref{screen})}
\end{deluxetable}

\clearpage
\thispagestyle{empty}
\begin{deluxetable}{lrrrrrrrrrr}
\rotate
\tabletypesize{\footnotesize}
\tablecaption{Coalsack Polarimetry Results. \label{pol_res}}
\tablewidth{0pt}
\tablehead{
\colhead{Star} & \colhead{p$_U$} & \colhead{$\theta_U$} & \colhead{p$_B$} & \colhead{$\theta_B$} & \colhead{p$_V$} & \colhead{$\theta_V$} & \colhead{p$_R$} & \colhead{$\theta_R$} & \colhead{p$_I$} & \colhead{$\theta_I$}\\
\colhead{ }  & \colhead{[\%]} & \colhead{[$^\circ$;~ E of N]} & \colhead{[\%]} & \colhead{[$^\circ$;~ E of N]} &\colhead{[\%]} & \colhead{[$^\circ$;~ E of N]} & \colhead{[\%]} & \colhead{[$^\circ$;~ E of N]} & \colhead{[\%]} &\colhead{[$^\circ$;~ E of N]}}
\startdata
HD 108417 & 0.49$\pm$0.05 & 78.7$\pm$1.4 & 0.59$\pm$0.03 & 78.1$\pm$0.9 & 0.64$\pm$0.03 & 78.1$\pm$0.7 & 0.64$\pm$0.03 & 72.7$\pm$0.7 & 0.69$\pm$0.03 & 76.1$\pm$0.8\\

HD 108639 &1.69$\pm$0.04 & 88.5$\pm$1.3 & 1.85$\pm$0.03 & 90.5$\pm$0.9 & 1.91$\pm$0.03 & 90.1$\pm$0.8 & 1.84$\pm$0.03 & 88.6$\pm$0.8 & 1.65$\pm$0.03 & 88.7$\pm$1.0\\

CPD -64 1950 & 0.63$\pm$0.09 & 76.7$\pm$2.6 & 0.89$\pm$0.05 & 76.6$\pm$1.3 & 0.92$\pm$0.04 & 72.8$\pm$1.2 & 1.00$\pm$0.04 & 70.8$\pm$1.1 & 0.94$\pm$0.04 & 71.9$\pm$1.2\\

HD 108939 & 0.90$\pm$0.05 & 70.1$\pm$1.4 & 0.99$\pm$0.03 & 71.8$\pm$0.9 & 1.02$\pm$0.03 & 71.1$\pm$0.9 & 1.03$\pm$0.03 & 70.2$\pm$0.9 & 1.01$\pm$0.04 & 72.5$\pm$1.0\\

HD 109065 & 0.82$\pm$0.03 & 68.1$\pm$0.9 & 0.88$\pm$0.02 & 72.7$\pm$0.5 & 0.87$\pm$0.01 & 71.5$\pm$0.5 & 0.94$\pm$0.02 & 67.9$\pm$0.5 & 0.80$\pm$0.02 & 72.1$\pm$0.6\\

HD 110245 & 0.91$\pm$0.08 & 112.8$\pm$2.2 & 1.23$\pm$0.04 & 121.8$\pm$1.1 & 1.26$\pm$0.04 & 120.2$\pm$1.0 & 1.07$\pm$0.04 & 119.6$\pm$1.1 & 0.95$\pm$0.04 & 119.5$\pm$1.2\\

HD 110432 & 1.40$\pm$0.03 & 81.0$\pm$0.8 & 1.59$\pm$0.03 & 92.1$\pm$0.9 & 1.70$\pm$0.03 & 82.6$\pm$0.9 & 1.73$\pm$0.03 & 77.3$\pm$1.0 & 1.66$\pm$0.03 & 77.2$\pm$0.9\\

HD 112045 & 1.46$\pm$0.11 & 66.9$\pm$3.1 & 1.86$\pm$0.05 & 69.1$\pm$1.5 & 1.82$\pm$0.04 & 68.0$\pm$1.2 & 1.87$\pm$0.04 & 66.8$\pm$1.2 & 1.73$\pm$0.04 & 68.3$\pm$1.1\\

HD 112661 & 1.49$\pm$0.08 & 72.4$\pm$2.2 & 1.80$\pm$0.04 & 73.7$\pm$1.1 & 1.86$\pm$0.04 & 73.0$\pm$1.0 & 1.90$\pm$0.03 & 70.3$\pm$0.9 & 1.71$\pm$0.03 & 71.4$\pm$0.9\\

HD 112954 & 2.07$\pm$0.06 & 41.7$\pm$1.7 & 2.37$\pm$0.04 & 42.2$\pm$1.0 & 2.42$\pm$0.03 & 40.8$\pm$0.9 & 2.51$\pm$0.03 & 40.2$\pm$0.8 & 2.29$\pm$0.03 & 40.2$\pm$0.9\\

HD 112999 & 1.66$\pm$0.04 & 68.8$\pm$1.2 & 1.90$\pm$0.03 & 69.8$\pm$0.7 & 1.99$\pm$0.02 & 71.1$\pm$0.7 & 1.95$\pm$0.02 & 70.1$\pm$0.6 & 1.80$\pm$0.03 & 70.6$\pm$0.8\\

HD 113034 & 4.03$\pm$0.11 & 80.0$\pm$3.0 & 4.64$\pm$0.05 & 80.8$\pm$1.3 & 5.05$\pm$0.04 & 81.7$\pm$1.0 & 4.97$\pm$0.03 & 81.8$\pm$0.8 & 4.49$\pm$0.03 & 82.7$\pm$0.7\\

HD 114012 & 1.05$\pm$0.16 & 53.8$\pm$4.7 & 1.32$\pm$0.07 & 56.4$\pm$2.0 & 1.41$\pm$0.05 & 53.0$\pm$1.5 & 1.53$\pm$0.05 & 49.1$\pm$1.3 & 1.20$\pm$0.05 & 54.0$\pm$1.5\\

HD 114720 & 0.83$\pm$0.08 & 77.7$\pm$2.1 & 0.88$\pm$0.04 & 83.3$\pm$1.2 & 0.97$\pm$0.04 & 85.5$\pm$1.2 & 0.92$\pm$0.04 & 79.2$\pm$1.2 & 0.80$\pm$0.05 & 86.3$\pm$1.3\\

HD 117111 & 1.28$\pm$0.03 & 76.2$\pm$0.9 & 1.18$\pm$0.03 & 77.8$\pm$0.9 & 1.40$\pm$0.03 & 77.1$\pm$0.8 & 1.40$\pm$0.03 & 71.5$\pm$0.8 & 1.38$\pm$0.03 & 75.5$\pm$0.8\\
\enddata
\end{deluxetable}

\clearpage
\begin{deluxetable}{lrrrrr}
\tablecaption{ Coalsack Polarization Curve Fits. \label{pol_fit_CS}}
\tablewidth{0pt}
\tablehead{
\colhead{Star} & \colhead{p$_{max}$} & \colhead{$\lambda_{max}$} & \colhead{K }\\
 \colhead{ } &  \colhead{[\%]} & \colhead{[$\mu$m]} &\colhead{ } & \colhead{$\chi^2/\nu_S$} & \colhead{$\chi^2/\nu_W$}}
\startdata
HD 108417 & 0.68$\pm$0.02 & 0.64$\pm$0.03 & \nodata  & 1.3 & \nodata\\

HD 108639 & 1.91$\pm$0.02 & 0.52$\pm$0.01 & 0.87$\pm$0.13  & 0.7 & 0.1\\

CPD -64 1950 & 0.99$\pm$0.02 & 0.63$\pm$0.03 & \nodata & 0.8 & \nodata\\

HD 108939 & 1.03$\pm$0.02& 0.62$\pm$0.05 & 0.42$\pm$0.27 & 2.6 & 0.2\\

HD 109065 & 0.90$\pm$0.01 & 0.54$\pm$0.01 & \nodata & 10.4 & \nodata \\

HD 110245 & 1.22$\pm$0.02 & 0.50$\pm$0.02 & \nodata & 3.2 & \nodata \\

HD 110432 & 1.73$\pm$0.02 & 0.62$\pm$0.02 & 0.69$\pm$0.12 & 4.2 & 0.1\\

HD 112045 & 1.92$\pm$0.02 & 0.57$\pm$0.01 & \nodata & 3.1 & \nodata\\

HD 112661 & 1.92$\pm$0.02 & 0.57$\pm$0.01 & \nodata & 0.9 & \nodata\\

HD 112954 & 2.52$\pm$0.02 & 0.57$\pm$0.01 & \nodata & 5.9 & \nodata\\

HD 112999 & 1.99$\pm$0.02 & 0.56$\pm$0.01 & 0.88$\pm$0.10 & 2.1 & 0.5\\

HD 113034 & 5.05$\pm$0.02 & 0.574$\pm$0.005 & \nodata & 0.35 & \nodata\\

HD 114012 & 1.44$\pm$0.03 & 0.57$\pm$0.02 & \nodata & 2.6 & \nodata\\

HD 114720 & 0.95$\pm$0.02 & 0.54$\pm$0.02 & \nodata & 0.3 & \nodata \\

HD 117111 & 1.45$\pm$0.01 & 0.58$\pm$0.01 & \nodata & 21.6 & \nodata \\
\enddata
\end{deluxetable}

\clearpage
\begin{deluxetable}{llrrrrrrrll}
\rotate
\tabletypesize{\footnotesize}
\tablecaption{Stars with data from archives and the literature. \label{stars_data}}
\tablewidth{0pt}
\tablehead{
\colhead{Star} & \colhead{Sp. Class\tablenotemark{a}} & \colhead{V} & \colhead{B-V} & \colhead{(B-V)$_0$} & \colhead{V-K} & \colhead{(V-K)$_0$} & \colhead{R$_V$} & \colhead{A$_V$} & \colhead{ref.\tablenotemark{b}} & 
\colhead{Screen? \tablenotemark{c}}\\
\colhead{ } & \colhead{ } & \colhead{[mag.]} & \colhead{[mag.]} & \colhead{[mag.]} & \colhead{[mag.]} & \colhead{[mag.]} & \colhead{ } & \colhead{[mag.]} & 
\colhead{ } & \colhead{}}
\startdata
Chamaeleon\\
F1 & K4 III (2) & 10.31$\pm$0.03 & 1.56$\pm$0.02 & 1.39$\pm$0.22 & 3.99$\pm$0.04 & 3.26$\pm$0.60 & 4.7$\pm$7.3 & 0.8$\pm$0.7 & VR84 & \\

F2 & B8 V (2) & 9.90$\pm$0.03 & 0.51$\pm$0.03 & -0.11$\pm$0.04 & 1.42$\pm$0.04 & -0.24$\pm$0.12 & 3.0$\pm$0.3 & 1.8$\pm$0.1 & VR84 & \\

F3 & B4 V (1) & 8.03$\pm$0.01 & 0.42$\pm$0.01 & -0.19$\pm$0.02 & 1.65$\pm$0.02 & -0.49$\pm$0.07 & 3.9$\pm$0.2 & 2.4$\pm$0.1 & MSS & \\

F6 & A2 V (1) & 10.85$\pm$0.08 & 0.74$\pm$0.11 & 0.05$\pm$0.03 & 1.63$\pm$0.08 & 0.14$\pm$0.07 & 2.4$\pm$0.4 & 1.6$\pm$0.1 & VR84 & s\\

F7 & B5 V (2) & 10.25$\pm$0.04 & 0.41$\pm$0.06 & -0.17$\pm$0.04 & 1.15$\pm$0.05 & -0.42$\pm$0.12 & 3.0$\pm$0.4 & 1.7$\pm$0.1  & VR84 & \\

F9 & K0 III (2) & 9.56$\pm$0.03 & 1.46$\pm$0.07 & 1.00$\pm$0.16 & 4.54$\pm$0.04 & 2.31$\pm$0.40 & 5.3$\pm$2.2 & 2.5$\pm$0.4  & VR84 & \\

F11 & B9 V (2) & 10.53$\pm$0.06 & 0.72$\pm$0.08 & -0.06$\pm$0.09 & 2.53$\pm$0.07 & -0.13$\pm$0.14 & 3.7$\pm$0.6 & 2.9$\pm$0.2 & VR84 & s\\

F16 & G2 IV (2) & 11.46$\pm$0.03 & 1.45$\pm$0.02 & 0.63$\pm$0.05 & 4.25$\pm$0.04 & 1.46$\pm$0.06 & 3.7$\pm$0.3 & 3.1$\pm$0.1 & VR84 & \\

F21 & K3 III (2) & 11.39$\pm$0.03 & 1.94$\pm$0.02 & 1.27$\pm$0.22 & 4.97$\pm$0.04 & 3.00$\pm$0.60 & 3.2$\pm$1.5 & 2.2$\pm$0.7 & VR84 & s\\

F23 & M5 III (3) & 12.93$\pm$0.06 & 2.55$\pm$0.06 & 1.63$\pm$0.03 & 9.09$\pm$0.08 & 5.96$\pm$1.60 & 3.7$\pm$1.9 & 3.4$\pm$1.8  & W87 & s\\

F24 & B6 IV/V (1) & 7.69$\pm$0.01 & 0.14$\pm$0.01 & -0.15$\pm$0.02 & 0.75$\pm$0.03 & -0.36$\pm$0.07 & 4.2$\pm$0.4 & 1.2$\pm$0.1 & MSS & \\

F25 & G8 (3) & 13.23$\pm$0.06 & 2.22$\pm$0.06 & 0.94$\pm$0.06 & 7.09$\pm$0.08 & 2.16$\pm$0.35 & 4.2$\pm$0.4 & 5.4$\pm$0.4  & W87 & s \\

F27 & M1.5 (2) & 14.80$\pm$0.10 & 1.26$\pm$0.10 & 1.47$\pm$0.03 & 7.52$\pm$0.10 & 3.99$\pm$0.12 & \tablenotemark{d} & \nodata  & VR84 & s \\

F28 & K4(3) & 15.14$\pm$0.10 & 2.50$\pm$0.12 & 1.39$\pm$0.33 & 8.82$\pm$0.10 & 3.26$\pm$0.90 & 5.5$\pm$2.0 & 6.1$\pm$1.0  & W87 & s \\

F29 & K4 III (2) & 13.33$\pm$0.03 & 1.74$\pm$0.02 & 1.39$\pm$0.22 & 5.67$\pm$0.04 & 3.26$\pm$0.60 & 7.6$\pm$5.1 & 2.7$\pm$0.7  & VR84 & s \\

F30 & K3 III (1) & 11.47$\pm$0.03 & 1.72$\pm$0.02 & 1.27$\pm$0.11 & 4.77$\pm$0.04 & 3.00$\pm$0.30 & 4.3$\pm$1.3 & 1.9$\pm$0.3  & VR84 & \\

F32 & F0 V (1) & 10.53$\pm$0.06 & 0.66$\pm$0.07 & 0.30$\pm$0.03 & 2.71$\pm$0.07 & 0.70$\pm$0.06 & 6.1$\pm$1.3 & 2.2$\pm$0.1  & VR84 & s \\

F36 & K0 (3) & 13.76$\pm$0.06 & 2.22$\pm$0.06 & 1.00$\pm$0.24 & 7.51$\pm$0.08 & 2.31$\pm$0.70 & 4.7$\pm$1.1 & 5.7$\pm$0.8  & W87 & s \\

F37 & G9 III (1) & 9.58$\pm$0.02 & 1.27$\pm$0.05 & 0.97$\pm$0.03 & 3.38$\pm$0.03 & 2.24$\pm$0.08 & 4.3$\pm$0.9 & 1.3$\pm$0.1 & VR84 & \\

F39 & K3 III (1) & 10.12$\pm$0.04 & 1.50$\pm$0.15 & 1.27$\pm$0.11 & 3.82$\pm$0.05 & 3.00$\pm$0.20 & 3.8$\pm$3.2 & 0.9$\pm$0.2  & VR84 & s\\

F40 & B8 III (1) & 7.61$\pm$0.01 & 0.52$\pm$0.01 & -0.11$\pm$0.02 & 1.71$\pm$0.02 & -0.24$\pm$0.09 & 3.4$\pm$0.2 & 2.1$\pm$0.1 & MSS & \\

F41 & B8 V (2) & 8.53$\pm$0.01 & 0.16$\pm$0.02 & -0.11$\pm$0.04 & 0.61$\pm$0.03 & 0.24$\pm$0.18 & 1.5$\pm$0.8 & 0.4$\pm$0.2 & MSS & s \\

F42 & A3/4 IV (2) & 8.37$\pm$0.01 & 0.42$\pm$0.02 & 0.05$\pm$0.07 & 1.35$\pm$0.02 & 0.22$\pm$0.16 & 3.4$\pm$0.8 & 1.2$\pm$0.2 & MSS & \\

F48 & B9.5 V (1) & 8.88$\pm$0.02 & 0.12$\pm$0.02 & 0.00$\pm$0.02 & 0.44$\pm$0.04 & -0.07$\pm$0.07 & 4.6$\pm$1.2 & 0.6$\pm$0.1 & MSS & \\

F50 & A8 IV (2) & 9.38$\pm$0.20 & 0.41$\pm$0.03 & 0.24$\pm$0.06 & 1.08$\pm$0.20 & 0.57$\pm$0.14 & 3.2$\pm$2.0 & 0.6$\pm$0.3 & MSS & \\

F52 & B9.5 V (1) & 8.51$\pm$0.01 & 0.23$\pm$0.02 & 0.00$\pm$0.02 & 1.01$\pm$0.03 & -0.07$\pm$0.07 & 5.2$\pm$0.7 & 1.2$\pm$0.1 & MSS & \\

F54 & G6 III/IV (1) & 8.40$\pm$0.01 & 1.09$\pm$0.02 & 0.89$\pm$0.03 & 2.89$\pm$0.02 & 2.15$\pm$0.05 & 4.0$\pm$0.7 & 0.8$\pm$0.1 & MSS & \\
\\
Musca\tablenotemark{e}\\
HD 109753 & B9 III/IV (1) & 8.58$\pm$0.01 & 0.16$\pm$0.02 & -0.07$\pm$0.05 & 0.73$\pm$0.03 & -0.13$\pm$0.13 & 4.15$\pm$1.16 & 0.95$\pm$0.15 & MSS & \\

HD109885 & B2 III (1) & 9.02$\pm$ 0.02 & 0.07$\pm$0.02 & -0.24$\pm$0.06 & 0.36$\pm$0.03 & -0.66$\pm$0.10 & 3.68$\pm$0.85 & 1.12$\pm$0.11 & MSS & \\

HD 110022 & B8 III/IV (1) & 7.84$\pm$0.01 & 0.10$\pm$0.02 & -0.11$\pm$0.05 & 0.54$\pm$0.03 & -0.24$\pm$0.09 & 4.12$\pm$1.15 & 0.86$\pm$0.10 & MSS & \\

HD107875 & B8 V (1) & 9.69$\pm$0.02 & 0.17$\pm$0.03 & -0.11$\pm$0.05 & 0.68$\pm$0.03 & -0.24$\pm$0.09 & 3.56$\pm$ 0.80 & 1.01$\pm$0.10 & MSS & \\

HD 109082 & B9 IV (1) & 8.12$\pm$0.01 & 0.19$\pm$0.02 & -0.07$\pm$0.07 & 0.81$\pm$0.03 & -0.13$\pm$0.13 & 3.95$\pm$1.22 & 1.04$\pm$0.15 & MSS & \\

HD109234 & B9.5/A0 V (1) & 9.67$\pm$0.02 & 0.14$\pm$ 0.03 & -0.07$\pm$0.10 & 0.79$\pm$0.03 & -0.13$\pm$0.20 & 4.81$\pm$2.60 & 1.01$\pm$0.22 & MSS & \\

HD 109565 & B8 II (1) & 9.94$\pm$0.03 & 0.16$\pm$0.03 & -0.11$\pm$0.05 & 0.43$\pm$0.03 & -0.24$\pm$0.09 & 2.75$\pm$0.72 & 0.74$\pm$0.11 & MSS & \\

HD 106328 & B4 IV (1) & 8.44$\pm$0.01 & 0.05$\pm$0.02 & -0.19$\pm$0.02 & 0.39$\pm$0.03 & -0.49$\pm$0.07 & 4.00$\pm$0.56 & 0.96$\pm$0.08 & MSS & \\

HD109399 & B0.5 III (1) & 7.63$\pm$0.01 & -0.03$\pm$0.02 & -0.28$\pm$0.04 & -0.01$\pm$0.03 & -0.79$\pm$0.06 & 3.43$\pm$0.66 & 0.86$\pm$0.07 & MSS & \\

HD 107983 & B9/9.5 V (1) & 8.09$\pm$0.01 & 0.13$\pm$0.02 & -0.07$\pm$0.07 & 0.60$\pm$0.03 & -0.13$\pm$0.13 & 4.01$\pm$1.61 & 0.80$\pm$0.15 & MSS & \\

HD 106147 & B9 IV (1) & 8.65$\pm$0.02 & 0.09$\pm$0.02 & -0.07$\pm$0.07 & 0.44$\pm$0.03 & -0.13$\pm$0.13 & 4.00$\pm$2.07 & 0.63$\pm$0.15 & MSS & \\

CD-70 925 & K5 V (1) & 9.16$\pm$0.02 & 1.10$\pm$0.04 & 1.15$\pm$0.08 & 3.07$\pm$ 0.03 & 2.85$\pm$0.20 & \tablenotemark{d} &\nodata & Y & \\

CPD-69 1677 & A0 (1)  & 10.06$\pm$0.03 & 0.27$\pm$0.04 & -0.02$\pm$0.06 & 0.67$\pm$0.04 & 0.00$\pm$0.07 & 2.60$\pm$0.70 & 0.74$\pm$0.09 & Y & \\

HD 110118 & B9 IV (1) & 8.44$\pm$0.01 & 0.10$\pm$0.02 & -0.07$\pm$0.07 & 0.41$\pm$0.04 & -0.13$\pm$0.13 & 3.44$\pm$1.66 & 0.60$\pm$0.15 & MSS & \\

CD-69 1024 & F0 (1)  & 9.75$\pm$0.02 & 0.44$\pm$0.03 & 0.30$\pm$0.03 & 1.38$\pm$0.04 & 0.70$\pm$0.07 & 5.30$\pm$1.71 & 0.74$\pm$0.09 & Y & \\

HD 107252 & A3 V (1) & 9.02$\pm$0.02 & 0.23$\pm$0.02 & 0.08$\pm$0.03 & 0.91$\pm$0.02 & 0.22$\pm$0.08 & 5.10$\pm$1.37 & 0.76$\pm$0.09 & MSS & \\

HD 109233 & A0 III (1) & 9.37$\pm$0.02 & 0.13$\pm$0.02 & -0.02$\pm$0.03 & 0.54$\pm$0.03 & 0.00$\pm$0.07 & 4.01$\pm$1.13 & 0.60$\pm$0.08 & MSS & \\

HD 110080 & A5 V (1) & 7.42$\pm$0.01 & 0.24$\pm$0.02 & 0.15$\pm$0.03 & 0.76$\pm$0.03 & 0.38$\pm$0.11 & 4.58$\pm$2.17 & 0.42$\pm$0.13 & MSS & \\
\\
Ophiuchus\\
001 & G2 V (1) & 8.72$\pm$0.02 & 0.69$\pm$0.03 & 0.63$\pm$0.03 & \nodata & 1.46$\pm$0.03 & \nodata & \nodata & VR93 & \\
005 & A2 (1) & 9.32$\pm$0.02 & 0.63$\pm$0.03 & 0.05$\pm$0.03 & 2.26$\pm$0.03 & 0.14$\pm$0.07 & 4.0$\pm$0.3 & 2.3$\pm$0.1 & VR93 & \\
006 & B9 (1) & 8.60$\pm$0.02 & 0.39$\pm$0.03 & -0.06$\pm$0.05 & 1.47$\pm$0.03 & -0.13$\pm$0.07 & 3.9$\pm$0.5 & 1.8$\pm$0.1 & VR93 & \\
010 & M4 III (2) & 13.42$\pm$0.02 & 2.53$\pm$0.03 & 1.62$\pm$0.02 & 11.18$\pm$0.04 & 5.10$\pm$1.70 & 7.3$\pm$2.1 & 6.7$\pm$1.9 & VR93 &s\\
014 & F7 III (1) & 12.03$\pm$0.02 & 1.32$\pm$0.03 & 0.49$\pm$0.03 & 4.41$\pm$0.04 & 1.32$\pm$0.11 & 4.1$\pm$0.3 & 3.4$\pm$0.1 & VR93 &s\\
015 & M4e III (2) & 13.59$\pm$0.02 & 2.31$\pm$0.03 & 1.62$\pm$0.02 & 8.19$\pm$0.04 & 5.10$\pm$1.70 & 4.9$\pm$2.7 & 3.4$\pm$1.9 & VR93 &\\
021 & G8 III (1) & 11.96$\pm$0.02 & 1.28$\pm$0.03 & 0.94$\pm$0.03 & 4.18$\pm$0.03 & 2.16$\pm$0.08 & 6.5$\pm$0.9 & 2.2$\pm$0.1 & VR93 &s \\
024 & F5 III (2) & 12.44$\pm$0.02 & 1.30$\pm$0.03 & 0.44$\pm$0.06 & 5.04$\pm$0.03 & 1.10$\pm$0.22 & 5.0$\pm$0.5 & 4.3$\pm$0.2 & VR93 &s\\
028 & G2 III (2) & 11.26$\pm$0.02 & 1.31$\pm$0.03 & 0.63$\pm$0.10 & 4.61$\pm$0.03 & 1.91$\pm$0.16 & 4.4$\pm$0.7 & 3.0$\pm$0.2 & VR93 & s \\
035 & \nodata & 13.66$\pm$0.02 & 1.49$\pm$0.03 & \nodata & 5.74$\pm$0.03 & \nodata & \nodata & \nodata & VR93 & s \\
036 & M2 III (2) & 12.21$\pm$0.02 & 2.52$\pm$0.03 & 1.60$\pm$0.04 & 7.38$\pm$0.03 & 4.30$\pm$0.80 & 3.7$\pm$1.0 & 3.4$\pm$0.9 & VR93 & s \\
041 & \nodata & 13.77$\pm$0.02 & 1.90$\pm$0.03 & \nodata & 5.16$\pm$0.03 & \nodata & \nodata & \nodata & VR93 & s \\
042 & M3 III (2) & 12.16$\pm$0.02 & 2.16$\pm$0.03 & 1.61$\pm$0.02 & 6.39$\pm$0.03 & 4.64$\pm$1.30 & 3.5$\pm$2.6 & 1.9$\pm$1.4 & VR93 & s \\
043 & K0 III (1) & 11.80$\pm$0.02 & 1.55$\pm$0.03 & 1.00$\pm$0.08 & 4.33$\pm$0.03 & 2.31$\pm$0.20 & 4.0$\pm$0.7 & 2.2$\pm$0.2 &VR93 & \\
049 & G3 + K2 (1) & 10.57$\pm$0.02 & 1.61$\pm$0.03 & \nodata & 4.09$\pm$0.04 & \nodata & \nodata & \nodata & VR93 &\\
050 & B9 (1) & 10.12$\pm$0.02 & 0.94$\pm$0.03 & -0.06$\pm$0.05 & 3.37$\pm$0.03 & -0.13$\pm$0.07 & 3.8$\pm$0.2 & 3.8$\pm$0.1 & VR93 &\\
052 & \nodata & 13.40$\pm$0.02 & 1.32$\pm$0.03 & \nodata & 4.19$\pm$0.03 & \nodata & \nodata & \nodata & VR93 &\\
053 & F8 III (1) & 12.58$\pm$0.02 & 1.23$\pm$0.03 & 0.52$\pm$0.03 & 4.24$\pm$0.03 & 1.35$\pm$0.03 & 4.5$\pm$0.3 & 3.2$\pm$0.1 & VR93 &\\
055 & \nodata & 13.94$\pm$0.02 & 2.26$\pm$0.03 & \nodata & 6.78$\pm$0.03 & \nodata & \nodata & \nodata & VR93 &\\
056 & \nodata & 12.94$\pm$0.02 & 1.00$\pm$0.03 & \nodata & 2.71$\pm$0.03 & \nodata & \nodata & \nodata & VR93 &\\
066 & \nodata & 13.34$\pm$0.02 & 1.95$\pm$0.03 & \nodata & 6.07$\pm$0.03 & \nodata & \nodata & \nodata & VR93 &\\
069 & A0 (1) & 12.74$\pm$0.02 & 0.45$\pm$0.03 & -0.02$\pm$0.03 & 3.21$\pm$0.03 & 0.00$\pm$0.07 & 7.5$\pm$0.7 & 3.5$\pm$0.1 & VR93 &\\
074 & \nodata & 14.08$\pm$0.02 & 1.57$\pm$0.03 & \nodata & 5.51$\pm$0.03 & \nodata & \nodata & \nodata & VR93 &\\
075 & K0 III (1) & 8.22$\pm$0.02 & 1.18$\pm$0.03 & 1.00$\pm$0.08 & 3.15$\pm$0.04 & 2.31$\pm$0.20 & 5.1$\pm$2.7 & 0.9$\pm$0.2 & VR93 &\\
080 & G5 III (2) & 12.10$\pm$0.02 & 1.45$\pm$0.03 & 0.86$\pm$0.06 & 4.44$\pm$0.03 & 2.10$\pm$0.10 & 4.4$\pm$0.5 & 2.6$\pm$0.1 & VR93 &\\
081 & K2 III (1) & 11.82$\pm$0.02 & 1.90$\pm$0.03 & 1.16$\pm$0.08 & 5.43$\pm$0.03 & 2.70$\pm$0.30 & 4.1$\pm$0.7 & 3.0$\pm$0.3 & VR93 &\\
082 & M2 III (2) & 11.15$\pm$0.02 & 1.85$\pm$0.03 & 1.60$\pm$0.04 & 9.25$\pm$0.04 & 4.30$\pm$0.80 & \tablenotemark{d} & \nodata & VR93 & s\\
084 & F2 (1) & 12.86$\pm$0.02 & 1.27$\pm$0.03 & 0.35$\pm$0.03 & 3.07$\pm$0.03 & 0.82$\pm$0.06 & 2.7$\pm$0.2 & 2.5$\pm$0.1 & VR93 &\\
085 & K0 III (2) & 12.48$\pm$0.02 & 2.14$\pm$0.03 & 1.00$\pm$0.16 & 6.12$\pm$0.03 & 2.31$\pm$0.40 & 3.7$\pm$0.7 & 4.2$\pm$0.4 & VR93 &\\
086 & \nodata & 13.51$\pm$0.02 & 1.83$\pm$0.03 & \nodata & 5.48$\pm$0.04 & \nodata & \nodata & \nodata & VR93 &\\
087 & K0 III (1) & 9.89$\pm$0.02 & 1.91$\pm$0.03 & 1.00$\pm$0.08 & 5.09$\pm$0.04 & 2.31$\pm$0.20 & 3.4$\pm$0.4 & 3.1$\pm$0.2 & VR93 &\\
094 & G5 III (2) & 12.68$\pm$0.02 & 2.21$\pm$0.03 & 0.86$\pm$0.06 & 6.29$\pm$0.04 & 2.10$\pm$0.10 & 3.4$\pm$0.2 & 4.6$\pm$0.1 & VR93 &s \\
096 & K0 III (2) & 12.93$\pm$0.02 & 2.17$\pm$0.03 & 1.00$\pm$0.16 & 6.51$\pm$0.03 & 2.31$\pm$0.40 & 3.9$\pm$0.7 & 4.6$\pm$0.4 & VR93 &s \\
097 & K3 III (2) & 10.99$\pm$0.02 & 2.04$\pm$0.03 & 1.24$\pm$0.16 & 5.55$\pm$0.03 & 3.00$\pm$0.60 & 3.5$\pm$1.1 & 2.8$\pm$0.7 & VR93 &\\
098 & \nodata & 14.01$\pm$0.02 & 2.85$\pm$0.03 & \nodata & 8.46$\pm$0.03 & \nodata & \nodata & \nodata & VR93 &\\
102 & \nodata & 14.44$\pm$0.02 & 1.88$\pm$0.03 & \nodata & 5.34$\pm$0.03 & \nodata & \nodata & \nodata & VR93 &\\
104 & K0 III (2) & 13.49$\pm$0.02 & 1.37$\pm$0.03 & 1.00$\pm$0.16 & 3.58$\pm$0.03 & 2.31$\pm$0.40 & 3.8$\pm$2.0 & 1.4$\pm$0.4 & VR93 &\\
105 & K-M & 12.63$\pm$0.02 & 1.62$\pm$0.03 & \nodata & 4.86$\pm$0.03 & \nodata &  & \nodata & VR93 &\\
106 & G5 III (1) & 13.22$\pm$0.02 & 1.53$\pm$0.03 & 0.86$\pm$0.03 & 4.60$\pm$0.03 & 2.10$\pm$0.05 & 4.1$\pm$0.3 & 2.7$\pm$0.1 & VR93 &\\
110 & \nodata & 14.14$\pm$0.02 & 1.16$\pm$0.03 & \nodata & 4.16$\pm$0.03 & \nodata & \nodata & \nodata & VR93 &\\
111 & \nodata & 14.10$\pm$0.02 & 2.70$\pm$0.03 & \nodata & 7.27$\pm$0.03 & \nodata & \nodata & \nodata & VR93 &\\
115 & \nodata & 14.04$\pm$0.02 & 1.14$\pm$0.03 & \nodata & 3.41$\pm$0.03 & \nodata & \nodata & \nodata & VR93 &\\
117 & \nodata & 14.33$\pm$0.02 & 1.71$\pm$0.03 & \nodata & 4.80$\pm$0.03 & \nodata & \nodata & \nodata & VR93 &\\
134 & A1 (1) & 12.16$\pm$0.02 & 1.04$\pm$0.03 & 0.02$\pm$0.03 & \nodata & 0.07$\pm$0.07 & \nodata & \nodata & VR93 &\\
136 & B8 (1) & 7.07$\pm$0.02 & 0.15$\pm$0.03 & -0.11$\pm$0.04 & 0.68$\pm$0.03 & -0.24$\pm$0.10 & 3.9$\pm$0.9 & 1.0$\pm$0.1 & VR93 &\\
143 & F-G & 14.10$\pm$0.02 & 0.95$\pm$0.03 & \nodata & \nodata & \nodata & \nodata & \nodata & VR93 &\\
144 & K5 III (2) & 10.88$\pm$0.02 & 1.91$\pm$0.03 & 1.50$\pm$0.16 & \nodata &  3.60$\pm$0.60 & \nodata & VR93 &\\
146 & B5 (2) & 11.00$\pm$0.02 & 0.87$\pm$0.03 & -0.17$\pm$0.04 & 3.25$\pm$0.03 & -0.42$\pm$0.12 & 3.9$\pm$0.2 & 4.0$\pm$0.1 & VR93 &\\
147 & \nodata & 13.77$\pm$0.02 & 1.88$\pm$0.03 & \nodata & \nodata & \nodata & \nodata & \nodata & VR93 &\\
148 & \nodata & 13.87$\pm$0.02 & 1.71$\pm$0.03 & \nodata & \nodata & \nodata & \nodata & \nodata & VR93 &\\
150 & K0 III (1) & 11.33$\pm$0.02 & 1.61$\pm$0.03 & 1.00$\pm$0.08 & \nodata &  2.31$\pm$0.20 & \nodata & \nodata & VR93 &\\
151 & K-M & 13.08$\pm$0.02 & 2.09$\pm$0.03 & \nodata & \nodata & \nodata & \nodata & \nodata & VR93 &\\
152 & M5 III (2) & 12.40$\pm$0.02 & 2.59$\pm$0.03 & 1.63$\pm$0.02 & 8.24$\pm$0.04 & 5.96$\pm$1.80 & 2.6$\pm$2.1 & 2.5$\pm$2.0 & VR93 & \\
153 & G & 11.52$\pm$0.02 & 1.48$\pm$0.03 & \nodata & 4.83$\pm$0.03 & \nodata & & & VR93 &\\
\\
R~CrA\\
02 & F7 III (1) & 12.84$\pm$0.03 & 0.99$\pm$0.02 & 0.49$\pm$0.03 & 3.60$\pm$0.04 & 1.32$\pm$0.11 & 5.0$\pm$0.4 & 2.5$\pm$0.1 & VR84 & s \\

10 & B8V (2) & 10.13$\pm$0.06 & 0.56$\pm$0.07 & -0.11$\pm$0.09 & 2.55$\pm$0.07 & -0.24$\pm$0.12 & 4.6$\pm$0.8 & 3.1$\pm$0.2 & VR84 & s \\

12 & G8 III (2) & 11.87$\pm$0.03 & 1.48$\pm$0.02 & 0.94$\pm$0.06 & 3.74$\pm$0.04 & 2.16$\pm$0.15 & 3.2$\pm$0.5 & 1.7$\pm$0.2 & VR84 & \\

13 & K1 III (1) & 12.70$\pm$0.03 & 1.43$\pm$0.02 & 1.08$\pm$0.08 & 4.07$\pm$0.04 & 2.50$\pm$0.20 & 4.9$\pm$1.3 & 1.7$\pm$0.2 & VR84 & \\

15 & G1 (5) & 13.27$\pm$0.03 & 0.72$\pm$0.02 & 0.60$\pm$0.10 & 2.40$\pm$0.04 & 1.43$\pm$0.10 & 8.8$\pm$7.6 & 1.1$\pm$0.1 & \nodata & \\

22 & K5 III (2) & 11.50$\pm$0.03 & 2.17$\pm$0.02 & 1.50$\pm$0.22 & 6.58$\pm$0.04 & 3.60$\pm$0.30 & 4.9$\pm$1.7 & 3.3$\pm$0.3 & VR84 & s\\

28 & M5 III (1) & 10.54$\pm$0.03 & 1.95$\pm$0.02 & 1.63$\pm$0.05 & 7.12$\pm$0.05 & 5.96$\pm$0.80 & 4.0$\pm$2.8 & 1.3$\pm$0.9 & VR84 & \\

30 & A0 V (2) & 10.92$\pm$0.03 & 0.56$\pm$0.02 & -0.02$\pm$0.09 & 2.36$\pm$0.04 & 0.00$\pm$0.14 & 4.5$\pm$0.8 & 2.6$\pm$0.2 & VR84 & s \\

43 & K0 III (1) & 9.55$\pm$0.03 & 1.46$\pm$0.02 & 1.00$\pm$0.08 & 4.02$\pm$0.04 & 2.31$\pm$0.20 & 4.1$\pm$0.9 & 1.9$\pm$0.2 & VR84 & s \\

46 & G8 III (2) & 11.96$\pm$0.03 & 1.49$\pm$0.02 & 0.94$\pm$0.06 & 5.11$\pm$0.04 & 2.16$\pm$0.15 & 5.9$\pm$0.8 & 3.3$\pm$0.2 & VR84 & s \\

50 & A6 V (1) & 10.69$\pm$0.03 & 0.510$\pm$0.02 & 0.18$\pm$0.03 & 1.65$\pm$0.04 & 0.44$\pm$0.06 & 4.0$\pm$0.5 & 1.3$\pm$0.1 & VR84 & s \\

52 & G5 III (2) & 12.56$\pm$0.03 & 1.34$\pm$0.02 & 0.86$\pm$0.06 & 3.59$\pm$0.04 & 2.10$\pm$0.10 & 3.4$\pm$0.5 & 1.6$\pm$0.1 & VR84 & s \\

56 & G5 IV (2) & 10.82$\pm$0.03 & 1.21$\pm$0.02 & 0.68$\pm$0.05 & 3.29$\pm$0.04 & 1.58$\pm$0.10 & 3.5$\pm$0.4 & 1.9$\pm$0.1 & VR84 & \\

58 & K1 III (2) & 11.47$\pm$0.03 & 1.66$\pm$0.02 & 1.08$\pm$0.16 & 4.62$\pm$0.04 & 2.50$\pm$0.40 & 4.0$\pm$1.4 & 2.3$\pm$0.4 & VR84 & s \\

71 & F6 V (1) & 12.13$\pm$0.03 & 0.81$\pm$0.02 & 0.47$\pm$0.03 & 2.38$\pm$0.04 & 1.21$\pm$0.11 & 3.8$\pm$0.6 & 1.3$\pm$0.1 & VR84 & \\

73 & G0 V (1) & 12.380$\pm$0.03 & 0.92$\pm$0.02 & 0.58$\pm$0.03 & 2.47$\pm$0.04 & 1.41$\pm$0.03 & 3.4$\pm$0.4 & 1.2$\pm$0.1 & VR84 & s\\

88 & G8 & \nodata & 1.24$\pm$0.10 & 0.94$\pm$0.20 & \nodata & 2.16$\pm$0.80 & \nodata & \nodata & \nodata & \\

TY CrA & B9 IV (1) & 9.43$\pm$0.04 & 0.50$\pm$0.04 & -0.07$\pm$0.04 & 2.80$\pm$0.05 & -0.13$\pm$0.11 & 5.7$\pm$0.6 & 3.2$\pm$0.1 & MSS & s\\

\\
Taurus\\
HD 28170 & A3 V (1) & 8.99$\pm$0.02 & 0.51$\pm$0.03 & 0.08$\pm$0.03 & 1.55$\pm$0.03 & 0.22$\pm$0.08 & 3.4$\pm$0.4 & 1.5$\pm$0.1 & KDH94 & \\

HD 28225 & A3 III (1) & 7.81$\pm$0.01 & 0.51$\pm$0.02 & 0.08$\pm$0.03 & 1.36$\pm$0.02 & 0.22$\pm$0.08 & 2.9$\pm$0.3 & 1.3$\pm$0.1 & KDH94 & \\

HD 28975 & A4 III (1) & 9.08$\pm$0.02 & 0.66$\pm$0.03 & 0.11$\pm$0.03 & 2.00$\pm$0.03 & 0.30$\pm$0.08 & 3.4$\pm$0.3 & 1.9$\pm$0.1 & KDH94 & \\

HD 29333 & A2 V (1) & 8.79$\pm$0.05 & 0.65$\pm$0.02 & 0.05$\pm$0.03 & 1.99$\pm$0.06 & 0.14$\pm$0.08 & 3.4$\pm$0.3 & 2.0$\pm$0.1 & W01 & \\

HD 29647 & B6-7 IV (1) & 8.47$\pm$0.01 & 0.82$\pm$0.02 & -0.14$\pm$0.04 & 3.15$\pm$0.023 & -0.33$\pm$0.08 & 4.0$\pm$0.2 & 3.8$\pm$0.1 & W01 & s\\

HD 29835 & K2 III (1) & 8.68$\pm$0.02 & 1.56$\pm$0.05 & 1.16$\pm$0.08 & 3.81$\pm$0.03 & 2.70$\pm$0.30 & 3.1$\pm$1.1 & 1.2$\pm$0.3 & SM80 & \\

HD 30168 & B8 V (1) & 7.71$\pm$0.01 & 0.28$\pm$0.02 & -0.11$\pm$0.03 & 0.95$\pm$0.02 & -0.24$\pm$0.08 & 3.3$\pm$0.4 & 1.3$\pm$0.1 & KDH94 & \\

HD 30675 & B3 V (1) & 7.55$\pm$0.02 & 0.32$\pm$0.02 & -0.22$\pm$0.03 & 0.95$\pm$0.03 & -0.56$\pm$0.01 & 3.1$\pm$0.2 & 1.7$\pm$0.1 & KDH94 & \\

HD 279652 & A2 V (2) & 9.92$\pm$0.04 & 0.35$\pm$0.05 & 0.05$\pm$0.06 & 1.21$\pm$0.04 & 0.14$\pm$0.16 & 3.9$\pm$1.2 & 1.2$\pm$0.2 & U85 & \\

HD 279658 & A7 V (1) & 9.95$\pm$0.04 & 0.52$\pm$0.05 & 0.21$\pm$0.03 & 1.58$\pm$0.05 & 0.50$\pm$0.06 & 3.8$\pm$0.8 & 1.2$\pm$0.1 & U85 & \\

HD 283367 & B9 V (1) & 10.45$\pm$0.07 & 0.58$\pm$0.08 & -0.06$\pm$0.05 & 1.81$\pm$0.07 & -0.13$\pm$0.09 & 3.3$\pm$0.5 & 2.1$\pm$0.1 & VR85 & s \\

HD 283637 & B9/A0 V (2) & 11.23$\pm$0.02 & 0.74$\pm$0.02 & -0.06$\pm$0.12 & 2.07$\pm$0.03 & -0.13$\pm$0.18 & 3.0$\pm$0.5 & 2.4$\pm$0.2 & VR85 & \\

HD 283642 & A3V (1) & 10.42$\pm$0.07 & 0.82$\pm$0.10 & 0.08$\pm$0.03 & 2.09$\pm$0.08 & 0.22$\pm$0.08 & 2.8$\pm$0.4 & 2.1$\pm$0.1 & SM80 & \\

HD 283643 & 2A V (1) & 10.88$\pm$0.12 & 0.61$\pm$0.15 & 0.05$\pm$0.03 & 1.63$\pm$0.12 & 0.14$\pm$0.08 & 2.9$\pm$0.8 & 1.6$\pm$0.2 & SM80 & \\

HD 283701 & B8 III (2) & 9.74$\pm$0.04 & 0.74$\pm$0.06 & -0.11$\pm$0.03 & 2.34$\pm$0.05 & -0.24$\pm$0.18 & 3.3$\pm$0.4 & 2.8$\pm$0.2 & M68/KDH94 & \\

HD 283725 & F5 III (1) & 10.01$\pm$0.04 & 0.98$\pm$0.07 & 0.44$\pm$0.03 & 2.59$\pm$0.05 & 1.10$\pm$0.16 & 3.0$\pm$0.5 & 1.6$\pm$0.2 & KDH94 & \\

HD 283757 & A5 V (1) & 10.85$\pm$0.10 & 0.66$\pm$0.13 & 0.15$\pm$0.03 & 1.93$\pm$0.10 & 0.38$\pm$0.06 & 3.3$\pm$0.9 & 1.7$\pm$0.1 & SM80 & \\

HD 283800 & B5 V (1) & 9.84$\pm$0.04 & 0.32$\pm$0.06 & -0.17$\pm$0.03 & 1.181$\pm$0.05 & -0.42$\pm$0.06 & 3.6$\pm$0.5 & 1.8$\pm$0.1 & SM80 & s\\ 

HD 283809 & B3 V (1) & 10.73$\pm$0.10 & 1.29$\pm$0.26 & -0.22$\pm$0.03 & 4.60$\pm$0.10 & -0.56$\pm$0.10 & 3.8$\pm$0.7 & 5.7$\pm$0.2 & SCH85 & s \\

HD 283812 & A1 V (1) & 9.56$\pm$0.04 & 0.72$\pm$0.07 & 0.01$\pm$0.03 & 1.94$\pm$0.05 & 0.07$\pm$0.07 & 2.9$\pm$0.4 & 2.1$\pm$0.1 & KDH94 & \\ 

HD 283815 & A5 V (1) & 9.90$\pm$0.04 & 0.69$\pm$0.05 & 0.15$\pm$0.03 & 2.33$\pm$0.05 & 0.38$\pm$0.06 & 4.0$\pm$0.5 & 2.1$\pm$0.1 & SM80 & \\ 

HD 283855 & A2/0 (2) & 11.40$\pm$0.10 & 0.10$\pm$0.10 & 0.05$\pm$0.06 & 2.21$\pm$0.10 & 0.07$\pm$0.14 & \nodata & 2.4$\pm$0.2 & W01/N95 & \\

HD 283877 & F5/8 V (3) & 9.95$\pm$0.06 & 0.57$\pm$0.07 & 0.44$\pm$0.09 & 1.79$\pm$0.07 & 1.20$\pm$0.20 & 4.9$\pm$4.5 & 0.6$\pm$0.2 & SSS80/N95 & \\

HD 283879 & B5 V (2) & 11.08$\pm$0.02 & 1.01$\pm$0.02 & -0.17$\pm$0.06 & 2.68$\pm$0.03 & -0.42$\pm$0.12 & 2.9$\pm$0.2 & 3.4$\pm$0.1 & W01 & \\

\enddata
\tablenotetext{a}{Estimated uncertainties, in subclasses, are given in parenthesis}
\tablenotetext{b}{Spectral classes from: KDH94: \citet{kenyon1994}; M68: \citet{metreveli1968}; MSS: \citet{mss1975}; N95: \citet{nesterov1995}; SSS80: \citet{slutskij1980}; SM80: \citet{straizys1980}; SCH85: \citet{straizys1985}; U85: \citet{ungerer1985}; VR84: \citet{vrba1984}; VR85: \citet{vrba1985}; VR93: \citet{vrba1993}; W84: \citet{whittet1987}; W87: \citet{whittet1987}; W01: \citet{whittet2001}; Y \citet{yale1997}}
\tablenotetext{c}{Sightlines that are screened out in the Hipparcos field star based analysis. (see section \ref{screen})}
\tablenotetext{d}{The derived value is unphysical and is ignored in the following}
\tablenotetext{e}{Based on comparisons of Figure 1 in \citet{arnal1993} with the Aladin tool \citep{aladin2000} we have made the following target identifications in Musca from \citet{arnal1993}(AMZ): AMZ16=CD-71 836, AMZ18=CD-71 832, AMZ28=CD-70 925, AMZ41=CPD-69 1677, AMZ45=CD-69 1024}
\end{deluxetable}

\clearpage
\begin{deluxetable}{lrrr}
\tablecaption{Polarization Curve Fits for stars from the literature. \label{pol_fit}}
\tablewidth{0pt}
\tablehead{
\colhead{Star} & \colhead{p$_{max}$} & \colhead{$\lambda_{max}$} & \colhead{K}\\
 \colhead{ } &  \colhead{[\%]} & \colhead{[$\mu$m]} & \colhead{ }}
\startdata
Chamaeleon\\
F1 & 3.35$\pm$0.04 & 0.547$\pm$0.007 & 0.82$\pm$0.05 \\

F2 & 3.85$\pm$0.03 & 0.625$\pm$0.007 & 1.03$\pm$0.05 \\

F3 & 5.45$\pm$0.04 & 0.655$\pm$0.005 & \nodata \\

F6 & 5.48$\pm$0.03 & 0.576$\pm$0.007 & 1.00$\pm$0.03 \\

F7 & 5.92$\pm$0.02 & 0.538$\pm$0.005 & 0.82$\pm$0.01 \\

F9 & 4.82$\pm$0.05 & 0.628$\pm$0.006 & 0.94$\pm$0.03 \\

F11 & 4.81$\pm$0.04 & 0.530$\pm$0.009 & 0.92$\pm$0.04 \\

F16 & 7.30$\pm$0.06 & 0.618$\pm$0.012 & 1.03$\pm$0.05
 \\

F21 & 5.41$\pm$0.11 & 0.460$\pm$0.020 & 0.71$\pm$0.05 \\

F23 & 7.18$\pm$0.06 & 0.680$\pm$0.010 & \nodata \\

F24 & 3.07$\pm$0.04 & 0.557$\pm$0.008 & \nodata \\

F25 & 8.01$\pm$0.13 & 0.600$\pm$0.030 & 1.01$\pm$0.13 \\

F27 & 6.09$\pm$0.06 & 0.613$\pm$0.005 & \nodata \\

F28 & 7.02$\pm$0.09 & 0.722$\pm$0.008 & \nodata \\

F29 & 5.05$\pm$0.04 & 0.650$\pm$0.020 & 0.94$\pm$0.07 \\

F30  & 4.41$\pm$0.06 & 0.570$\pm$0.020 & 0.80$\pm$0.05 \\

F32 & 2.34$\pm$0.02 & 0.560$\pm$0.010 & 0.81$\pm$0.06 \\

F36 & 12.19$\pm$0.07 & 0.661$\pm$0.007 & 1.05$\pm$0.03 \\

F37 & 3.18$\pm$0.04 & 0.570$\pm$0.010 & \nodata \\

F39 & 3.22$\pm$0.07 & 0.480$\pm$0.030 & 0.77$\pm$0.12 \\

F40 & 8.01$\pm$0.02 & 0.569$\pm$0.005 & 0.96$\pm$0.02 \\

F41 & 2.60$\pm$0.03 & 0.590$\pm$0.020 & 0.87$\pm$0.11 \\

F42 & 2.87$\pm$0.03 & 0.600$\pm$0.010 & 0.87$\pm$0.10 \\

F48 & 2.29$\pm$0.04 & 0.570$\pm$0.020 & 0.99$\pm$0.15 \\

F50 & 1.61$\pm$0.05 & 0.570$\pm$0.020 & \nodata \\

F52 & 2.99$\pm$0.04 & 0.610$\pm$0.010 & 0.99$\pm$0.11 \\

F54 & 2.68$\pm$0.03 & 0.520$\pm$0.020 & 0.77$\pm$0.11 \\

\\
Musca\\
HD 109753 & 2.12$\pm$0.02 & 0.60$\pm$0.02 & 0.65$\pm$0.15 \\

HD109885 & 3.82$\pm$0.02 & 0.592$\pm$0.005 & \nodata \\

HD 110022 & 2.59$\pm$0.01 & 0.587$\pm$0.005 & \nodata \\

HD107875 & 2.65$\pm$0.03 & 0.505$\pm$0.009\tablenotemark{a}  & \nodata \\

HD 109082 & 3.41$\pm$0.02 & 0.580$\pm$0.006 & \nodata \\

HD109234 & 3.67$\pm$0.04 & 0.577$\pm$0.007 & 1.51$\pm$0.18 \\

HD 109565 & 3.17$\pm$0.04 & 0.56$\pm$0.01 & 0.93$\pm$0.21 \\

HD 106328 & 2.43$\pm$0.02 & 0.602$\pm$0.008 & \nodata \\

HD109399 & 2.14$\pm$0.02 & 0.560$\pm$0.007 & \nodata \\

HD 107983 & 2.36$\pm$0.02 & 0.56$\pm$0.01 & \nodata \\

HD 106147 & 1.83$\pm$0.03 & 0.58$\pm$0.01 & 0.73$\pm$0.18\\

CD-70 925 & 2.06$\pm$0.02 & 0.54$\pm$0.01 & \nodata \\

CPD-69 1677 & 2.48$\pm$0.02 & 0.59$\pm$0.01 & \nodata \\

HD 110118 & 1.86$\pm$0.02 & 0.58$\pm$0.01 & 0.55$\pm$0.16\\

CD-69 1024 & 2.49$\pm$0.03 & 0.58$\pm$0.02 & 0.54$\pm$0.20 \\

HD 107252 & 2.42$\pm$0.03 & 0.54$\pm$0.01 & \nodata \\

HD 109233 & 2.10$\pm$0.02 & 0.58$\pm$0.01 & \nodata \\

HD 110080 & 1.42$\pm$0.02 & 0.56$\pm$0.01 & \nodata \\

\\
Ophiuchus\\
005 & 0.50$\pm$0.04 & 0.50$\pm$0.05 & \nodata \\
006 & 0.86$\pm$0.02 & 0.65$\pm$0.03 & \nodata \\
010 & 3.83$\pm$0.14 & 0.64$\pm$0.04 & \nodata \\
014 & 1.42$\pm$0.21 & 0.93$\pm$0.20 & \nodata \\
015 & 6.45$\pm$0.15 & 0.89$\pm$0.04 & \nodata \\
021 & 2.59$\pm$0.07 & 0.66$\pm$0.03 & \nodata \\
024 & 4.31$\pm$0.10 & 0.74$\pm$0.03 & \nodata \\
036 & 4.11$\pm$0.07 & 0.77$\pm$0.04 & \nodata \\
041 & 4.87$\pm$0.13 & 0.66$\pm$0.04 & \nodata \\
042 & 3.65$\pm$0.06 & 0.70$\pm$0.03 & \nodata \\
043 & 2.73$\pm$0.07 & 0.71$\pm$0.03 & \nodata \\
049 & 2.49$\pm$0.07 & 0.73$\pm$0.02 & \nodata \\
050 & 2.34$\pm$0.04 & 0.69$\pm$0.02 & \nodata \\
052 & 4.56$\pm$0.14 & 0.71$\pm$0.05 & \nodata \\
053 & 2.63$\pm$0.09 & 0.67$\pm$0.05 & \nodata \\
055 & 5.53$\pm$0.13 & 0.72$\pm$0.05 & \nodata \\
056 & 1.37$\pm$0.41 & 1.05$\pm$0.22 & \nodata \\
066 & 2.95$\pm$0.09 & 0.74$\pm$0.06 & \nodata \\
069 & 1.96$\pm$0.13 & 0.57$\pm$0.07 & \nodata \\
074 & 7.94$\pm$0.17 & 0.72$\pm$0.04 & \nodata \\
075 & 0.97$\pm$0.03 & 0.51$\pm$0.04 & \nodata \\
080 & 3.99$\pm$0.05 & 0.60$\pm$0.02 & \nodata \\
081 & 4.55$\pm$0.10 & 0.50$\pm$0.02 & \nodata \\
082 & 3.05$\pm$0.02 & 0.67$\pm$0.01 & \nodata \\
084 & 5.47$\pm$0.09 & 0.72$\pm$0.02 & \nodata \\
085 & 2.72$\pm$0.06 & 0.76$\pm$0.04 & \nodata \\
086 & 6.54$\pm$0.19 & 0.79$\pm$0.04 & \nodata \\
087 & 4.26$\pm$0.06 & 0.55$\pm$0.01 & \nodata \\
094 & 5.33$\pm$0.09 & 0.69$\pm$0.02 & \nodata \\
096 & 3.42$\pm$0.08 & 0.82$\pm$0.06 & \nodata \\
097 & 3.07$\pm$0.04 & 0.66$\pm$0.02 & \nodata \\
098 & 5.97$\pm$0.09 & 0.78$\pm$0.04 & \nodata \\
102 & 8.24$\pm$0.34 & 0.66$\pm$0.06 & \nodata \\
104 & 5.17$\pm$0.14 & 0.65$\pm$0.03 & \nodata \\
105 & 4.86$\pm$0.21 & 0.65$\pm$0.05 & \nodata \\
106 & 8.36$\pm$0.12 & 0.61$\pm$0.01 & \nodata \\
110 & 9.31$\pm$0.38 & 0.75$\pm$0.06 & \nodata \\
111 & 9.07$\pm$0.20 & 0.62$\pm$0.02 & \nodata \\
115 & 3.60$\pm$0.25 & 0.55$\pm$0.07 & \nodata \\
134 & 1.55$\pm$0.05 & 0.72$\pm$0.05 & \nodata \\
136 & 0.84$\pm$0.02 & 0.71$\pm$0.02 & \nodata \\
144 & 1.69$\pm$0.07 & 0.61$\pm$0.03 & \nodata \\
146 & 4.88$\pm$0.08 & 0.54$\pm$0.01 & \nodata \\
147 & 7.53$\pm$0.30 & 0.65$\pm$0.04 & \nodata \\
150 & 3.40$\pm$0.07 & 0.69$\pm$0.03 & \nodata \\
153 & 7.69$\pm$0.05 & 0.67$\pm$0.01 & \nodata \\

\\
R~CrA\\
02 & 1.79$\pm$0.04 & 0.87$\pm$0.01 & \nodata \\

10 & 0.67$\pm$0.02 & 0.85$\pm$0.03 & \nodata \\

12 & 0.82$\pm$0.02 & 0.77$\pm$0.02 & \nodata \\

13 & 2.77$\pm$0.04 & 0.83$\pm$0.01 & \nodata \\

15 & 2.99$\pm$0.04 & 0.77$\pm$0.01 & \nodata \\

22 & 1.25$\pm$0.00 & 0.50$\pm$0.02 & \nodata \\

28 & 2.10$\pm$0.02 & 0.78$\pm$0.01 & \nodata \\

30 & 1.86$\pm$0.02 & 0.80$\pm$0.01 & \nodata \\

43 & 1.81$\pm$0.02 & 0.74$\pm$0.01 & \nodata \\

46 & 2.61$\pm$0.03 & 0.81$\pm$0.01 & \nodata \\

50 & 1.12$\pm$0.02 & 0.76$\pm$0.02 & \nodata \\

52 & 2.01$\pm$0.04 & 0.70$\pm$0.01 & \nodata \\

56 & 2.09$\pm$0.02 & 0.67$\pm$0.01 & \nodata \\

58 & 0.84$\pm$0.03 & 0.68$\pm$0.02 & \nodata \\

71 & 1.04$\pm$0.04 & 0.70$\pm$0.03 & \nodata \\

73 & 1.83$\pm$0.03 & 0.68$\pm$0.01 & \nodata \\

88 & 4.73$\pm$0.05 & 0.84$\pm$0.01 & \nodata \\

TY CrA & 0.86$\pm$0.02 & 0.62$\pm$0.04 & \nodata \\

\\
Taurus\\
HD 28170 & 1.92$\pm$0.02 & 0.55$\pm$0.01 & 0.96$\pm$0.04 \\

HD 28225 & 1.85$\pm$0.02 & 0.56$\pm$0.01 & 0.98$\pm$0.05 \\

HD 28975 & 3.16$\pm$0.02 & 0.54$\pm$0.01 & 0.85$\pm$0.04 \\

HD 29333 & 5.26$\pm$0.04 & 0.55$\pm$0.01 & 0.88$\pm$0.04 \\

HD 29647 & 2.33$\pm$0.03 & 0.74$\pm$0.01 & \nodata \\

HD 29835 & 4.07$\pm$0.03 & 0.50$\pm$0.01 & 0.87$\pm$0.04 \\

HD 30168 & 4.07$\pm$0.03 & 0.53$\pm$0.01 & 0.83$\pm$0.03 \\

HD 30675 & 3.90$\pm$0.04 & 0.53$\pm$0.01 & 0.95$\pm$0.06 \\

HD 279652 & 1.28$\pm$0.05 & 0.48$\pm$0.06 & 0.77$\pm$0.15 \\

HD 279658 & 2.82$\pm$0.03 & 0.54$\pm$0.01 & 0.97$\pm$0.05 \\

HD 283367 & 1.41$\pm$0.02 & 0.70$\pm$0.01 & \nodata \\

HD 283637 & 2.72$\pm$0.03 & 0.58$\pm$0.02 & 0.87$\pm$0.09 \\

HD 283642 & 1.97$\pm$0.01 & 0.61$\pm$0.01 & 1.01$\pm$0.04 \\

HD 283643 & 1.37$\pm$0.02 & 0.64$\pm$0.01 & \nodata \\

HD 283701 & 3.17$\pm$0.04 & 0.60$\pm$0.01 & 0.84$\pm$0.06 \\

HD 283725 & 4.71$\pm$0.03 & 0.49$\pm$0.01 & 0.71$\pm$0.03 \\

HD 283757 & 2.89$\pm$0.01 & 0.60$\pm$0.01 & \nodata \\

HD 283800 & 3.95$\pm$0.03 & 0.53$\pm$0.01 & 0.85$\pm$0.04 \\ 

HD 283809 & 6.69$\pm$0.04 & 0.59$\pm$0.01 & 0.91$\pm$0.03 \\

HD 283812 & 6.29$\pm$0.03 & 0.55$\pm$0.01 & 0.93$\pm$0.02 \\ 

HD 283815 & 2.82$\pm$0.01 & 0.60$\pm$0.01 & \nodata \\ 

HD 283855 & 5.12$\pm$0.04 & 0.52$\pm$0.01 & 0.87$\pm$0.03 \\

HD 283877 & 1.06$\pm$0.02 & 0.63$\pm$0.01 & 0.96$\pm$0.06 \\

HD 283879 & 4.14$\pm$0.02 & 0.63$\pm$0.01 & \nodata \\
\enddata
\tablenotetext{a}{For HD~107875 in Musca \citet{arnal1993} report $\lambda_{max}$=0.641$\pm$0.009.  This is only case of a significant discrepancy in $\lambda_{max}$ for our polarization curve fits compared to the literature.  This sightline is excluded from the analysis.}
\end{deluxetable}

\clearpage
\begin{deluxetable}{lrrrr}
\tablecaption{B stars affecting the dust \label{app1tab}}
\tablewidth{0pt}
\tablehead{
\colhead{Cloud} & \colhead{Star} & \colhead{Sp. Class} & \colhead{V} & \colhead{d\tablenotemark{1}}\\
\colhead{} & \colhead{} &\colhead{} &\colhead{[pc]}}
\startdata
Chamaeleon & HD 97300 & B9 V & 9.0 & 188$\pm$36 \\
Musca & HD 109668 & B2 IV & 2.7 & 94$\pm$4\\
Ophiuchus & $\sigma$~Sco  & B1 III  & 2.9 & 225$\pm$41 \\
Ophiuchus & $\rho$~Oph~D  & B3/4 V  & 6.8 & 136$\pm$25 \\
R~CrA & HD 175362 & B3 V  & 5.4 & 130$\pm$16 \\
Taurus & HD 30122 & B5 III & 6.3 & 216$\pm$34 \\
\enddata
\tablenotetext{1}{Based on Hipparcos trigonometric parallaxes}
\end{deluxetable}

\clearpage

\clearpage

\clearpage

\clearpage

\clearpage

\clearpage

\clearpage

\clearpage

\clearpage

\clearpage
\begin{figure}
\plotone{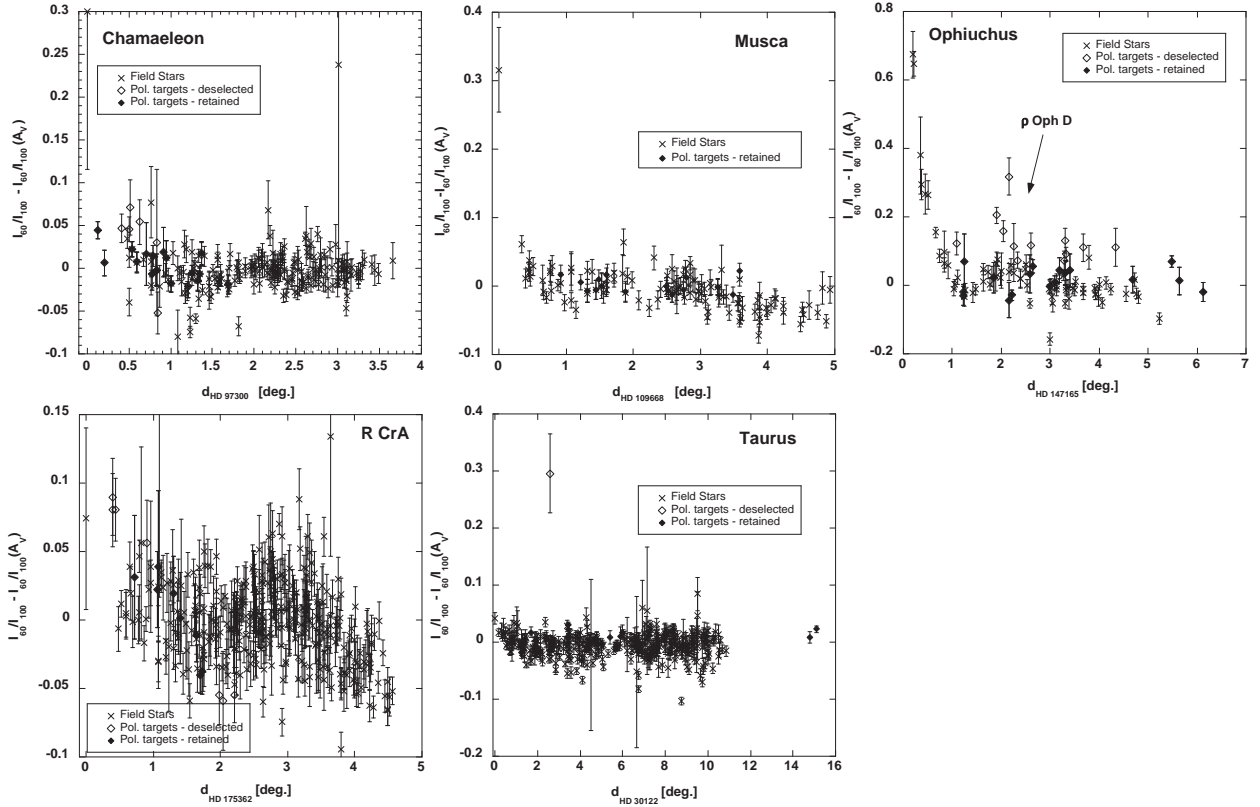}
\caption{The offset from the best-fit I$_{60}$/I$_{100}$ vs. A$_V$ relations (Tycho selected field stars) are plotted as functions of the 2-dimensional distance from nearby hot stars.  For lines of sight through Chamaeleon the dominant star is HD~97300 (top left); for Musca HD~109668; for Ophiuchus, $\sigma$~Sco and $\rho$~Oph~D both yield noticeable effects.  For lines of sight through R~CrA and Taurus the dominant stars are HD~175362 and HD~30122, respectively.  Stars from the Tycho field star samples are plotted with x:es while diamonds indicate stars in our polarization samples.  Open diamonds indicate sightlines classified as "anomalous".  The polarization sample stars with the largest offsets tend to be located close to the hot stars, indicating that many of the "anomalous" I$_{60}$/I$_{100}$ ratios are due to the additional irradiation from these stars.}
\label{DFIR_all}
\end{figure}

\clearpage
\begin{figure}
\epsscale{0.70}
\plotone{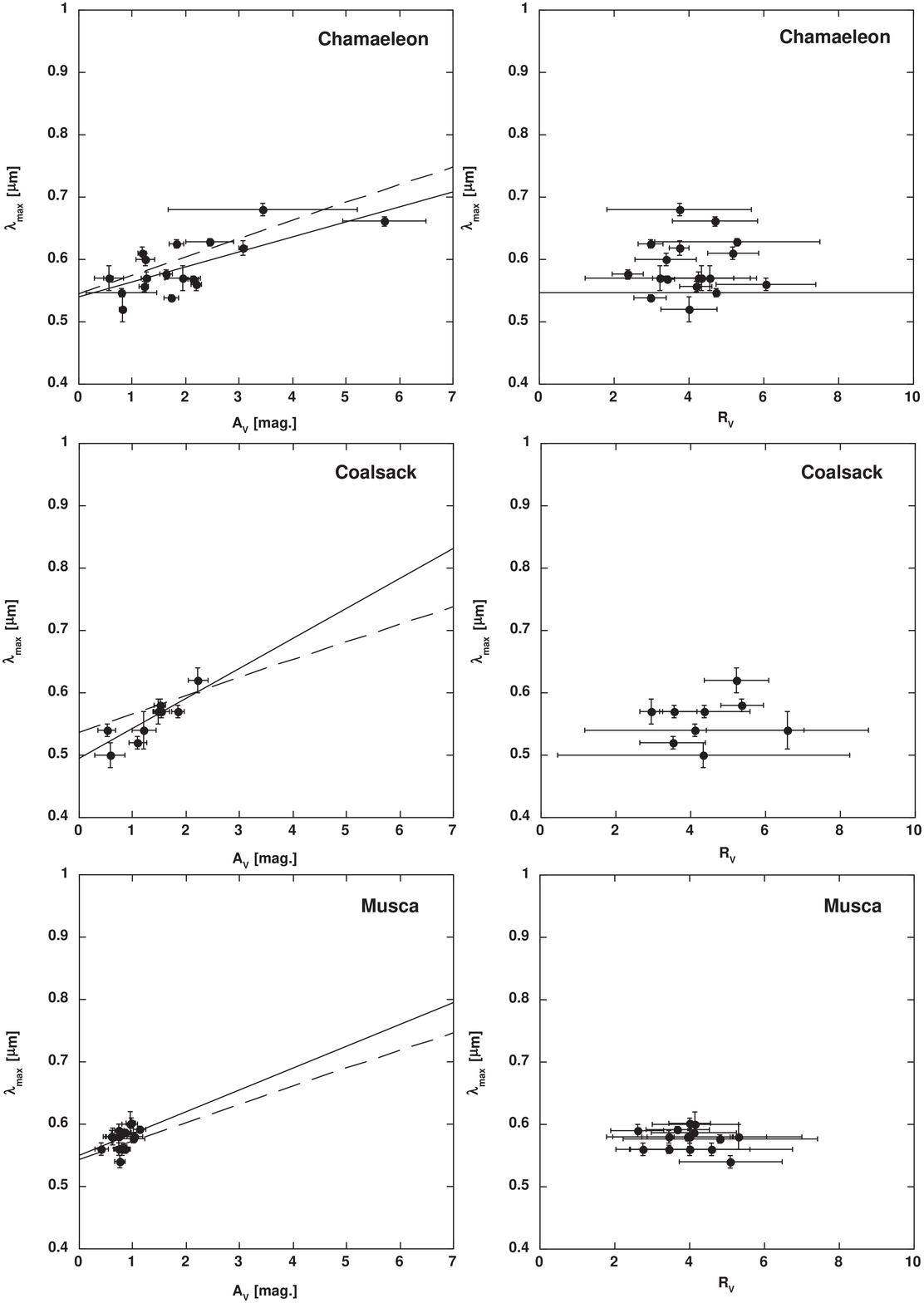}
\caption{Plots of $\lambda_{max}$ vs. A$_V$ for the debiased sample for Chamaeleon, Coalsack and Musca.  For each cloud, the best fit (using a robust linear fit) is over plotted as a full drawn line.  The dashed lines represent the "universal slope" of 0.028 and zero intercepts determined from the average R$_V$ for the cloud. The Hipparcos screened sample yield similar plots}
\label{lmax_vs_av_masked_TYC}
\end{figure}

\clearpage
\begin{figure}
\plotone{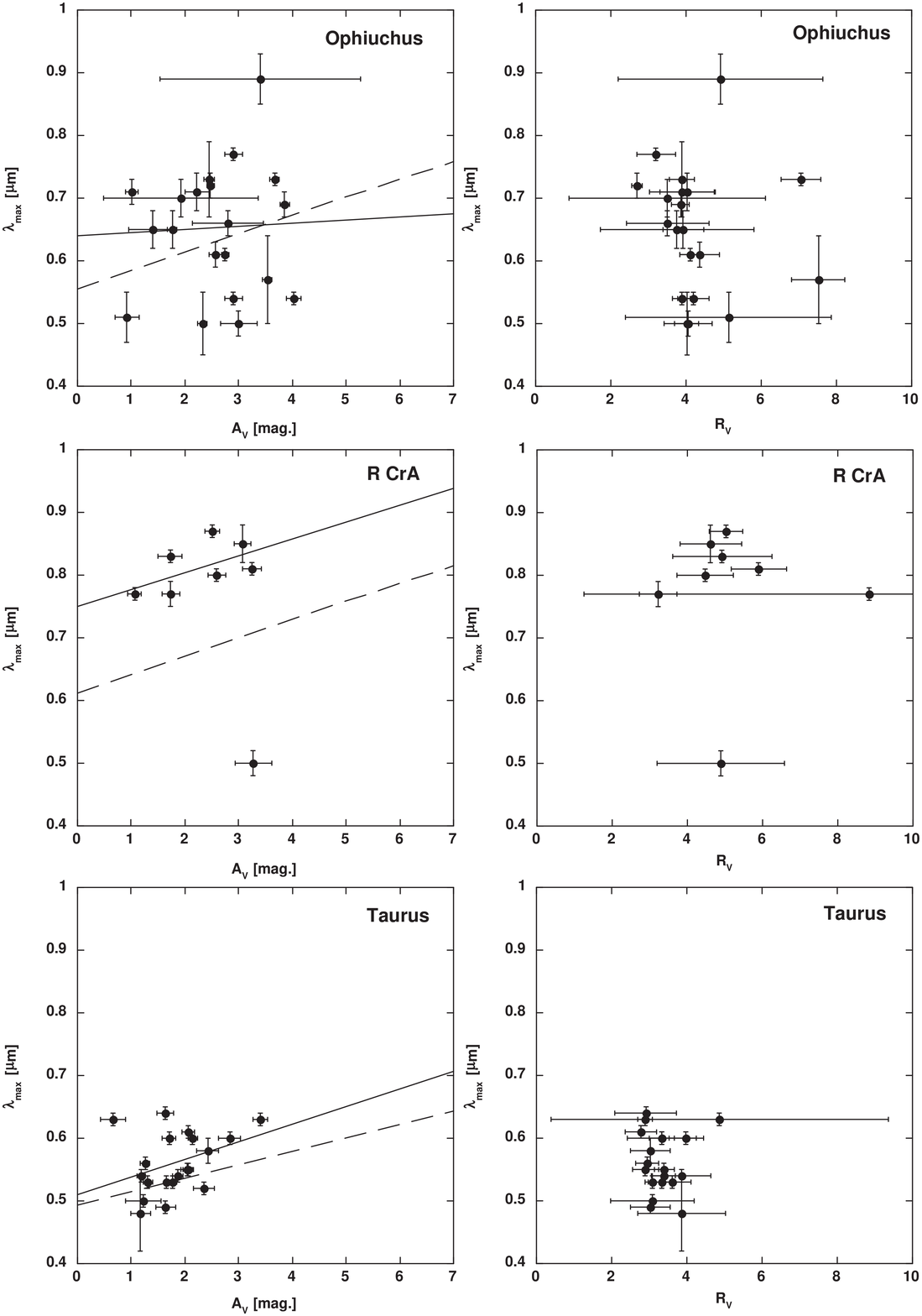}
\caption{Same as Figure \ref{lmax_vs_av_masked_TYC} for Ophiuchus, R~CrA and Taurus.  For Ophiuchus and Taurus the Hipparcos screened samples yield similar plots.  For R~CrA a significantly different selection is encountered which is illustrated in Figure \ref{lmax_vs_av_masked_RCrA}}
\end{figure}

\clearpage
\begin{figure}
\epsscale{1.00}
\plotone{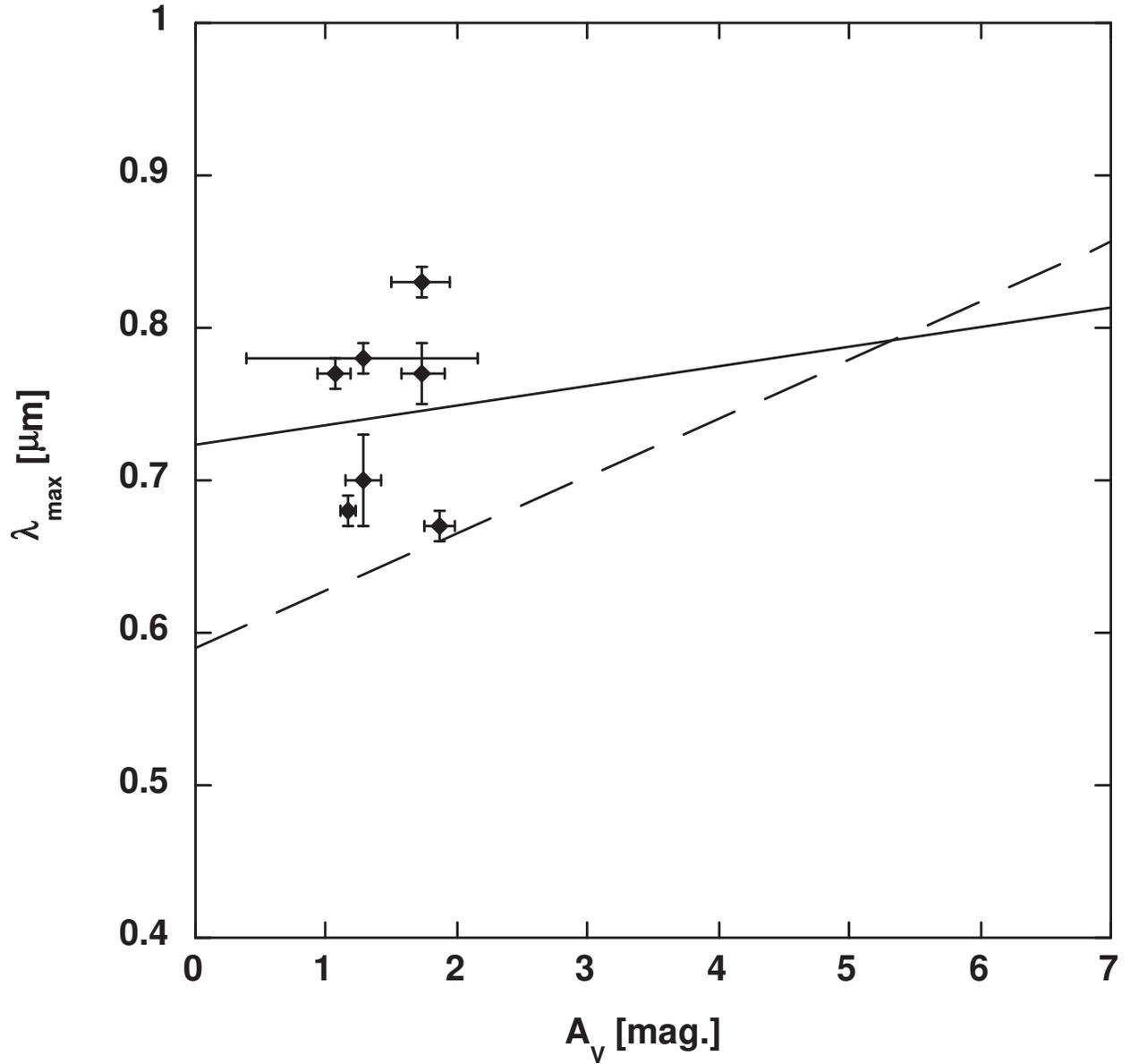}
\caption{Plot of $\lambda_{max}$ vs. A$_V$ for the debiased R~CrA sample screened based on the Hipparcos field stars.  As in Figure \ref{lmax_vs_av_masked_TYC}, the best fit (using a robust linear fit) is over plotted as a full drawn line.  The dashed line represents the "universal slope" of 0.038 and zero intercept determined from the average R$_V$ for the cloud.}
\label{lmax_vs_av_masked_RCrA}
\end{figure}

\clearpage
\begin{figure}
\plotone{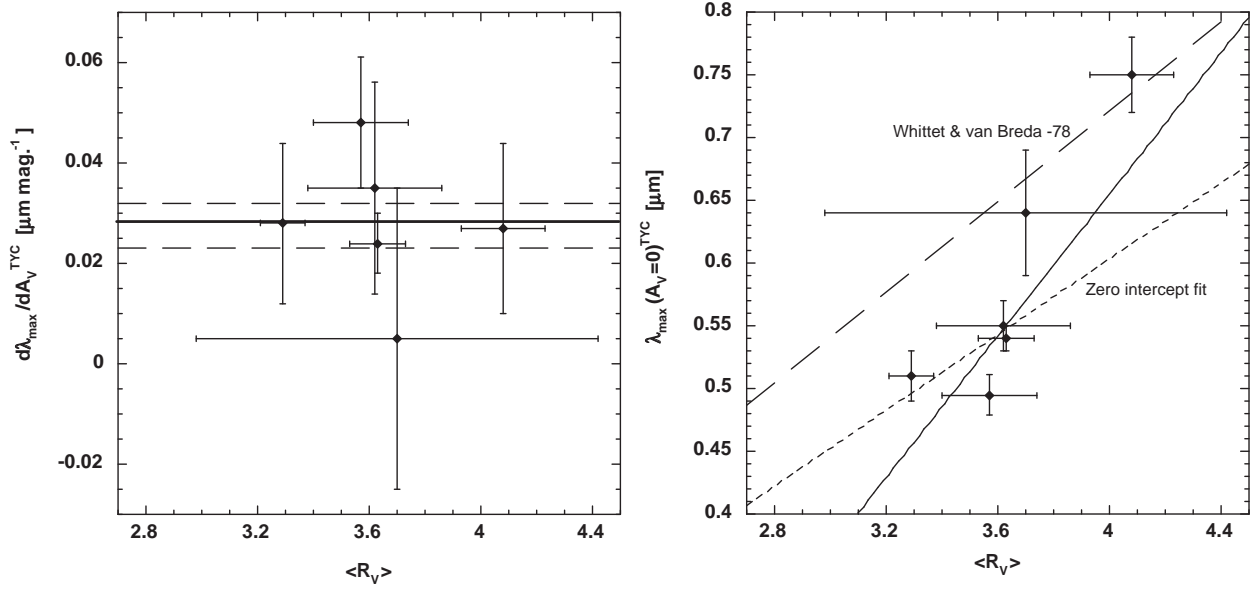}
\caption{The slope (left) and zero intercepts of the $\lambda_{max}$ vs. A$_V$ (right) for the six clouds under study are compared to the average of the total-to-selective extinction, using the Tycho screened samples.  The full drawn and dashed lines in the left-hand panel illustrate the weighted average of the slope and its uncertainty.  The full drawn line (right-hand panel) shows the best fit for $\lambda_{max}$ vs. A$_V$, allowing a non-zero intercept.  The short-dashed line shows the best fit assuming a zero intercept.  The long-dashed line corresponds to the R$_V$ vs. $\lambda_{max}$ correlation found by \citet{whittet1978}.}
\label{rvvslmax}
\end{figure}

\clearpage
\begin{figure}
\plotone{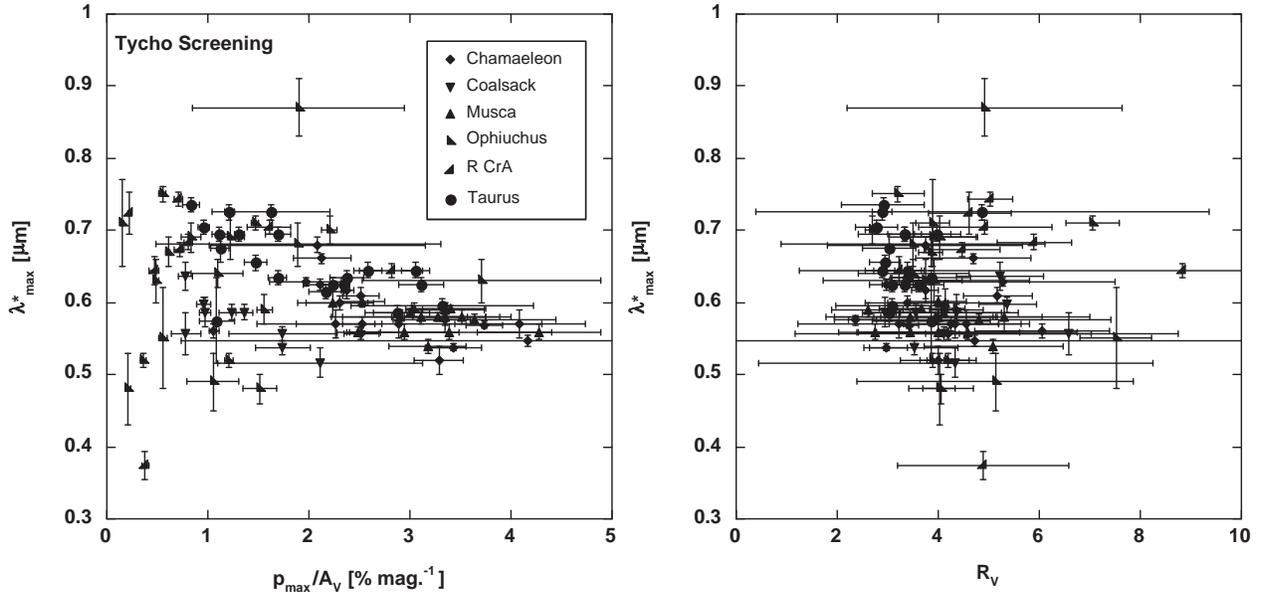}
\caption{The adjusted location of the peak of the polarization curve is shown as functions of alignment efficiency (p$_{max}$/A$_V$; left panel) and R$_V$ (right panel) for the Tycho screened sample.  In both cases the measured $\lambda_{max}$ has been adjusted to the Chamaeleon value by subtracting the difference in derived $\lambda_{max}(A_V=0)$ between the cloud and that for Chamaeleon for the sightlines in each cloud.  I.e. $\lambda_{max}^*$=$\lambda_{max}$-($\lambda_{max}$(A$_V$=0)$^{cloud}$-$\lambda_{max}$(A$_V$=0)$^{Cham}$).}
\label{lmax_vs_pdAmax_Rv}
\end{figure}

\clearpage
\begin{figure}
\plotone{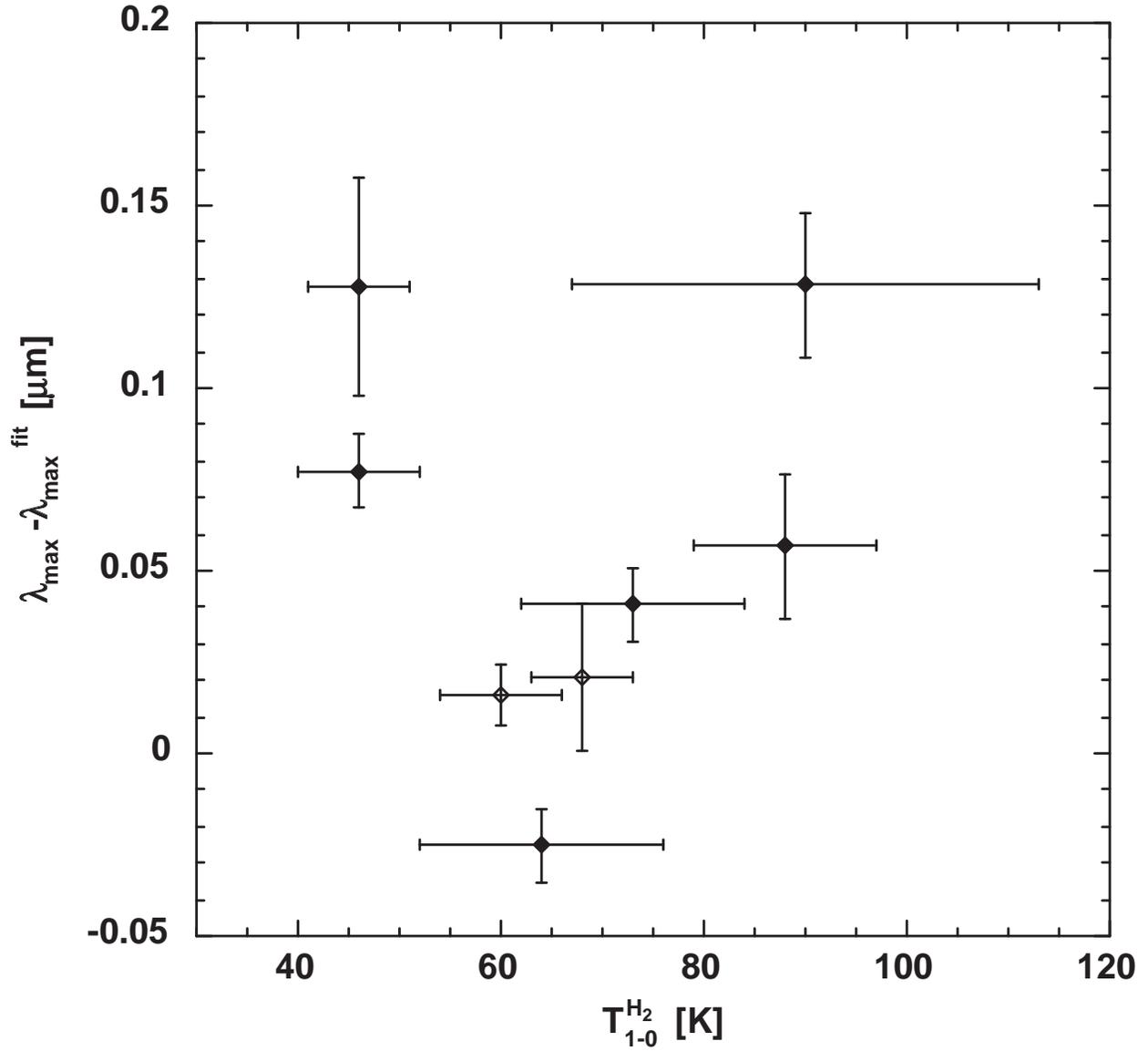}
\caption{The offsets from the $\lambda_{max}$ vs. A$_V$ fits are plotted as a function of the 1/0 H$_2$ temperatures for six stars in Ophiuchus (filled diamonds) and one each in Chamaeleon (open gray diamond) and the Coalsack (open black diamond).  No correlation is evident.}
\label{Dlmax_vs_T}
\end{figure}

\clearpage
\begin{figure}
\epsscale{0.70}
\plotone{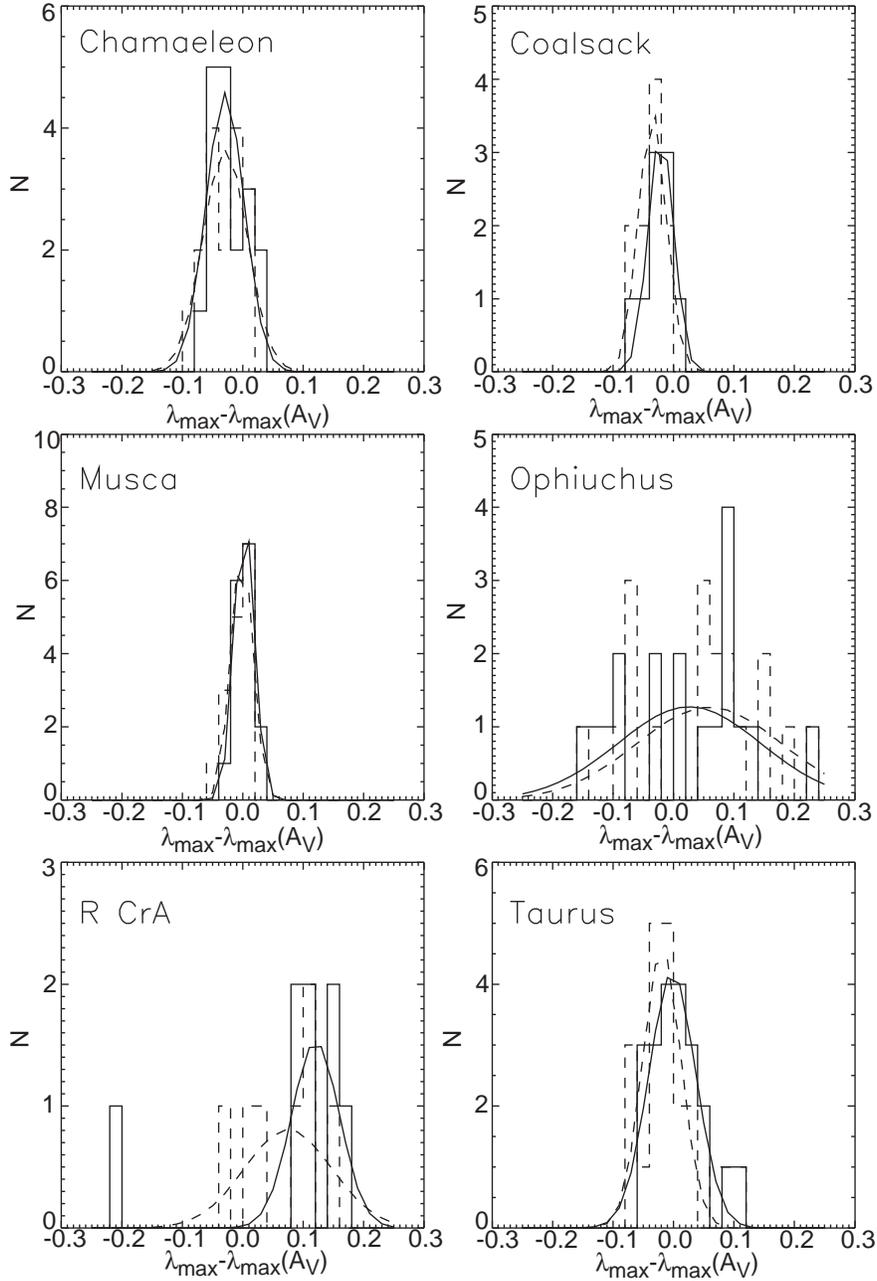}
\caption{The offsets from the fits in $\lambda_{max}$ vs. A$_V$ are shown in histogram form (full drawn histograms and lines for the Tycho screening, dashed histograms and lines for the Hipparcos screening).  The offsets are from the "universal slope" relations. }
\label{scatterhist}
\end{figure}

\clearpage
\begin{figure}
\epsscale{1.0}
\plotone{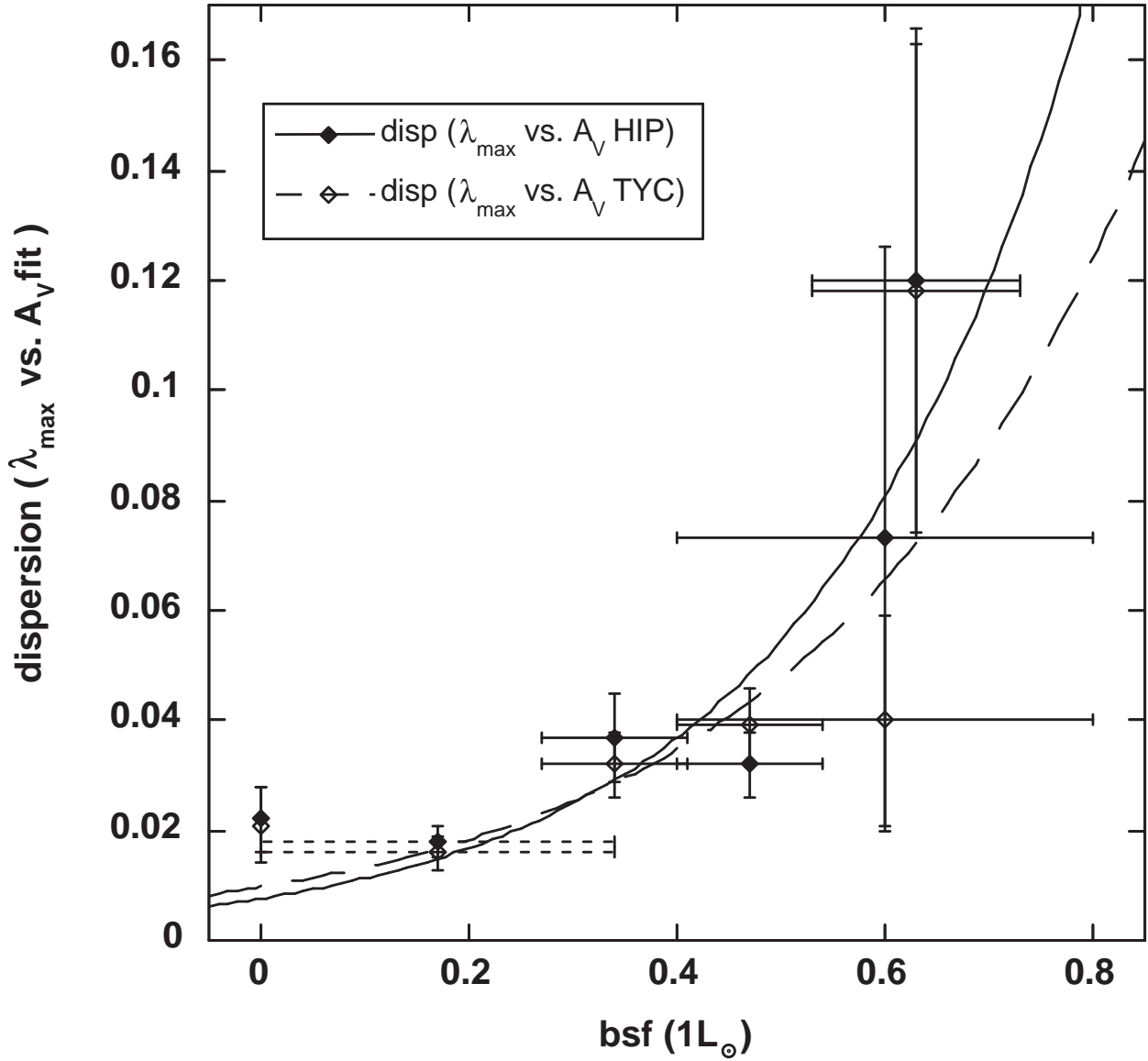}
\caption{The scatter in the fit of $\lambda_{max}$ vs. A$_V$ is plotted against the bright source fraction of the embedded young stellar objects (Open symbols for the Hipparcos screened sample and filled diamonds for the Tycho screened sample).  The curves are the best fit exponentials.}
\label{scattervsbsf}
\end{figure}

\clearpage
\begin{figure}
\plotone{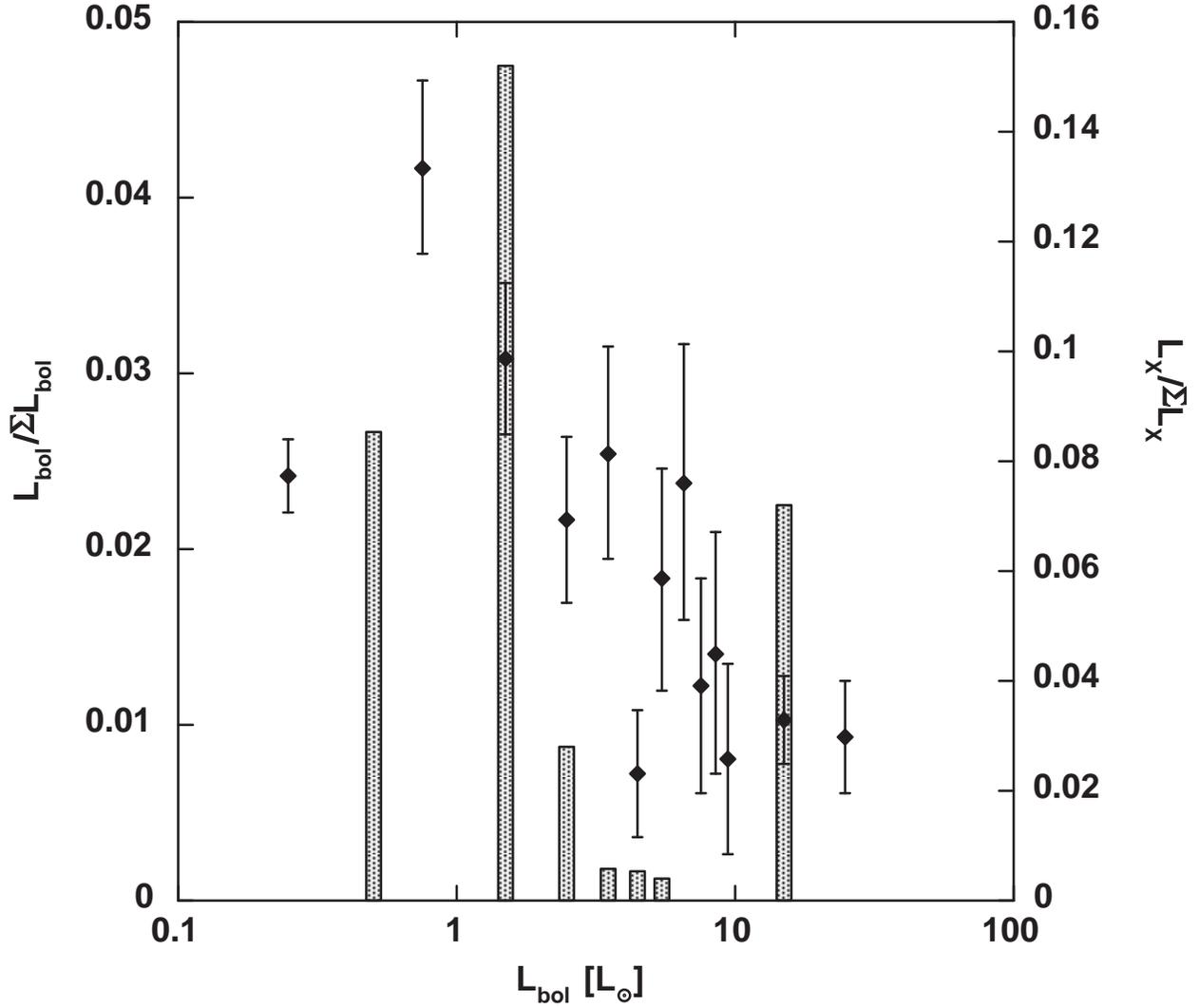}
\caption{The combined binned fractional bolometric luminosity function of YSOs in Chamaeleon Ophiuchus and Taurus is shown as diamonds \citep{chen1997}.  This shows the fraction of the total YSO luminosity, per unit luminosity, originating from a given sub sample of YSOs.  The binned fractional X-ray luminosity function of Ophiuchus YSOs is plotted as histograms.  The binned fractional luminosity functions peak at $\sim$1-2 L$_\odot$}
\label{YSO_lum}
\end{figure}

\clearpage
\begin{figure}
\plotone{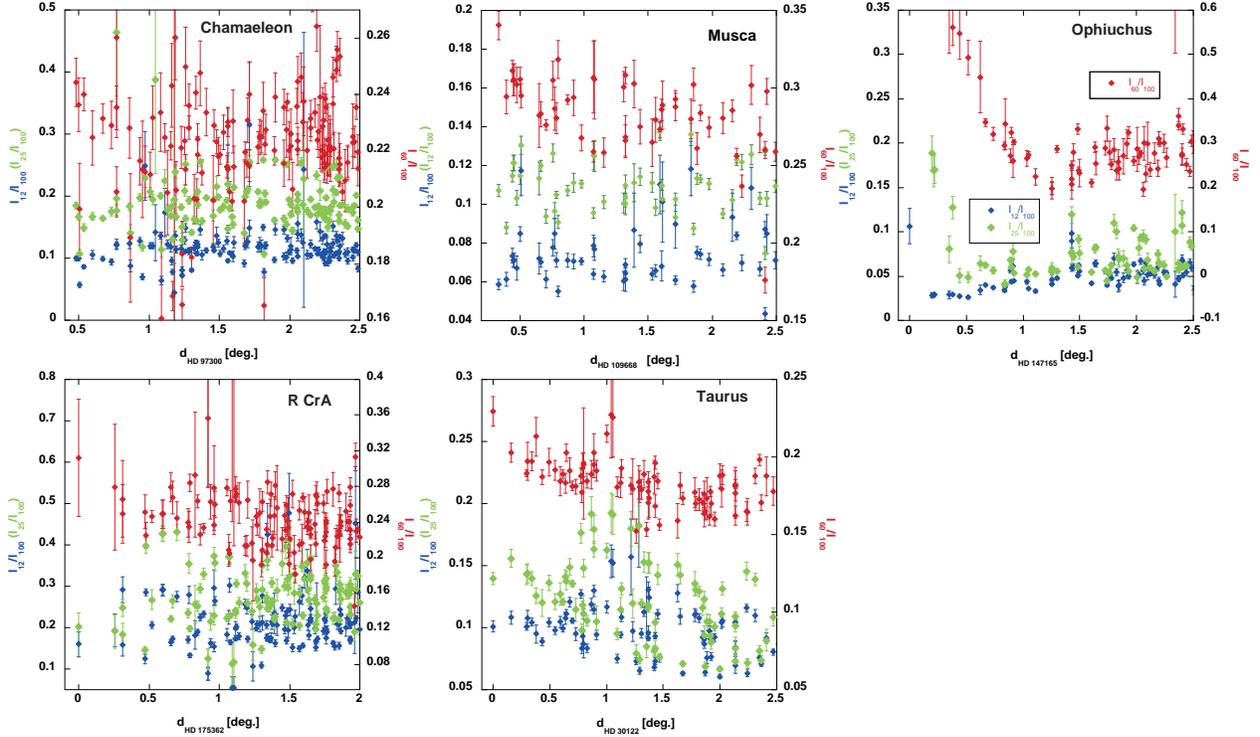}
\caption{The ratios of I$_{12}$/I$_{100}$ (blue), I$_{25}$/I$_{100}$ (green) and I$_{60}$/I$_{100}$ (red) are plotted for Chamaeleon, Musca, Ophiuchus, R~CrA and Taurus, against the on-the-sky distance to the dominant nearby hot star.  As discussed in the main part of the text, the 60 $\mu$m to 100$\mu$m ratio responds noticeably to the presence of the hot, high luminosity star.  The 25 to 100 $\mu$m ratio shows a smaller response, while the 12 to 100 $\mu$m ratio shows very little response to the presence of the hot stars.  This is consistent with an origin in dust heating through irradiation of the grains, but would be difficult to explain in terms of changes in the relative abundances of the different grain populations.}
\label{FIR_both_vs_Av}
\end{figure}

\clearpage
\begin{figure}
\plotone{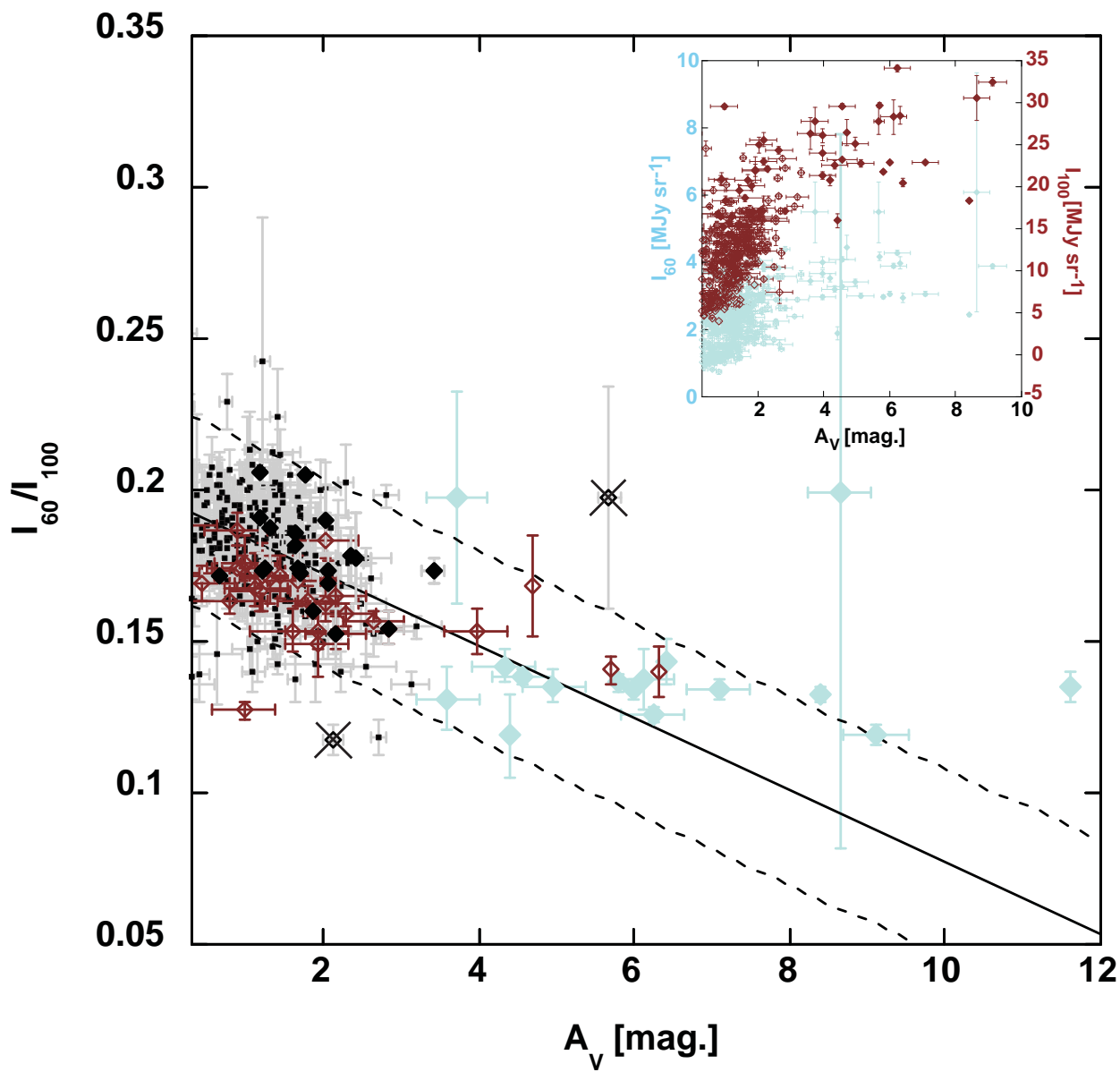}
\caption{The ratio I$_{60}$/I$_{100}$ vs. A$_V$ is plotted for Taurus, with the sightlines from \citet{murakawa2000} and \citet{teixeira1999} added.  Black dots with grey error-bars represent the Tycho field stars, black diamonds the polarization targets use in this paper (sightlines screened out as being anomalous are indicated with an X), red open diamonds represent sightlines with measured $\tau$(H$_2$O)$<$0.05 and filed blue diamonds $\tau$(H$_2$O)$>$0.1.  The insert shows the 60 and 100$\mu$m intensities for each sightline.}
\label{tau_deep}
\end{figure}

\clearpage
\begin{figure}
\plotone{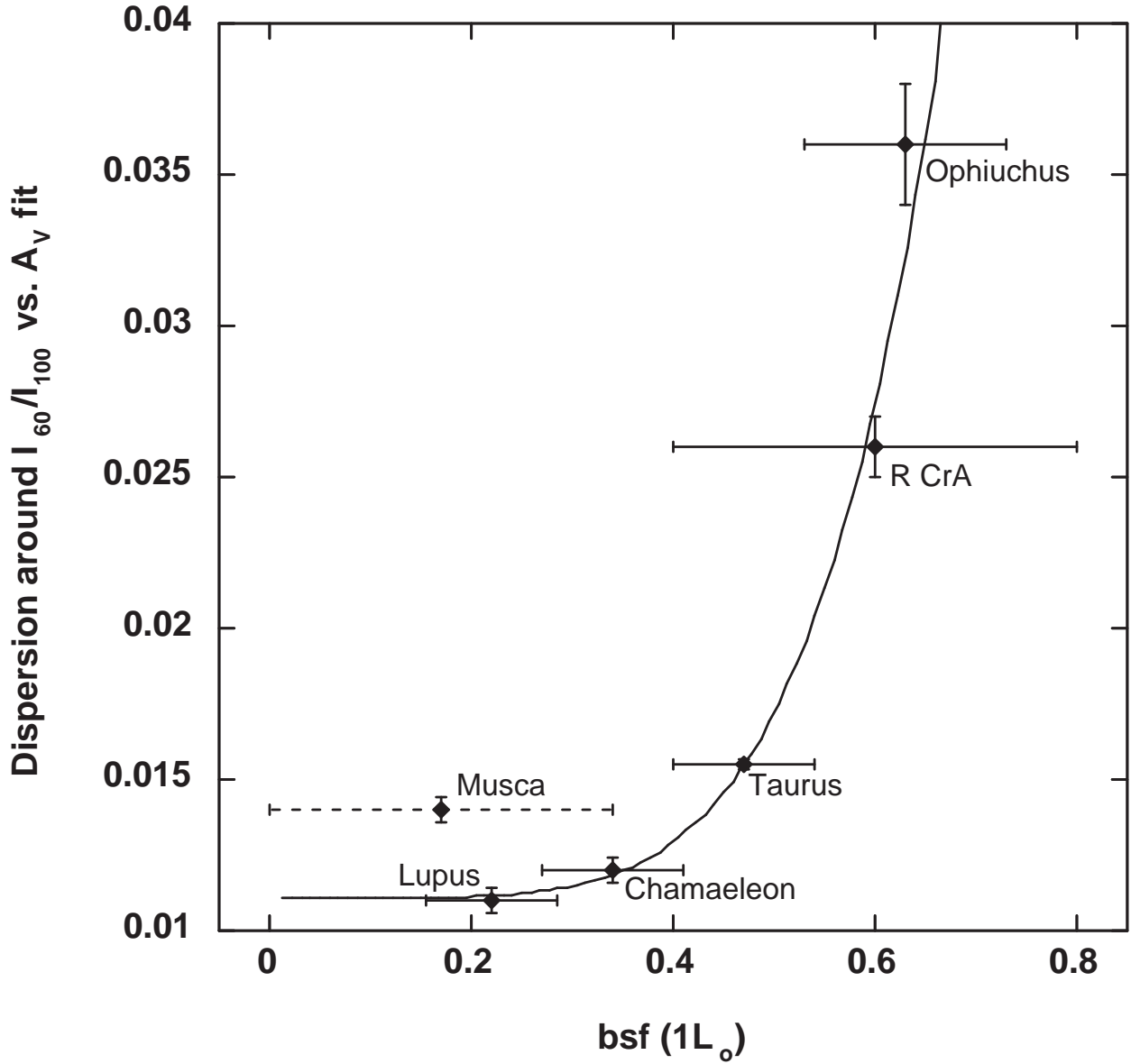}
\caption{The dispersions around the best I$_{60}$/I$_{100}$ fits (Figure \ref{fir_vs_av}) using the Tycho based screening are plotted as a function of the bright star fraction for each cloud (we have also added the results for a similar analysis of the Lupus I cloud).  A very well defined sequence is found, well fit by a power law function.
\label{disp_FIR_vs_Av}}
\end{figure}

\end{document}